\documentclass[%
 aip,
 jmp,%
 amsmath,amssymb,
 reprint,%
]{revtex4-1}

\usepackage{graphicx}
\usepackage{dcolumn}
\usepackage{bm}
\usepackage{graphics}
\usepackage{graphicx}
\usepackage{dcolumn}
\usepackage{epsfig}
\usepackage{bm}
\usepackage{color}

\usepackage{amssymb}
\usepackage{lineno}
\usepackage{epsfig}
\usepackage{subfigure}

\begin{document}

\preprint{AIP/123-QED}

\title{Electronic topological transitions in Nb$_3$X (X = Al, Ga, In, Ge and Sn) under compression investigated by first principles calculations}

\author{P. V. Sreenivasa Reddy}
\author{V. Kanchana}
  \email{kanchana@iith.ac.in}

\affiliation{%
Department of Physics, Indian Institute of Technology Hyderabad, Kandi-502285, Sangareddy, Telangana, India.}
\author{G. Vaitheeswaran}
\affiliation{
Advanced Centre of Research in High Energy Materials (ACRHEM), University of Hyderabad, Prof. C. R. Rao Road, Gachibowli, Hyderabad - 500 046, Telangana, India.
}
\author{P. Modak}
\author{Ashok K. Verma}
\affiliation{
High Pressure and Synchrotron Radiation Physics Division, Bhabha Atomic Research Centre, Trombay, Mumbai - 400 085, India.
}%

\date{\today}

\begin{abstract}
First principles electronic structure calculations of A-15 type Nb$_3$X (X = Al, Ga, In, Ge and Sn) compounds are performed at ambient and high pressures. Mechanical stability is confirmed in all the compounds both at ambient as well as under compression from the calculated elastic constants. We have observed four holes and two electron Fermi surfaces (FS) for all the compounds studied and FS nesting feature is observed at M and along X - $\Gamma$ in all the compounds. A continuous change in the FS topology is observed under pressure in all the compounds which is also reflected in the calculated elastic constants and density of states under pressure indicating the Electronic topological transitions (ETT). The ETT observed at around 21.5 GPa, 17.5 GPa in Nb$_3$Al and Nb$_3$Ga are in good agreement with the anomalies observed by the experiments.

\end{abstract}

\maketitle

\section{\label{sec:level1}Introduction}

Eversince the discovery of superconductivity in V$_3$Si with T$_c$ $\sim$17 K in the year 1953 by Hardy and Hulm\cite{Hardy}, the family of compounds with composition A$_3$B (A= V, Nb, Cr, Ti, Mo, Zr, Ta, W to Hf and B= Al, Ga, Ge, In, Sn etc) had attracted considerable attention of researcher as some of them possess quite high superconducting transition temperature (T$_c$). The interest in these compounds are not only due to the rather high T$_c$ but also their high critical current density and critical magnetic field, along with acceptable mechanical properties make them viable for applications. These compounds crystallize in the A15 type crystal structure (see Fig.1), where A atoms form three mutually orthogonal chain like structure parallel to the edges of the unit cell.

Some of these compounds undergo cubic to tetragonal martensitic transformation near to their superconducting transition temperatures T$_c$\cite{Batterman}. For example, the martensitic transition temperatures of V$_3$Si (21 K) and Nb$_3$Sn (45 K) are close to their respective superconducting transition temperatures 17 and 18 K. Acoustic phonon instabilities were found to be responsible for martensitic transition in previous studies\cite{Testardi}. A similar behaviour was also seen in the Nb$_3$Al$_x$Ge$_{1-x}$\cite{ref1}, V$_3$Ga\cite{ref2,ref3}, V$_3$Ge\cite{ref3} and Nb$_3$Al\cite{ref4} compounds. Experiments\cite{Shirane} also indicated a dimerization of the transition-metal chains accompanied by a tetragonal distortion of the lattice during the transformation. It has been proposed that the tetragonal transformation is driven by band Jahn-Teller like mechanism. These A15 compounds exhibits different behaviour in electronically derived properties at low temperatures such as knight shifts, electrical resistivity etc \cite{Klein}. This unusual behaviour of various properties of A15 family compounds has been related to the sharp peak in electronic density of states near to the Fermi level arising from the $d$ states of the transition metal atoms\cite{Klein}. Hence it is clear that many properties of these compounds are related to their electronic structures.

Recent experiments\cite{Tanaka} have explored the possible relationship between superconductivity and martensitic transition in V$_3$Si and Nb$_3$Sn compounds by measuring electrical resistance and specific heat under high pressure. They have observed that initially T$_c$ increases with pressure and merges with martensitic transition temperature at high pressure where further T$_c$ increase is ceased and concluded that the martensitic transition play an important role in superconductivity of these compounds. So to understand this, a thorough understanding of electronic structures is necessary. So far only a few electronic structure studies are available\cite{Pickett,Rajagopalan,Paduani,Sundareswari} but a systematic study especially under high pressure is still lacking. Some elements like Cd\cite{Srinivasan} and Co\cite{Kvashnin} show a continuous change in FS topology under pressure which highlights the ETT in these elements. In this work we have carried out a systematic study of electronic properties under pressure for Nb based A15 compounds namely Nb$_3$X (X = Al, Ga, In, Ge and Sn).

Since any property that involves the conduction electrons in a metal must depend on the shape of the Fermi surface and on the wave functions of the electrons at or near the Fermi surface of that metal\cite{Fermi surface} it will be interesting to study the Fermi surface topology and their pressure variations for these compounds. Further, Bilbro and McMillan\cite{Bilbro} have studied the interaction of charge density wave (CDW) and superconductivity (SC) in A15 materials within mean field approximation and predicted that both states compete with each other for developing their respective gaps in the same Fermi surface. In Nb$_3$Sn, opening of a charge-density wave gap is observed and expected that it would be due to the dimerization of Nb atoms and nesting at the Fermi surface\cite{Escudero}, and this further indicate the possibility of nesting feature in these type of compounds. Charge density wave, Fermi surface nesting and Peierls instability may be inter-related in these compounds. Calculation of the susceptibility\cite{Gruner}  $\chi(\vec{q}, \omega)$ for a given system is one of the way to understand the possibility of Fermi surface nesting and formation of CDW. Zero energy value of the Lindhard response function $\chi_{0}(\vec{q})\equiv \chi_{0}(\vec{q}, \omega=0)$ can be used to determine the presence of Fermi surface nesting. There should be a peak in the imaginary part of the response function Im\big[$\chi_{0}(\vec{q})$\big] at the Fermi surface nesting vector. The formation of charge density wave requires a large real part of the susceptibility, Re\big[$\chi_{0}(\vec{q})$\big]. Hence in this work we have also calculated the Lindhard response functions $\chi_{0}(\vec{q})\equiv \chi_{0}(\vec{q}, \omega=0)$ for these compounds to study Fermi surface nesting and possible CDW instability in these systems. 

We have predicted Fermi surface nesting for all the compounds at ambient as well as under pressure. The Fermi surface nesting is found to enhance under pressure for all the compounds except Nb$_3$Sn. We have also predicted an electronic topological transition (ETT) at different compressions for all the compounds without equation of state (EOS) anomaly.

\section{\label{sec:level1}Computational details}

Density functional calculations have been performed in the present work to calculate the electronic structure and elastic constants. The Full-Potential Linearized Augmented Plane Wave (FP-LAPW) method as implemented in WIEN2k \cite{Blaha} code is used. We have used PBE-GGA \cite{JPPE} (Perdew-Burke-Ernzerhof parametrization of the Generalized Gradient Approximation) approximation for the exchange correlation potential. Throughout the calculations, the R$_{MT}$ (radius of muffin tin spheres) value for each atom was fixed as 1.90 $a.u$ for Nb, 1.80 $a.u$ for Al and 2.10 $a.u$ for both Ga and Ge, 2.20 $a.u$ for In and 2.25 $a.u$ for Sn atoms. For the energy convergence, the criterion R$_{MT}$*K$_{max}$=9 was used, where K$_{max}$ is the plane wave cut-off. The potential and charge density were Fourier expanded up to G$_{max}$=12 $a.u^{-1}$. For band structure and density of states calculations we have taken total 20000 k points in the Monkhorst-Pack\cite{Monkhorst} scheme which gives 560 k-points in the irreducible part of the Brillouin Zone (BZ). Tetrahedron method \cite{tetra} was used to integrate the Brillouin zone. Energy convergence up to 10$^{-5}$ $Ry$ is used to get the proper convergence of the self consistent calculation. All Fermi surface and Lindhard functions are calculated with $44\times 44\times 44$ k-mesh to get smoother Fermi surfaces and accurate Lindhard function. Birch-Murnaghan \cite{FBIR} equation of state was used to fit the total energies as a function of primitive cell volume to obtain the bulk modulus and the equilibrium lattice parameter for the investigated compounds. We have checked the effect of spin-orbit coupling (SOC) and have not found any significant changes at the Fermi level with the inclusion of SOC. Further calculations are performed without including SOC. Real and imaginary part of the Lindhard response function $Re[\chi(q)]$ for $\omega=0$ is calculated directly from the energy eigenvalues using,

$Re[\chi(q)] = \sum_{kjj'}\frac{f_{kj}-f_{k+qj'}}{\in_{kj}-\in_{qj'}}$

and 

$Im[\chi(q)] = \sum_{kjj'}\delta(\in_{F}-\in_{kj})\delta(\in_{F}-\in_{k+qj'})$

Where $f$ is Fermi-Dirac distribution function, $\in_{kj}$ and $\in_{qj'}$ are the energy eigenvalues for band indices $j$ and $j'$, and $\in_{F}$ is the Fermi energy. We have evaluated these functions within constant matrix element approximation and considered only those bands which cross the Fermi level.

\section{\label{sec:level1}Results and Discussions}
Calculated ground state properties of Nb$_3$X (X = Al, Ga, In, Ge and Sn) are listed in Table I together with available experimental and theoretical results. Calculated values are in good agreement with the experiments\cite{Book,Viswanathan,Mkrtcheyan,Kawamura} as well as with earlier theoretical results\cite{Paduani,Klein,Kawamura}. Calculated bulk moduli for Nb$_3$Al and Nb$_3$Ga are underestimated as compared to experimental\cite{Yu,Mkrtcheyan} results but for Nb$_3$Ge the calculated bulk modulus is over estimated. However our calculated bulk moduli are in good agreement with earlier theoretical results\cite{Paduani,Sundareswari,Rajagopalan}. Nb$_3$Ge has the highest bulk modulus and Nb$_3$In has the lowest bulk modulus. Calculated bulk modulus increases as we move from compounds containing IIIA group elements (X = Al, Ga, In) to IVA group (X = Ge, Sn). 
Single crystal elastic constants were calculated to establish the mechanical stability of these compounds and the calculated values are given in the Table II together with available data\cite{Rajagopalan,Sundareswari}. Calculated values satisfy the Born mechanical stability criteria \cite{BORN} i.e. $C_{11}>0$, $C_{44}> 0$, $C_{11} > C_{12}$, and $C_{11} + 2C_{12}> 0$. polycrystalline elastic constants were also calculated from the single crystals elastic constants\cite{V.KANCHANA,VGAAM,VGAA,VGA} and these values are listed in the same table. Calculated Young's modulus is highest for Nb$_3$Sn indicating the stiffness of this among other compounds. The presence of elastic anisotropy\cite{V.KANCHANA} in the present compounds is also confirmed by calculating the anisotropy factor(A). Calculated positive values of Cauchy's pressure($C_{12}-C_{44}$) indicate the ductile nature of the present compounds. This ductility is also confirmed from the calculated Pugh's ratio ($\frac{G_H}{B}$) \cite{Pugh} value. These Pugh's ratio values are less than 0.57 which is known as critical number to separate brittle and ductile nature. The Poisson's \cite{Wortman} ratio indicate the stability of the crystal against shear and takes the values in between -1 to 0.5, where -1 and 0.5 serve as lower and upper limits respectively. From the calculated Poisson's ratio values, we observed that all the compounds have the value closer to the upper limit indicating the stiffness of the present compounds.

  The calculated density of states (DOS) are plotted in Fig. 2 along with partial density of states for all compounds. We have also tabulated the total DOS at F$_F$ (N(E$_F$)) for each compound in Table-I and compared with earlier calculations. The calculated N(E$_F$) and overall DOS features agree well with available data\cite{Paduani,Klein,Paduani1}. In all the compounds there is a pseudo gap on both the sides of the Fermi level which originates mainly due to crystal field splitting of Nb-$d$ states. The Fermi level shifts towards right (i.e from shoulder to peak) in total DOS for compounds containing group IVA elements compared to those containing group IIIA elements and is consistent with the fact that group IVA elements have one extra electrons. The total DOS at Fermi level is found to increase in both IIIA and IVA group as we move from top to bottom of the periodic table and is evident from Fig.2 as the Fermi level shifts towards the peak in total DOS. From partial DOS it is clearly seen that there is strong hybridisation between Nb-$d$ and X-$p$ states close to Fermi level which causes splitting of X-$p$ DOS exactly similar to t$_{2g}$ and e$_g$ splitting of Nb-$d$ states. In addition the covalent nature between Nb and X atom is also observed from the DOS plots.
We have also calculated Sommerfeld coefficient $\gamma$ and given in Table-I along with available data. The calculated $\gamma$ values are in good agreement with available data\cite{Rajagopalan} for Nb$_3$Al and are proportional to N(E$_F$).

	Calculated band structures for Nb$_3$Sn along some high symmetry directions in the Brillouin zone (high symmetry directions are given in Fig. 1(b)) with and without including spin orbit coupling (SOC) effect is shown in Fig. 1(c). As we did not find significant effect of SOC on bands close to the Fermi level (E$_F$), further calculations are performed without including the SOC, and the electronic band structure for all the compounds without SOC are shown in Fig. 3. Overall band profile is similar for all the compounds. In all the compounds we have observed multiple degenerate bands at R point. Close to E$_F$, the bands originate from the hybridization of Nb-$d$ and X-$p$ states and is consistent with our findings from partial DOS. From the keen observation of band structure, shifting of bands above the E$_F$ around R point is observed when we move from compounds possessing IIIA group elements (X = Al, Ga, In) to IVA group (X = Ge, Sn). In all the investigated compounds, total six bands are crossing the E$_F$ and are indicated in different colour with their numbers. In that first four bands are of hole nature as the bands are crossing from valence to conduction band and the remaining two bands are having electron nature. In the case of Nb$_3$Al, first four bands which are of hole nature and indicated with numbers 45 to 48 are crossing E$_F$ at M point and in addition to this, the bands 47 and 48 are crossing E$_F$ along R-X as shown in Fig. 3(a). In the case of Nb$_3$Ga and Nb$_3$In these bands are indicated with numbers from 49 to 52, as in Fig. 3(b, c) and the scenario at M point is the same as Nb$_3$Al but at R point these bands are shifted above the E$_F$. This shifting is more in Nb$_3$In when compared with Nb$_3$Ga which is evident from band structure in Fig. 3(c). The last two bands, which are having electronic nature indicated with numbers 49 and 50 in Nb$_3$Al are crossing E$_F$ at $\Gamma$, along R-X and at X point. In addition to this, band number 49 is crossing E$_F$ along $\Gamma$-M. In the case of Nb$_3$Ga and Nb$_3$In these two bands are indicated with numbers 53 and 54. For these two bands the scenario is same as in Nb$_3$Al except along R-X, where the band crossing is absent due to shifting of bands above the E$_F$. 

The band structure for the compounds Nb$_3$Ge and Nb$_3$Sn is given in Fig. 3(d, e) where X is replaced with IVA group elements. These compounds are having one extra electron per formula unit compared to the compounds which are having X = Al, Ga and In. If we move from IIIA to IVA elements containing compounds, i.e from Nb$_3$Ga to Nb$_3$Ge and Nb$_3$In to Nb$_3$Sn, we can observe the shifting of bands below E$_F$ due to one extra electron in latter compounds. In the case of Nb$_3$Ge, first four hole natured bands are crossing the E$_F$ only at M point. In Nb$_3$Sn, behaviour of these bands at M is similar as Nb$_3$Ge and  are also crossing the E$_F$ at R point due to shifting of bands above E$_F$ when compared to Nb$_3$Ge. The next two electronic natured bands in Nb$_3$Ge are crossing E$_F$ at $\Gamma$, along R-X and at X. In addition to this band 53 is crossing E$_F$ along $\Gamma$-M and near R. In Nb$_3$Sn, these bands have similar behaviour as Nb$_3$Ge except near R point due to shifting of bands above the E$_F$ at that point. In Nb$_3$Al, the calculated band structure properties agree well with reports by Rajagopalan\cite{Rajagopalan}, and for the remaining compounds our study is in good agreement with that of Paduani\cite{Paduani1}.

Fermi surface for the investigated compounds are given in Fig. 4 to Fig. 8. In the case of Nb$_3$Al first four FS are having hole nature and among them first two FS have pockets at M point (Fig. 4(a,b)) due to bands crossing the E$_F$ only at this point. The next two FS (Fig. 4(c,d)) are due to the bands (47, 48) crossing E$_F$ both at M and along R-X. Due to this we have sheet like FS along M-R and the last two FS are having electron nature. In these last two FS we have pockets at $\Gamma$, R, X and along R-X due to the crossing of the two bands (49, 50) at these high symmetry points as discussed in band structure. In the case of Nb$_3$Ga and Nb$_3$In, we have observed the presence of pocket/sheet at/near R point in the first three FS and absence of pockets at same R point in the remaining three FS as compared to Nb$_3$Al is due to the shifting of bands above the E$_F$ as discussed before and the FS are shown in Fig. 5 and 6. In these two compounds the first FS is similar to Nb$_3$Al except having an extra pocket at R point. The next two FS are having ribbon like sheet along R-M which is due to the bands (50, 51) continuously residing in the conduction band. Remaining three FS in these two compounds are similar to Nb$_3$Al except for the absence of pockets at R points (Fig. 5(d, e, f) and Fig. 6(d, e, f)). When we move from IIIA to IVA group elements (Nb$_3$Ga to Nb$_3$Ge and Nb$_3$In to Nb$_3$Sn), decrease in the width and length of sheet near M point in hole natured FS (first four FS) is observed which is due to the shifting of bands below E$_F$ which will cause the reduction in the area of bands in the conduction region. In the case of electron natured FS, an opposite trend is observed due to increase in the area of bands in valence region. A drastic change in FS topology is revealed in these two FS, resulting from an extra electron of IVA elements in these compounds. In the case of Nb$_3$Ge, hole natured FS have pockets only at M point due the bands (49 to 52) as discussed in band structure. In the case of Nb$_3$Sn, these FS have extra pockets at R due to the shifting of bands above E$_F$. Last two FS in both of the compounds have electronic nature due to bands (53, 54) crossing E$_F$ from conduction to valence band as discussed in the band structure. The calculated FS are in good agreement with the Paduani's \cite{Paduani1} study but in the case of  Nb$_3$Ge, the author's found only five Fermi surfaces, but in our calculations we have six FS corresponding to six bands crossing the E$_F$\cite{Paduani1}. From the FS of the investigated compounds, we observed hole pockets at M point in the first three FS of Nb$_3$Al, Nb$_3$Ge and Nb$_3$Sn and in the remaining compounds it is observed only in the first FS. In addition to this, we have also observed parallel sheet like structures in the last two FS of all the compounds. This might indicate the probability of nesting at M and along X - $\Gamma$ in all the compounds. Particularly in Nb$_3$Al (FS for band no. 49 (Fig. 4(e))), Nb$_3$Ga (FS for band no. 53 (Fig. 5(e))) and Nb$_3$In (FS for band no. 53 (Fig. 6(e))) we observed the nesting along X - $\Gamma$. 
To confirm this nesting feature we have calculated the Lindhard susceptibility for all the investigated compounds and are given in Fig. 14. From the imaginary part of susceptibility plots, we have observed peaks at M point and along X - $\Gamma$ in all the compounds indicating the nesting feature at these points. In all the compounds, the peak at M point in Im($\chi_{0}$) is due to the hole pockets at M point in the FS. The remaining peaks along X - $\Gamma$ are due to the last two FS which are having electronic nature in all the compounds. From the calculated real part of susceptibility plots, we have not observed any peaks except for Nb$_3$Sn where one sharp peak is observed at R point. One can speculate this as the probability for the CDW formation in Nb$_3$Sn as reported experimentally by Escudero et. al.\cite{Escudero}.

\section{\label{sec:level1}Pressure effect}

Pressure effect on the band structure, FS topology are given in supplementary data along with the manuscript for all the compounds. We have observed a continuous FS topology change in all the investigated compounds under compression. In the case of Nb$_3$Al\cite{Supplement}, major change in the band structure is found along R-X direction which is due to continuous shifting of bands below the $E_F$ with applied pressure. Due to this a continuous change in the FS topology is found in Nb$_3$Al up to the applied pressure. The changes are also reflecting in the calculated elastic constants and density of states under compression. Among those topology changes, we have given the band structure and maximum topology changed FS of Nb$_3$Al at V/V$_0$ = 0.90 in Fig 9. The bands related to these changes are 47, 48 and 49.
The major changes occurred along the X-R direction of the BZ where band no 47 and 48 descend through Fermi level under pressure causing destruction of Fermi sheets along this direction. Above this compression FS at X point is completely vanishing in this compound. 
From the calculated imaginary part of the susceptibility at this compression, we observe huge peaks at M point and at q=(0,0.2,0) indicating the pronounced nesting feature at this compression compared to ambient where the height of these peaks is very less. At M point it is due to the FS change observed in band 47 at this compression where we find flat sheet at M point in the FS, but this flatness of this sheet is absent at ambient. 
In the case of Nb$_3$Ga\cite{Supplement}, maximum change in the band structure is observed along M-R and R-X continuously with applied pressure due to the bands 50, 51 and 53 shifting below E$_F$ with applied pressure upto V/V$_0$=0.90 which results in maximum change in the FS topology. Above this pressure band 52 is also found to contribute more for the FS topology change. For Nb$_3$Ga, complete band structure at V/V$_0$=0.92 is given in Fig. 10.
From these figure, we observed that the bands 50 and 51 descend through Fermi level along M-R direction of the BZ under compression causing an ETT. The ribbon like single Fermi sheets corresponding to band 50 and 51 parallel to M-R direction of BZ, now splitted into two due to destruction of its part under compression. The calculated susceptibility plots at this compression for Nb$_3$Ga show the similar behaviour as Nb$_3$Al except, for the point that at M point the FS nesting decreased and at q=(0,0.13,0) it is found to increase.
In the same way, pressure effect on the band structure of Nb$_3$In\cite{Supplement} is also found to have more effect at M and along M-R. Upto V/V$_0$=0.90, pressure effect is more on the bands 50 and 51. Above this pressures remaining bands are also effected more, especially at M point, where we find the first two bands to shift completely down leading to the absence of the FS sheets at M point. In Fig. 11 we have given the band structure of Nb$_3$In at V/V$_0$=0.92, where we observe the change in band structure along M-R. As in Nb$_3$Ga the ribbon like single FS parallel to M-R correspond to band 50 is splitted in to two. The calculated imaginary part of susceptibility is found to increase a little at M point and more at q=(0,0.33,0).

 In Nb$_3$Ge\cite{Supplement}, effect of pressure on the band structure is found to be more along R-X where last two bands is found to shift continuously below the E$_F$. Due to these two bands (53 and 54) a continuous change in the FS topology is observed under pressure. To compare with ambient we have given the change in FS topology at  V/V$_0$ = 0.90 in Fig. 12.
At higher compression the FS nesting decreases in this compound as is evident from the imaginary part of susceptibility plot.
In the FS corresponding to the band 53, we observed change in FS topology near R point at this compression which is due to the increase in the electron concentration at that point, which is evident from the band structure given in Fig. 12(b).
In the case of Nb$_3$Sn\cite{Supplement}, band structure topology is found to change continuously along M-R and R-X due to the bands 51, 52, 53 and 54 under pressure. Due to this a continuous change in the FS topology is found under pressure upto V/V$_0$ = 0.90. Above this pressures the number of bands crossing E$_F$ is found to decrease from six to four. Band structure and FS of this compound at V/V$_0$ = 0.90 is given in Fig. 13.
From this, band 51 and 52 give rise to ETT near R point of the BZ around this compression as they descend through Fermi level and causing destruction of Fermi sheets around R. Here also like Nb$_3$Ge, the decrease of nesting property of the Fermi surfaces is evident from the imaginary part of the susceptibility plot. 
Under compression, real part of susceptibility is found to decrease in all compounds. In case of Nb$_3$Sn, at R point, the peak observed at ambient is found to be absent under compression possibly indicating the absence of CDW at this compression in this compound. Due to this FS topology change, we have observed the non monotonic variation in the DOS under compression in all the compounds as shown in Fig.15.

Now to study the effect of ETT on EOS and elastic properties we have  calculated EOS for Nb$_3$Al and Nb$_3$Ga and pressure effect on the elastic constants for all the compounds. The study of P-V relation for Nb$_3$Al and Nb$_3$Ga is particularly important because earlier experiments\cite{Yu,Mkrtcheyan} observed an anomaly near 19.2 and 15 GPa respectively. Figure 16 shows the calculated P-V relation for these two compounds along with earlier experimental data. It is clear from the figure that the calculated P-V curve varies smoothly and there is no anomaly for both of these compounds. Since ETT is an subtle electronic transition, its effect on EOS is expected to be very weak and may be washed out during fitting\cite{Godwal}. However its effect may be pronounced in the pressure variation of elastic constants. Hence we have calculated elastic constants for these compounds at different pressures and are shown in Fig.17. The calculated elastic shear modulus (C$_s$ = (C$_{11}$-C$_{12}$)/2) is also given under pressure. From the plots it is observed that values of elastic constants increase with pressure. The effect of pressure is observed to be more in C$_{11}$ than in C$_{44}$ in all the compounds. Under pressure non linear nature in C$_{11}$ and C$_s$ is observed in all the compounds. Calculated $C_{44}$ and C$_s$ is found to have a non-linear variation and can be correlated with the observed ETT's under pressure and are shown along with the FS topology change in Fig. 1 to 5 in supplementary data.


\section{\label{sec:level1}Conclusions}
Electronic structure of Nb$_3$X (X = Al, Ga, In, Ge and Sn) compounds was studied both at ambient and under compression by using density functional theory calculations. In all the compounds it is observed that $d$ states of Nb atoms has dominant nature at E$_F$ with admixture of $p$ sates of X atom. All the compounds are found to posses both hole and electron FS. Parallel sheets are observed along X-$\Gamma$ in the last two FS, which indicate the nesting property in these compounds which is also confirmed from the calculated Lindhard susceptibility plots, where sharp peaks are observed along X-$\Gamma$ and at M point in the imaginary part of susceptibility plots in all the compounds at ambient condition. Under compression continuouss change in the FS topology is observed in all the compounds.  
For the given ETT's corresponding changes are observed under compression in the imaginary part of susceptibility and huge peaks are found along X-$\Gamma$ and at M point in Nb$_3$Al. In Nb$_3$Ga and Nb$_3$In it is observed only along X-$\Gamma$. In Nb$_3$Ge, we have observed peaks at M, R and along X-$\Gamma$ under compression. But in Nb$_3$Sn the peak is observed to decrease along X-$\Gamma$. 
The change in the band structure and FS topology under compression lead to the non-monotonic variation in density of states. Mechanical stability of these compounds is also confirmed both at ambient as well as under compression and non linear nature in C$_{44}$ and C$_s$ is observed in all the compounds under pressure. Further experiments are needed to realise the continuous FS topology changes observed in these compounds. 

\section{\label{sec:level1}Acknowledgement}
The authors P.V.S.R and V.K would like to thank Department of Science and Technology (DST) for the financial support through SR/FTP/PS-027/2011. The authors would also like to acknowledge IIT-Hyderabad for providing the computational facility. G.V would like to acknowledge CMSD-UoH for providing the computational facility.

\begin{table*}
\caption{\label{tab:table3}Ground state properties of Nb$_3$X (X= Al, Ga, In, Ge and Sn). $\gamma_{exp}$ and $\gamma_{th}$ are experimental and theoretical Sommerfeld coefficient in the units of $mJ/mol K^{2}$, N(E$_F$) is density of states at E$_F$ in the units of states/eV/f.u.}

\begin{ruledtabular}
\begin{tabular}{cccccc} 

 Parameters 			&Nb$_3$Al  &Nb$_3$Ga  &Nb$_3$In  &Nb$_3$Ge  &Nb$_3$Sn\\
\hline

$a_{exp}$ (\AA)		     	&5.187\cite{Book}, 5.185\cite{Viswanathan}	&5.171\cite{Book}, 5.1674\cite{Mkrtcheyan} &5.303	&5.166\cite{Book}, 5.161\cite{Kawamura}	&5.289\\

$a_{the}$ (\AA)                 &5.198, 5.210\cite{Paduani}, 	&5.183, 5.200\cite{Paduani}, 	&5.328 &5.165, 5.185\cite{Paduani},	&5.322\\
                                &5.187\cite{Klein}, 5.164\cite{Kawamura}   &5.171\cite{Klein}              &      &5.160\cite{Klein}                &\\

$B_{exp}$ (GPa)		&177\cite{Yu}		&198\cite{Mkrtcheyan}		&-	&115\cite{Kawamura}		&-\\

$B_{the}$ (GPa)		&158, 156.33\cite{Paduani},        &162, 156.93\cite{Paduani},	&152,152.71\cite{Sundareswari}	&172, 168.04\cite{Paduani} &160, 160.51\cite{Sundareswari}\\
                                &166.815\cite{Rajagopalan}, 165.49\cite{Sundareswari}  &169.97\cite{Sundareswari}		&		&	         &\\

$\gamma_{exp}$ 	&31.81\cite{Book}, 30\cite{Viswanathan}	&-	&-	&-	&-\\

$\gamma_{th}$ 	&37.63, 31.231\cite{Rajagopalan}	&39.34	&41.08	&32.95	&45.68\\

N(E$_F$) 	&15.96, 16.60\cite{Paduani},	&16.70, 18.23\cite{Paduani},	&17.44 	&13.98, 14.40\cite{Paduani},	&19.39\\	 
                &14.64\cite{Klein}              &14.10\cite{Klein}, 15.36\cite{Paduani1}   &	&7.84\cite{Klein}, 14.39\cite{Paduani1}	&\\
\end{tabular}
\end{ruledtabular}

\end{table*}

\begin{table*}
\caption{\label{tab:table3}Calculated single crystalline and poly crystalline elastic constants at ambient condition for Nb$_3$X (X = Al, Ga, Ge, In and Sn). Where E is Young's modulus, G$_H$ is Voigt-Reuss-Hill modulus, $\sigma$ is Poisson's ratio, A is Anisotropy factor, CP = Cauchy's pressure ($C_{11}-C_{44}$), PR = Pugh's ratio.
}
\begin{ruledtabular}
\begin{tabular}{cccccc}
Parameters &Nb$_3$Al &Nb$_3$Ga &Nb$_3$In &Nb$_3$Ge &Nb$_3$Sn\\
\hline
$C_{11}$ (GPa)		&273, 310.530\cite{Rajagopalan}, 310.53\cite{Sundareswari}	&298, 305.41\cite{Sundareswari}	&289	&297	&297\\

$C_{12}$ (GPa)		&97, 92.976\cite{Rajagopalan}, 92.98\cite{Sundareswari} 	&102, 104.09\cite{Sundareswari}	&97	&108	&110\\

$C_{44}$ (GPa)		&49, 49.119\cite{Rajagopalan}, 57.12\cite{Sundareswari}  &50, 48.78\cite{Sundareswari}		&57	&63	&69\\

$E$ (GPa)		&171, 179.486\cite{Rajagopalan}, 193.54\cite{Sundareswari} 	&175, 174.11\cite{Sundareswari}	&184	&194	&203\\

$A$			&0.49, 0.451\cite{Rajagopalan}, 0.525\cite{Sundareswari}	&0.51, 0.485\cite{Sundareswari}	&0.59	&0.51	&0.74\\

CP 			&48.81, 43.857\cite{Rajagopalan}, 35.857\cite{Sundareswari} 	&51.84, 55.314\cite{Sundareswari}&39.86	&45.74	&40.68\\

PR			&0.39			&0.39			&0.44	&0.43	&0.45	\\

$\sigma$		&0.36, 0.286\cite{Rajagopalan}, 0.27\cite{Sundareswari} 	&0.33, 0.30\cite{Sundareswari}	&0.31	&0.66	&0.30\\

$G_{H}$ (GPa)		&64.53, 67.953\cite{Rajagopalan}, 74.15\cite{Sundareswari} 	&65.85, 65.49\cite{Sundareswari} &70.27	&73.87	&78.10\\









\end{tabular}
\end{ruledtabular}
\end{table*}


\bibliography{apssamp}

\begin{figure*}
\begin{center}
\subfigure[]{\includegraphics[width=50mm,height=50mm]{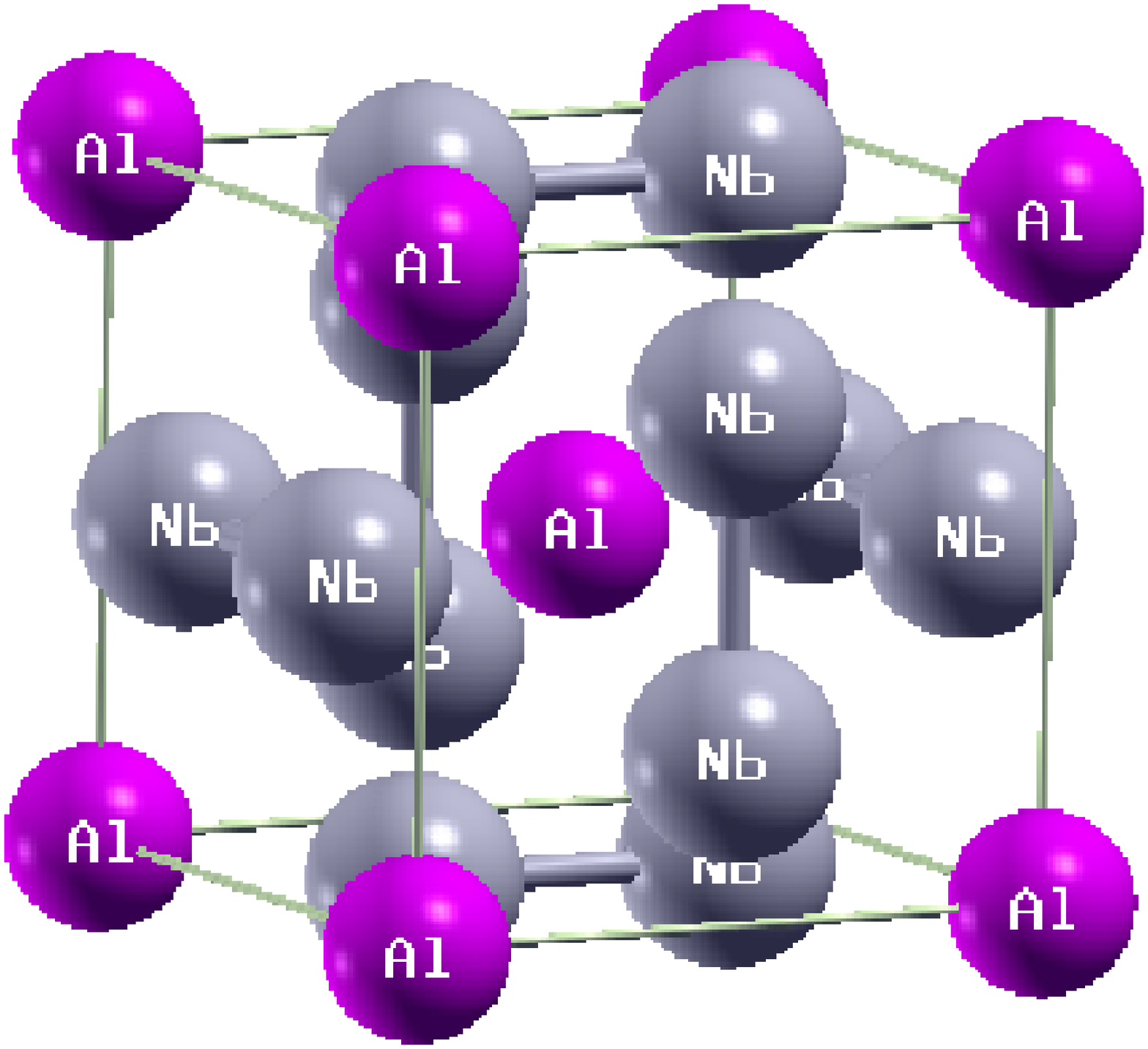}}
\subfigure[]{\includegraphics[width=50mm,height=50mm]{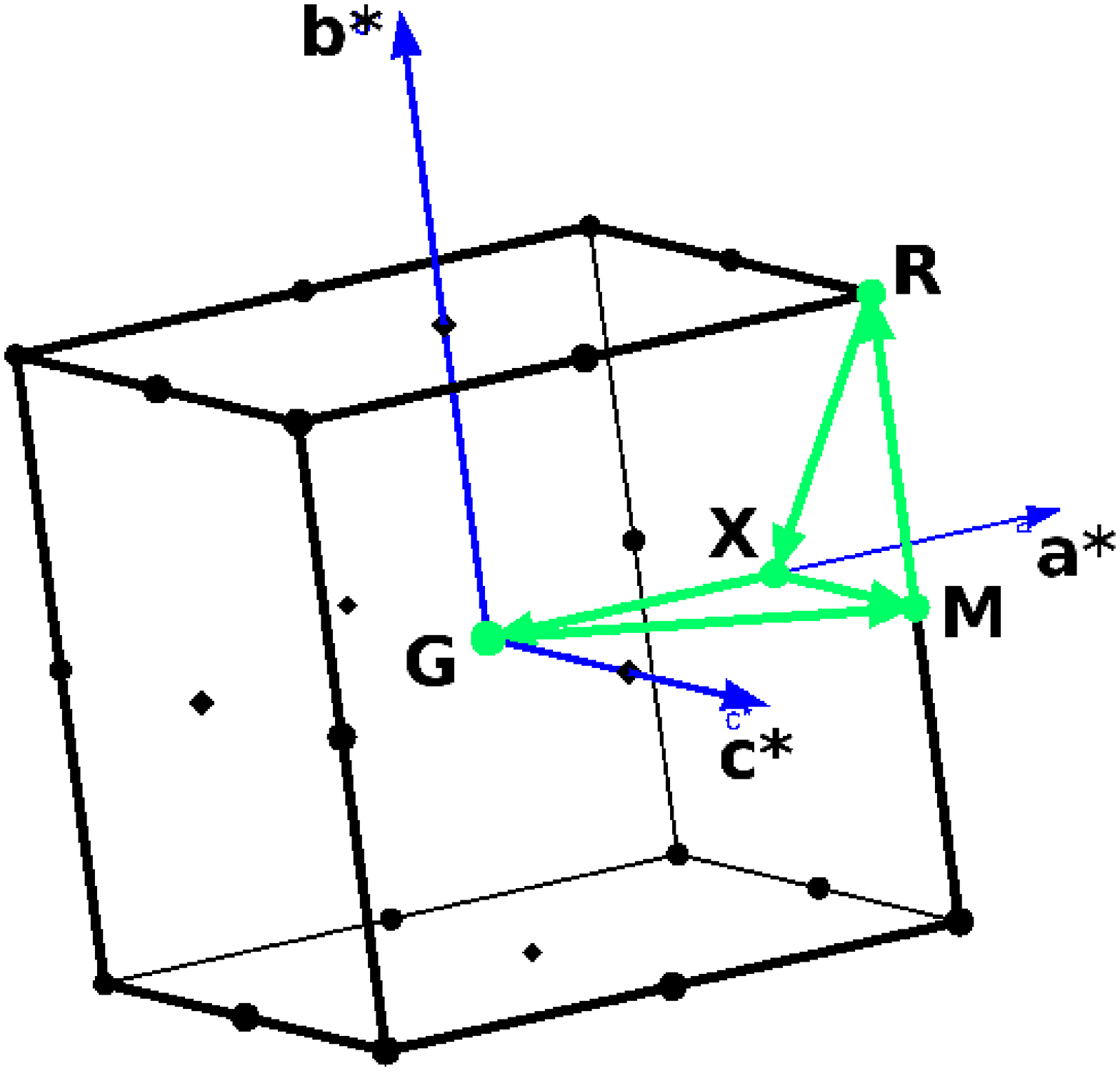}}
\subfigure[]{\includegraphics[width=50mm,height=70mm]{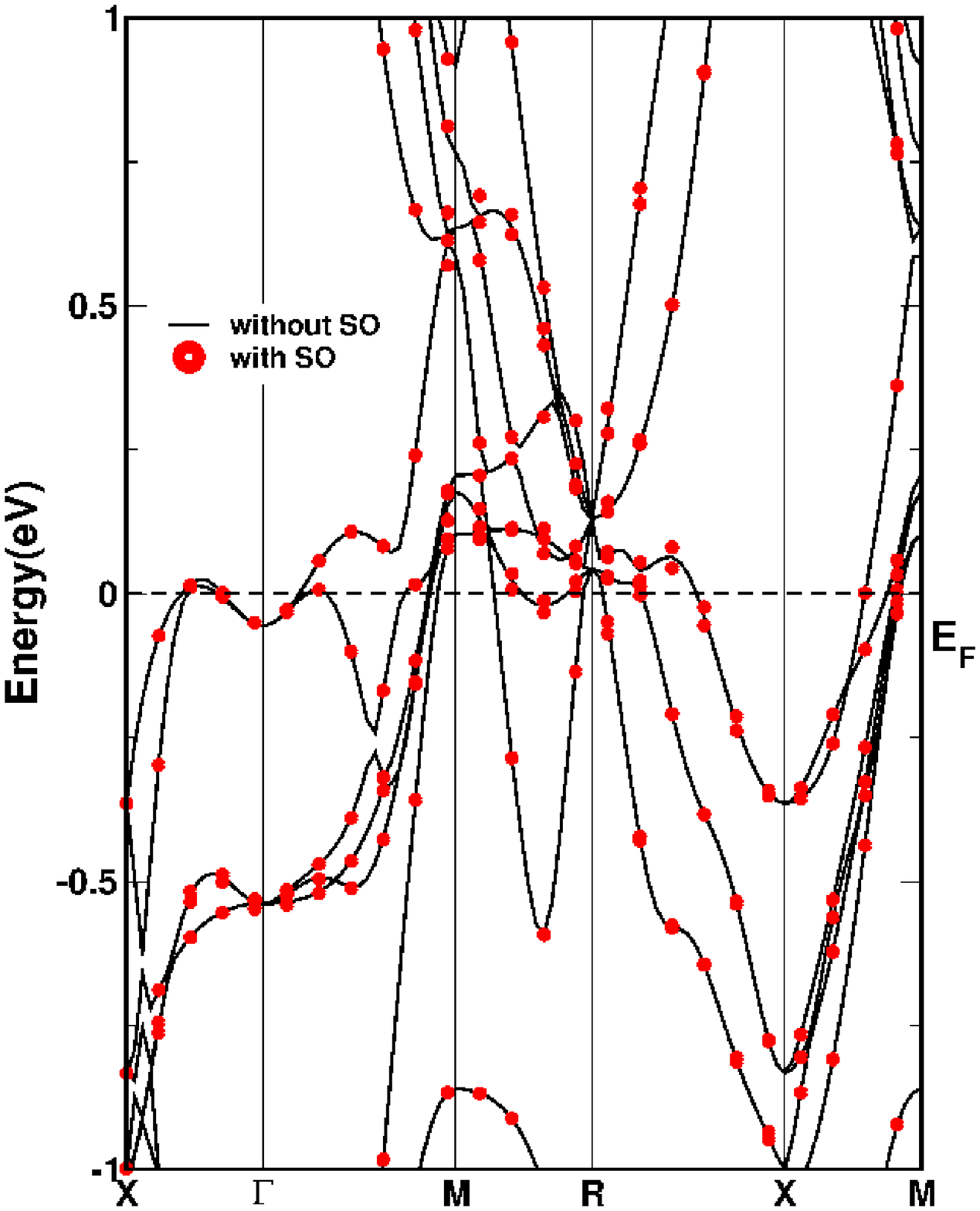}}
\caption{(a)Crystal structure of Nb$_3$X (X= Al, Ga, In, Ge and Sn) and (b)Brillouin zone high symmetry points and (c) band structure of Nb$_3$Sn with and without spin orbit coupling.}
\end{center}
\end{figure*}

\begin{figure*}
\begin{center}
\includegraphics[width=180mm,height=100mm]{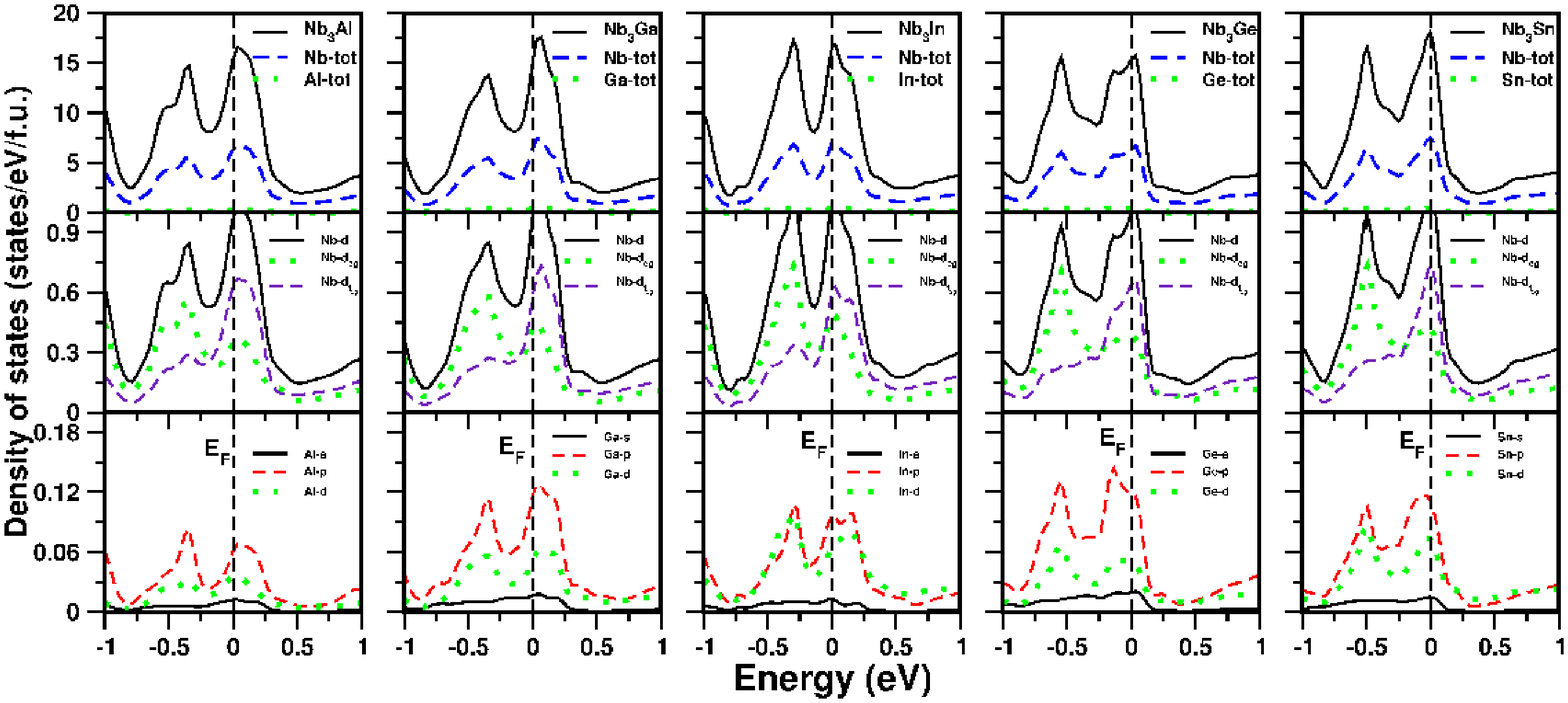}
\caption{Density of states at ambient condition for all compounds.}
\end{center}
\end{figure*}


\begin{figure*}
\begin{center}
\subfigure[]{\includegraphics[width=50mm,height=70mm]{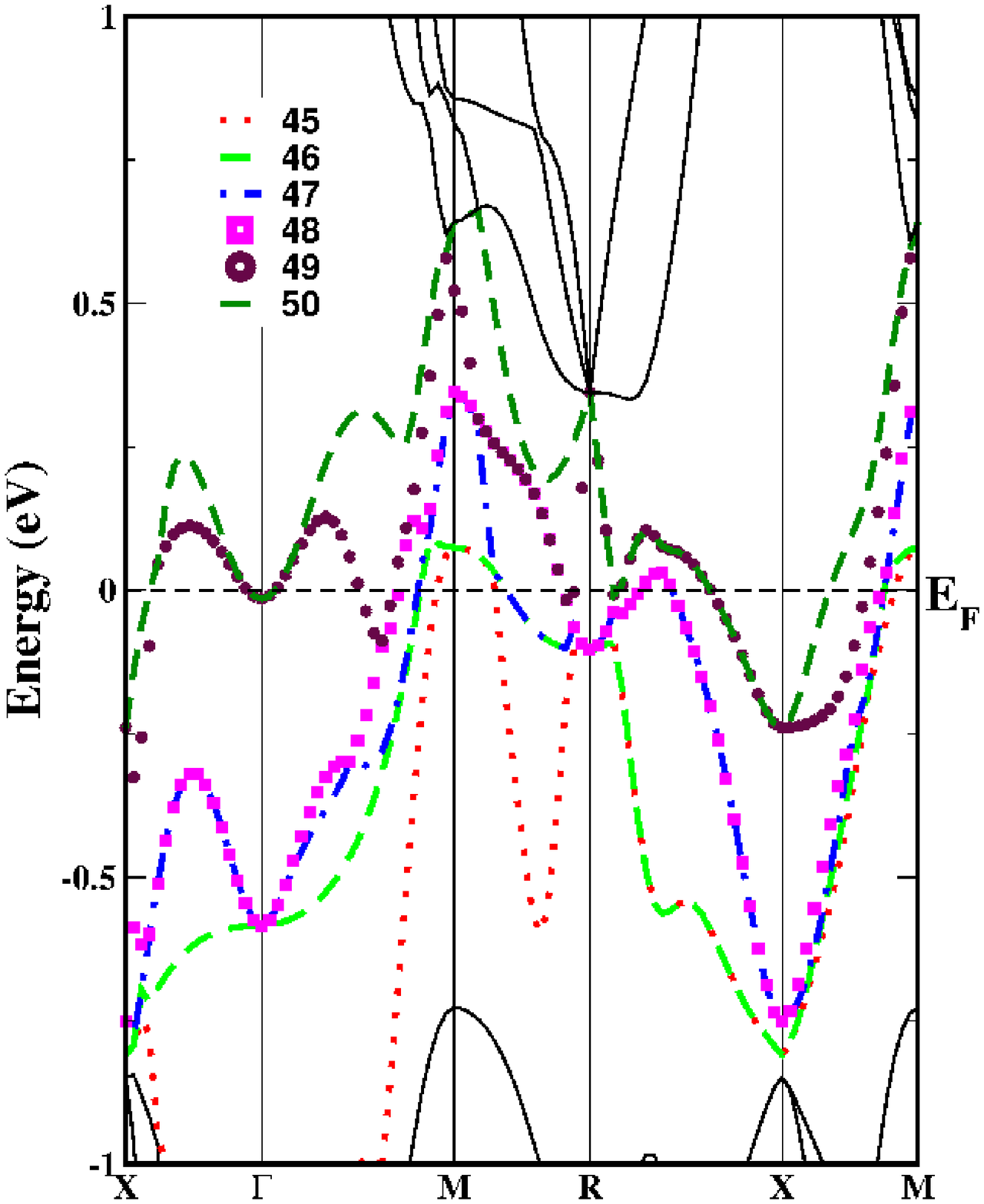}}
\subfigure[]{\includegraphics[width=50mm,height=70mm]{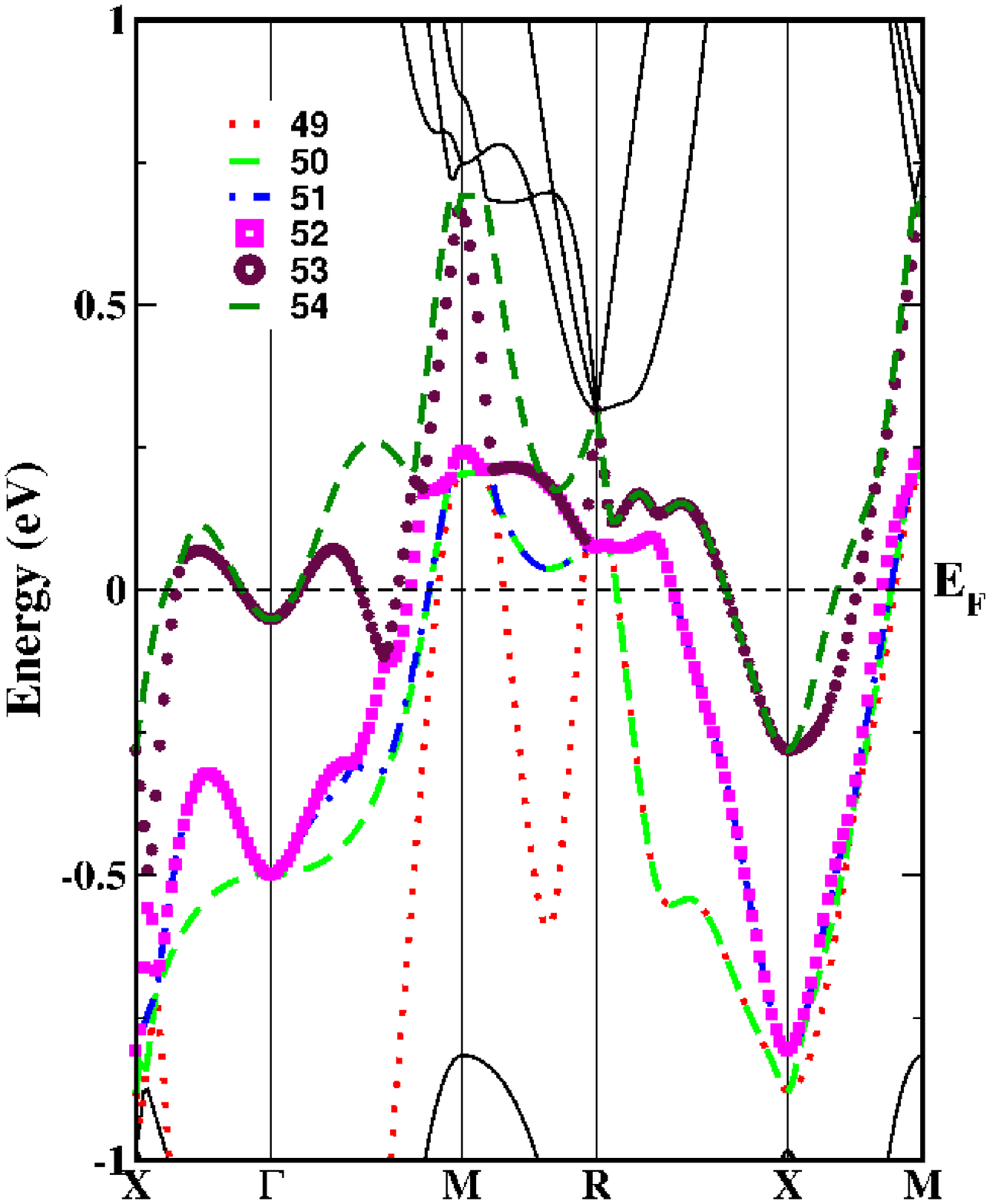}}
\subfigure[]{\includegraphics[width=50mm,height=70mm]{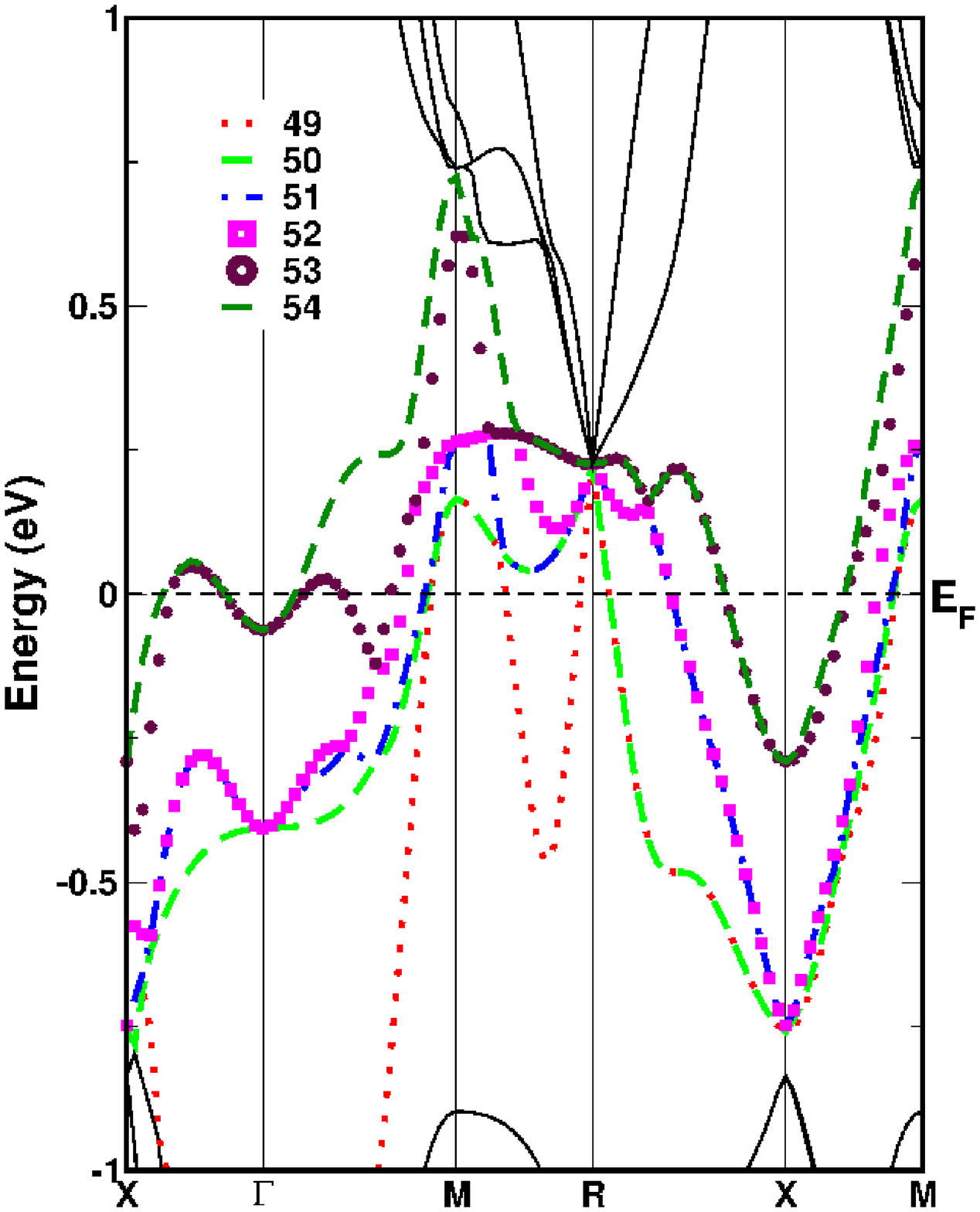}}\\
\subfigure[]{\includegraphics[width=50mm,height=70mm]{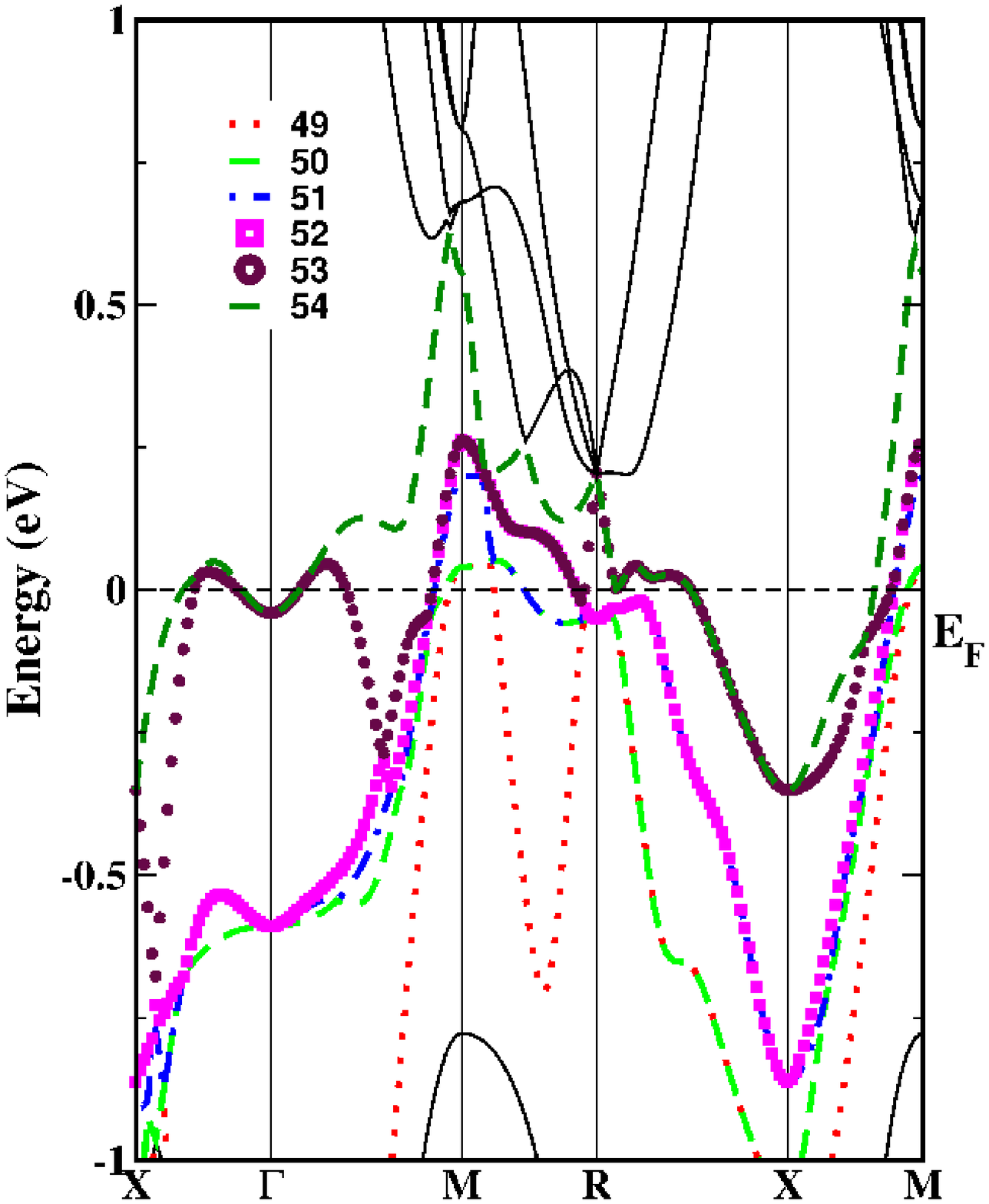}}
\subfigure[]{\includegraphics[width=50mm,height=70mm]{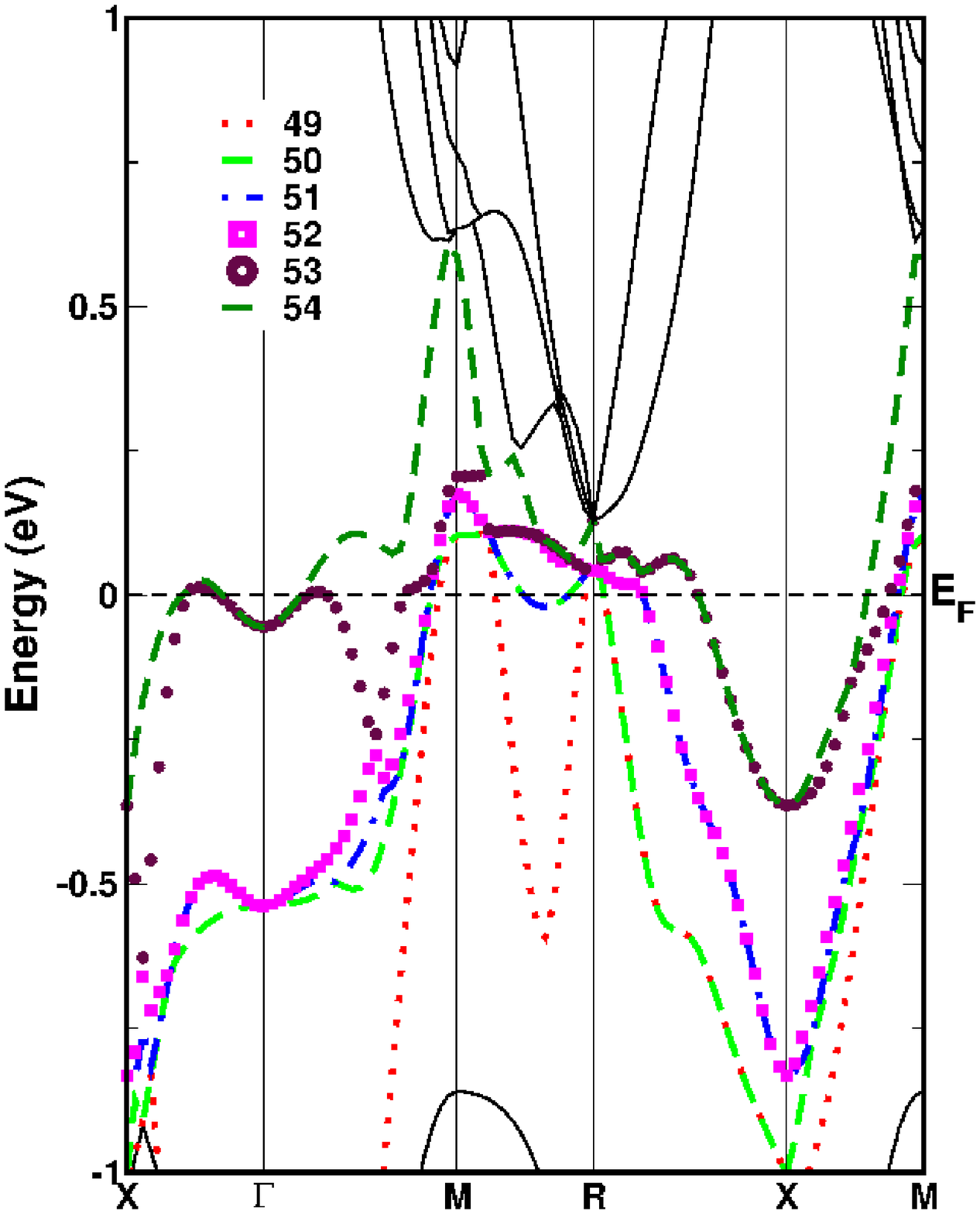}}
\caption{Band structure without spin orbit coupling (a)Nb$_3$Al, (b)Nb$_3$Ga, (c)Nb$_3$In, (d)Nb$_3$Ge and (e)Nb$_3$Sn at ambient condition.}
\end{center}
\end{figure*}

\begin{figure*}
\begin{center}
\subfigure[]{\includegraphics[width=35mm,height=35mm]{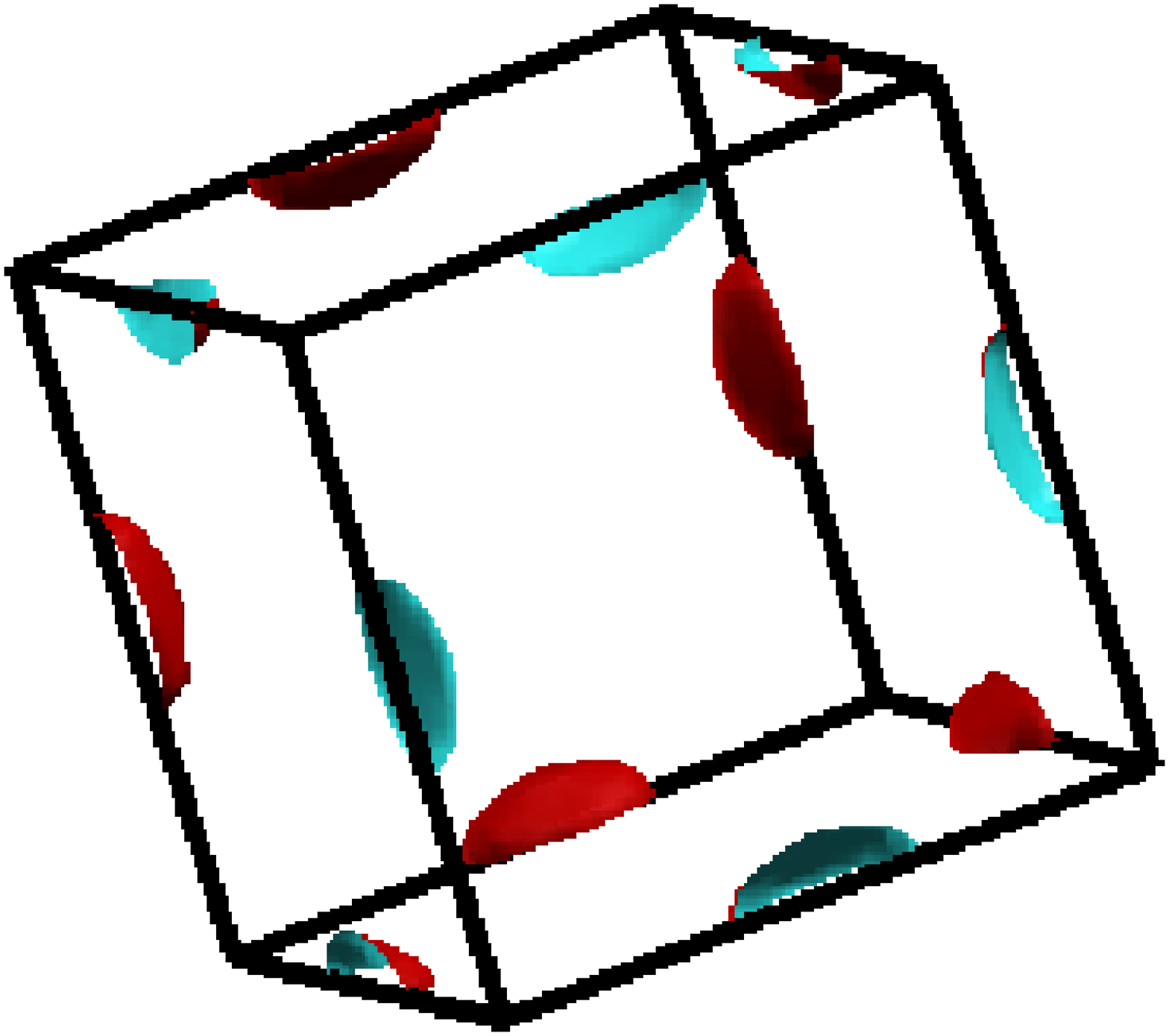}}
\subfigure[]{\includegraphics[width=35mm,height=35mm]{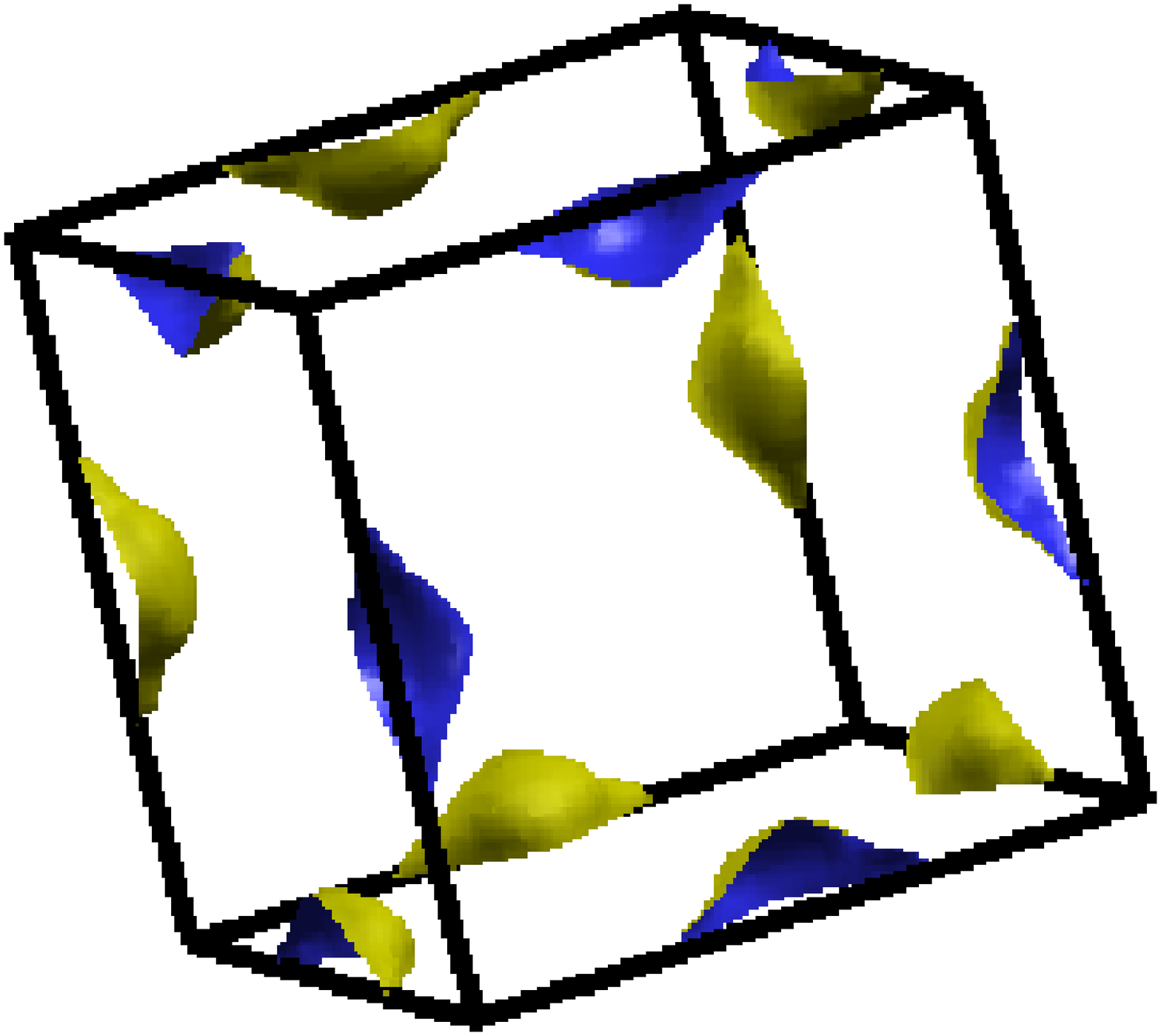}}
\subfigure[]{\includegraphics[width=35mm,height=35mm]{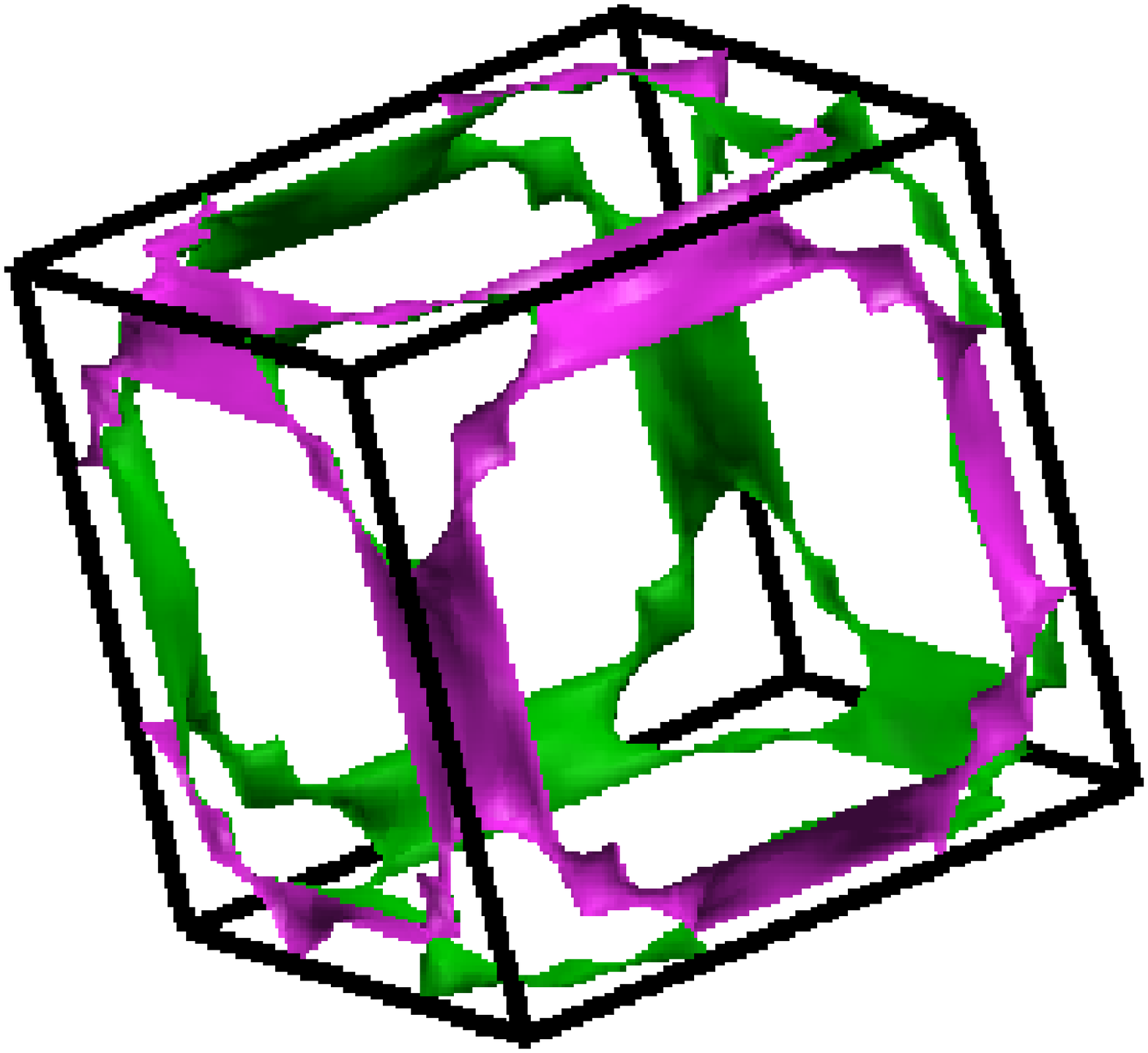}}\\
\subfigure[]{\includegraphics[width=35mm,height=35mm]{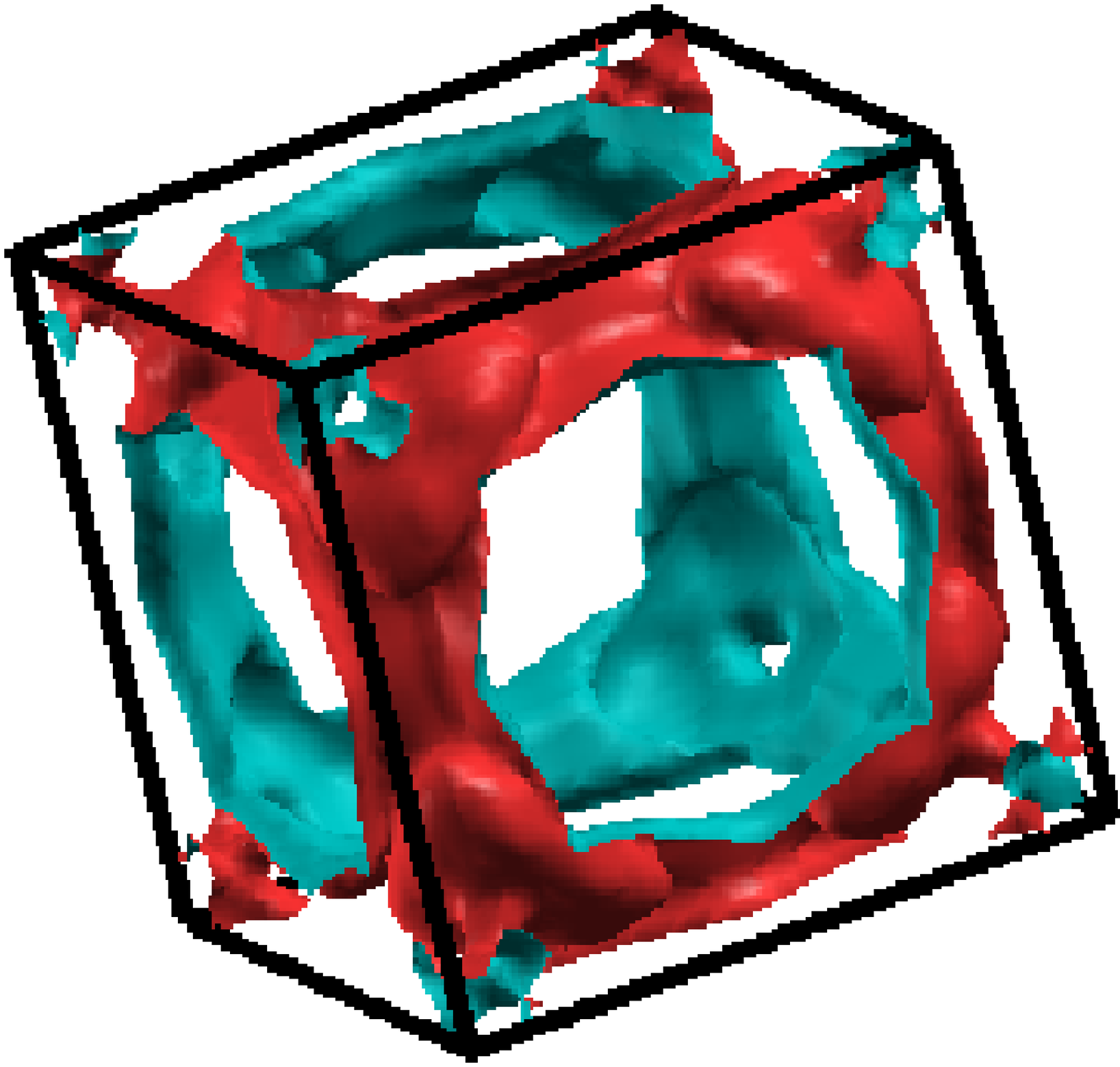}}
\subfigure[]{\includegraphics[width=35mm,height=35mm]{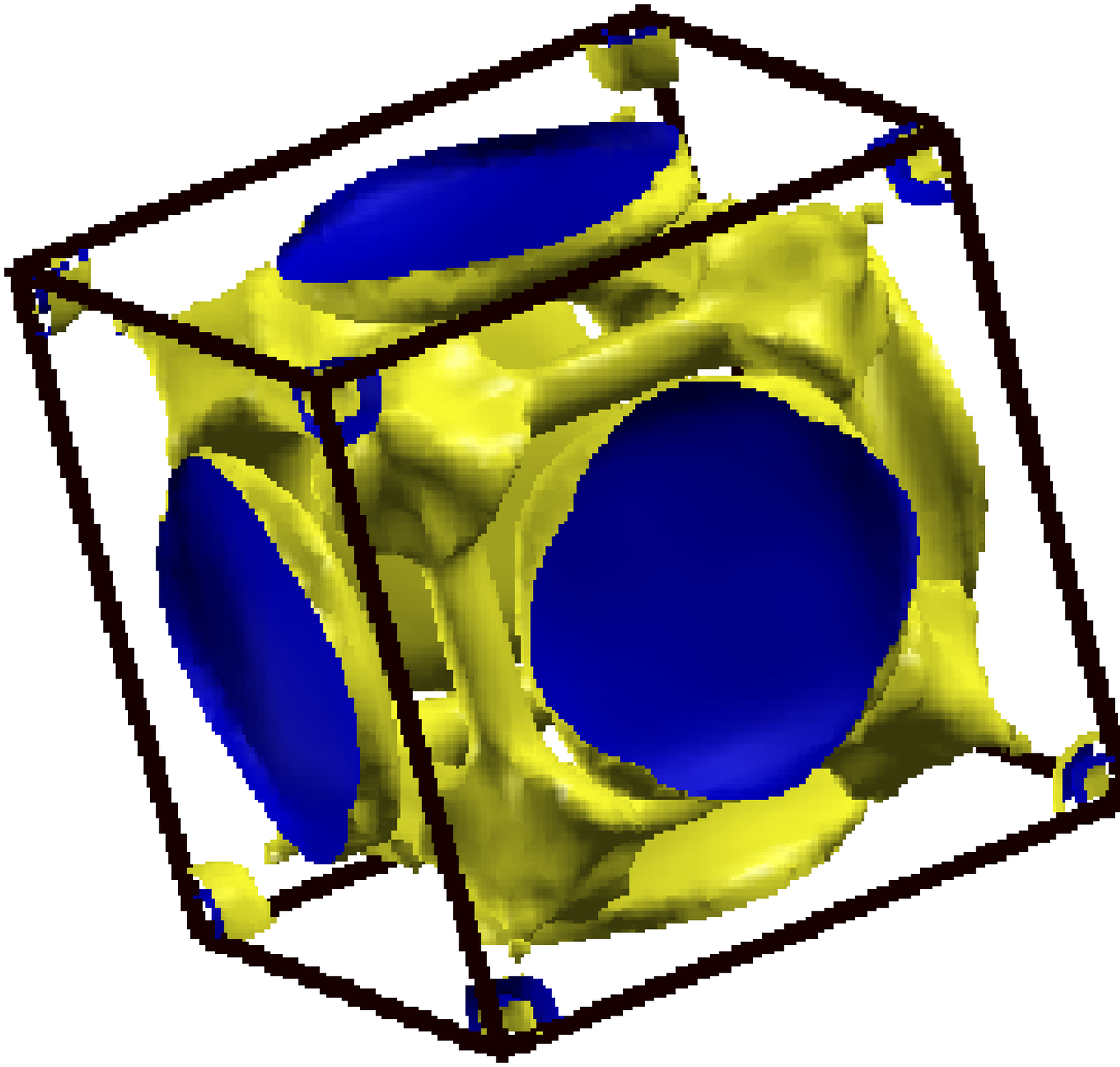}}
\subfigure[]{\includegraphics[width=35mm,height=35mm]{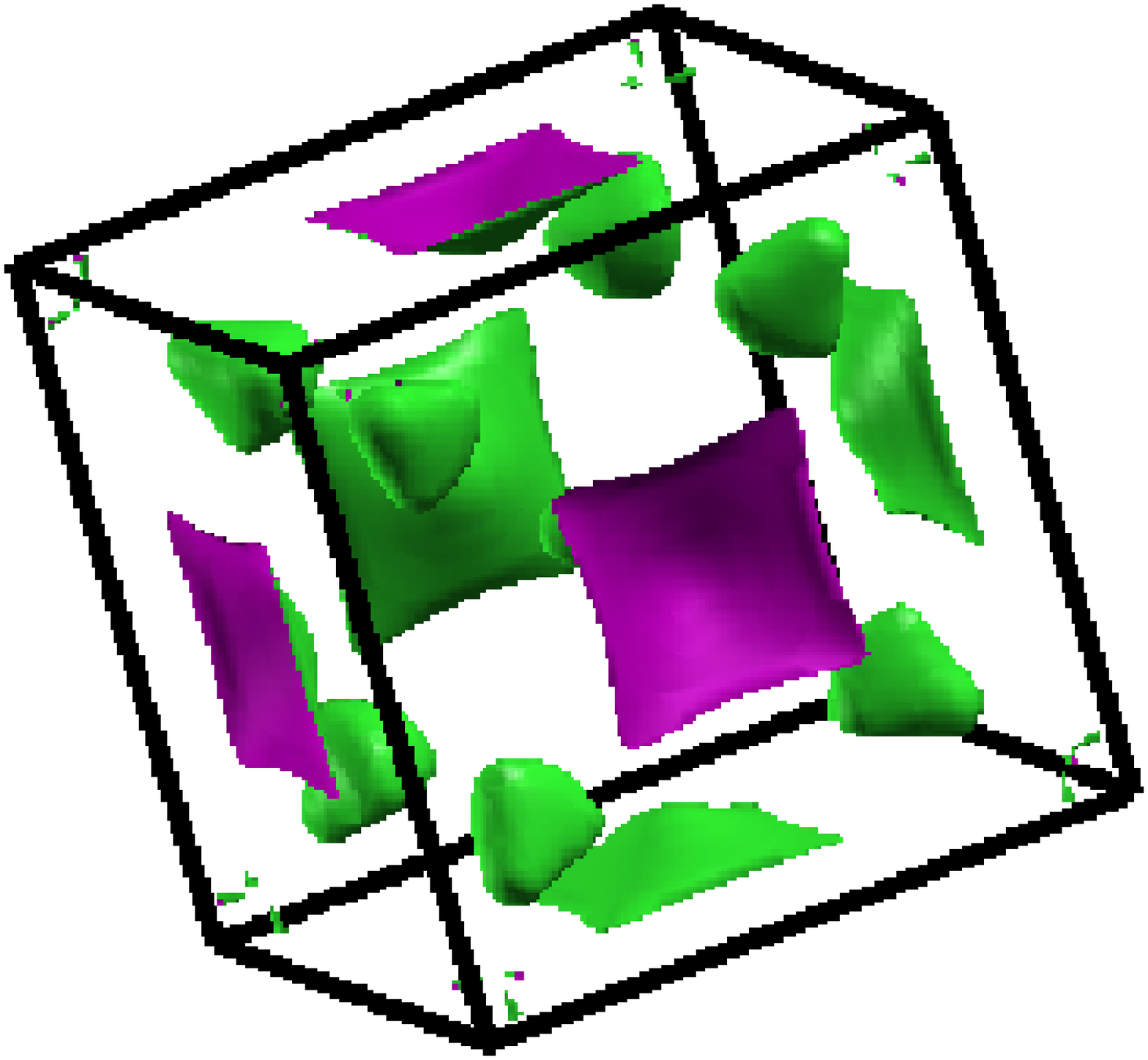}}\\
\caption{Fermi surface for Nb$_3$Al at ambient (a)band no. 45, (b)band no. 46, (c)band no. 47, (d)band no. 48, (e)band no. 49 and  (f)band no. 50. The first four FS are having hole nature and remaining two are having electronic nature. The first two FS are having pockets at M points. The next two FS are having ribbon like sheets along BZ edges. The last two FS are having sheets in middle of the BZ faces with a pocket at $\Gamma$ point.}
\end{center}
\end{figure*}

\begin{figure*}
\begin{center}
\subfigure[]{\includegraphics[width=35mm,height=35mm]{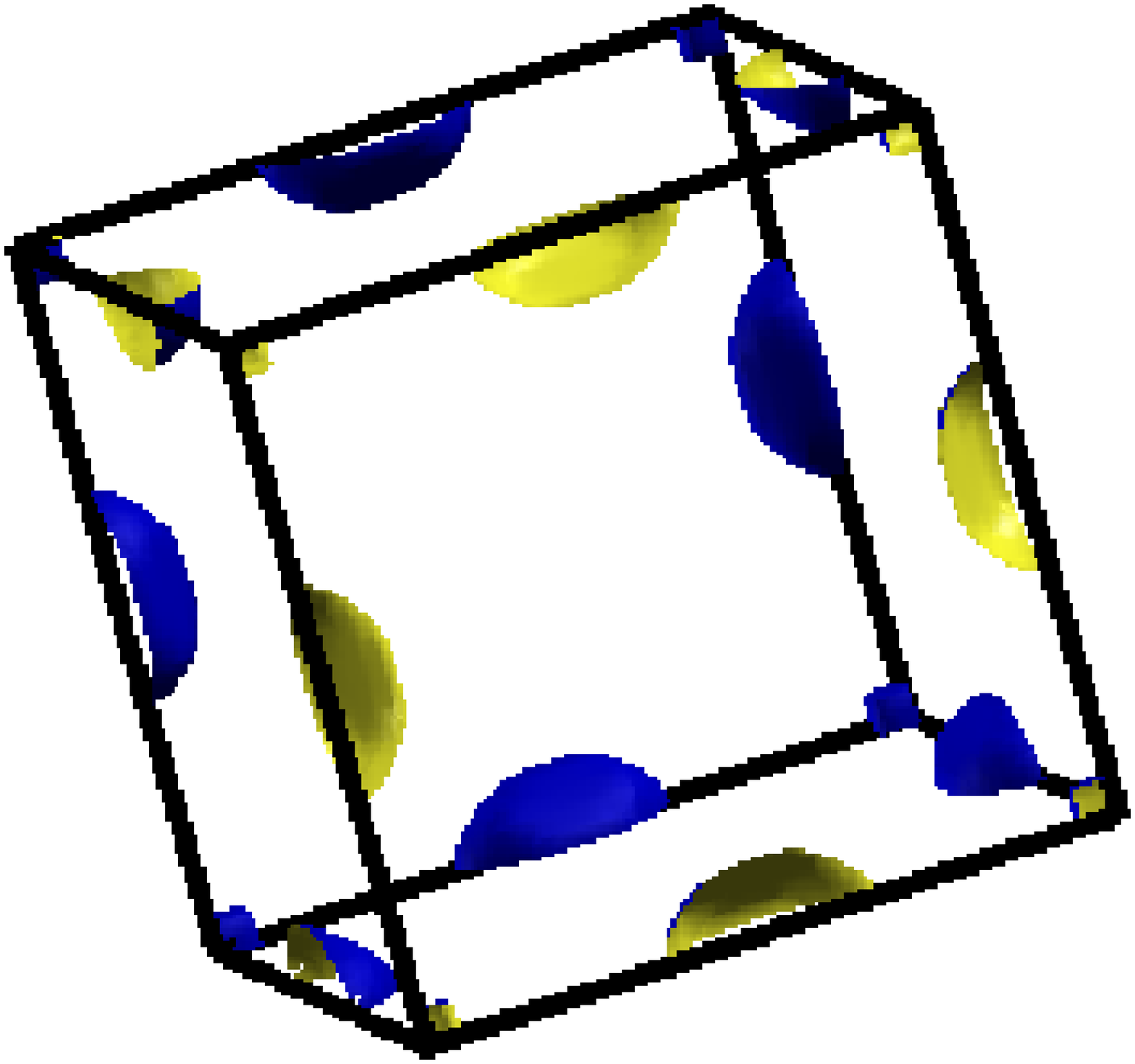}}
\subfigure[]{\includegraphics[width=35mm,height=35mm]{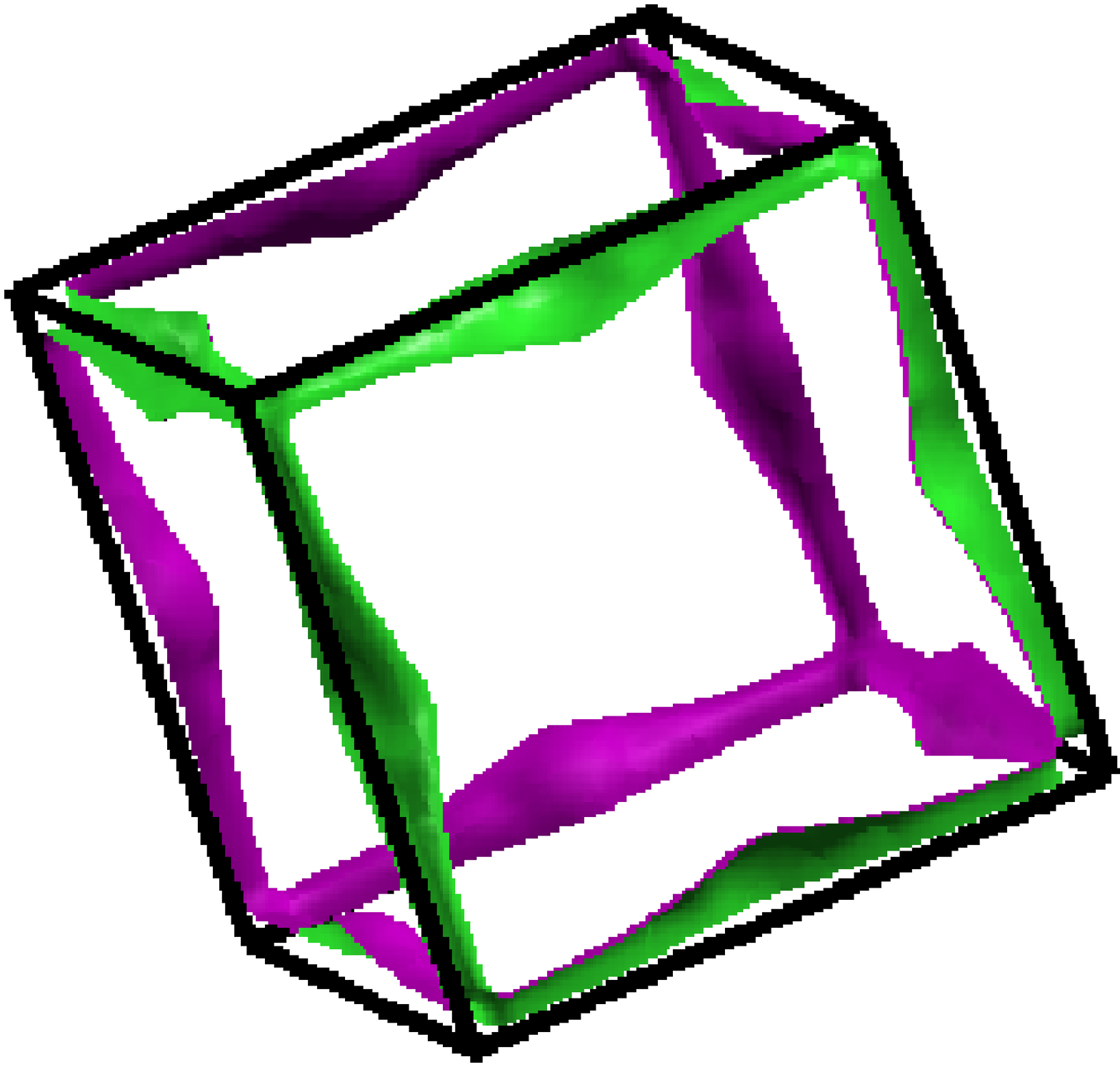}}
\subfigure[]{\includegraphics[width=35mm,height=35mm]{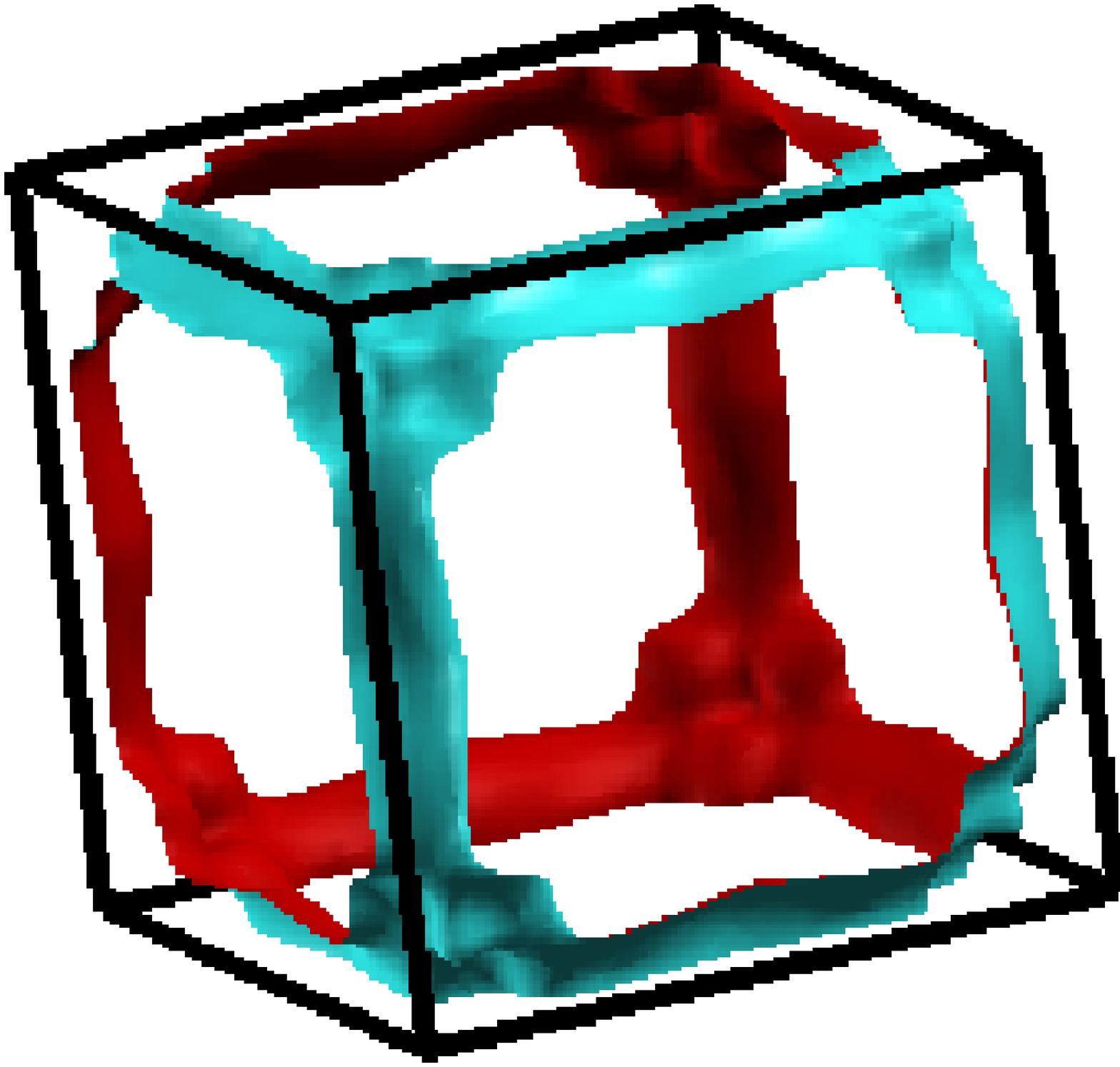}}\\
\subfigure[]{\includegraphics[width=35mm,height=35mm]{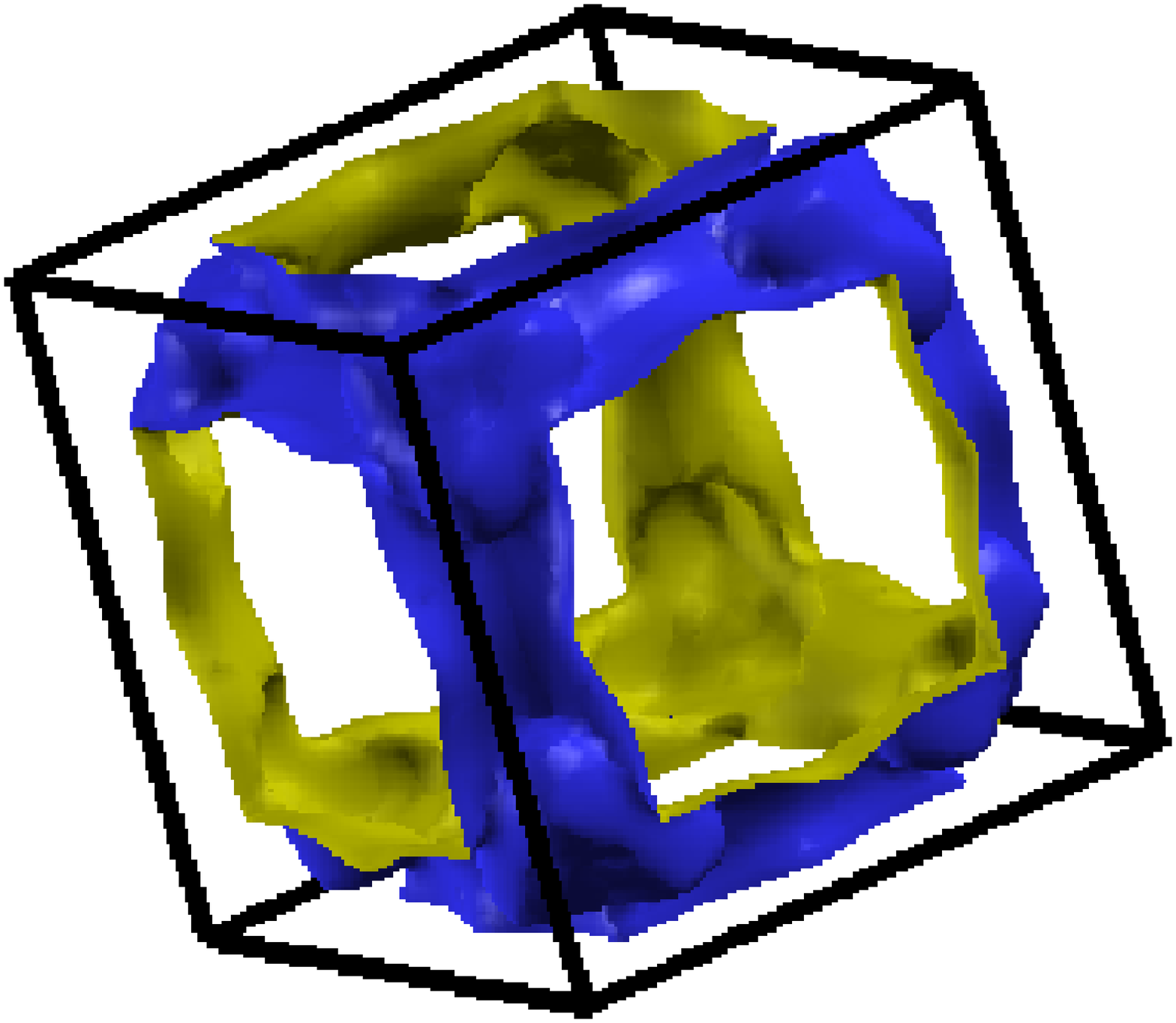}}
\subfigure[]{\includegraphics[width=35mm,height=35mm]{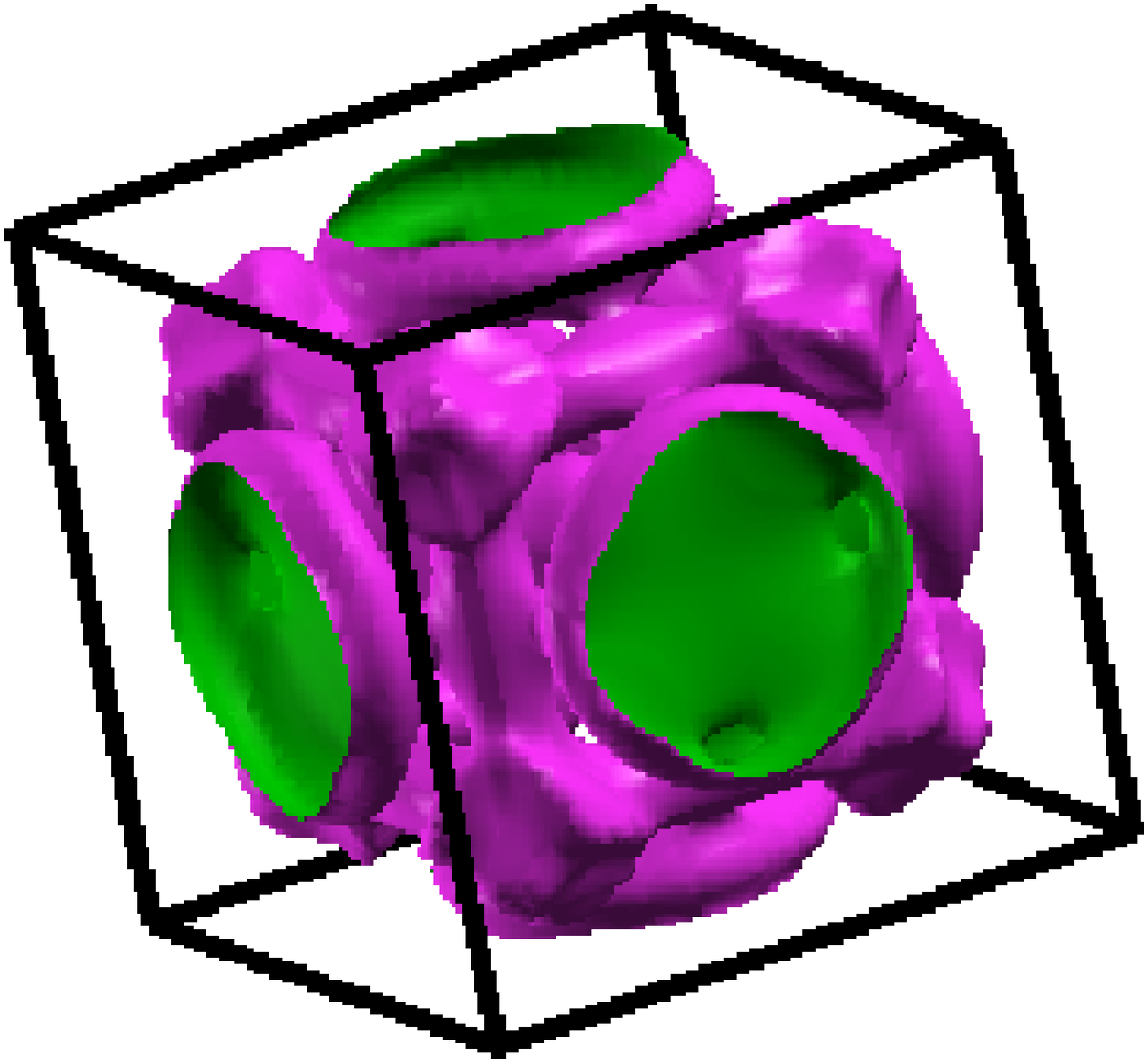}}
\subfigure[]{\includegraphics[width=35mm,height=35mm]{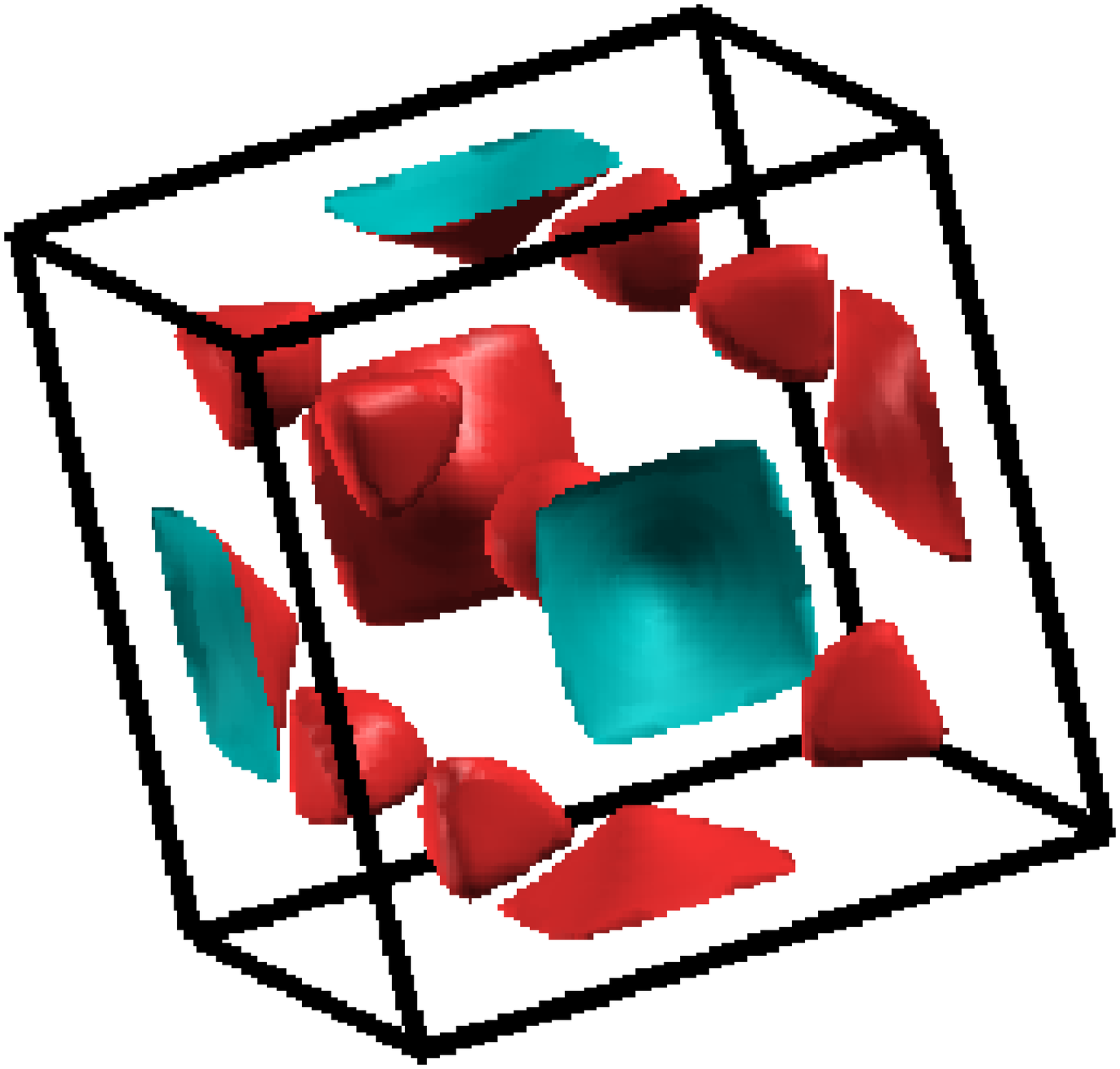}}\\
\caption{Fermi surface for Nb$_3$Ga at ambient (a)band no. 49, (b)band no. 50, (c)band no. 51, (d)band no. 52, (e)band no. 53 and  (f)band no. 54. The first four FS are having hole nature and remaining two are having electronic nature. The first FS having pockets at M points and BZ corners. The next three FS are having ribbon like sheets along BZ edges. The last two FS are having sheets in middle of the BZ faces with a pocket at $\Gamma$ point.}
\end{center}
\end{figure*}

\begin{figure*}
\begin{center}
\subfigure[]{\includegraphics[width=35mm,height=35mm]{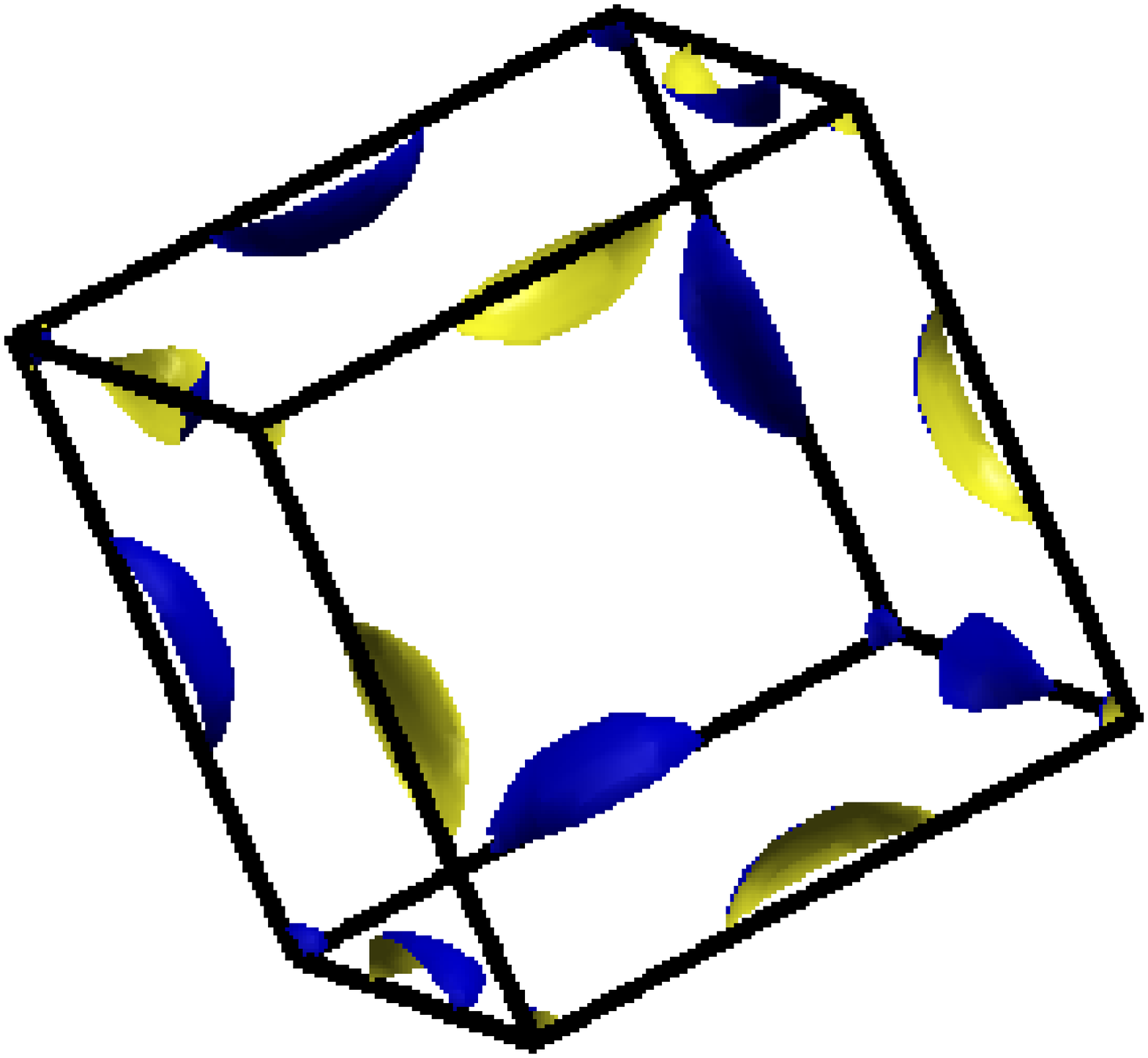}}
\subfigure[]{\includegraphics[width=35mm,height=35mm]{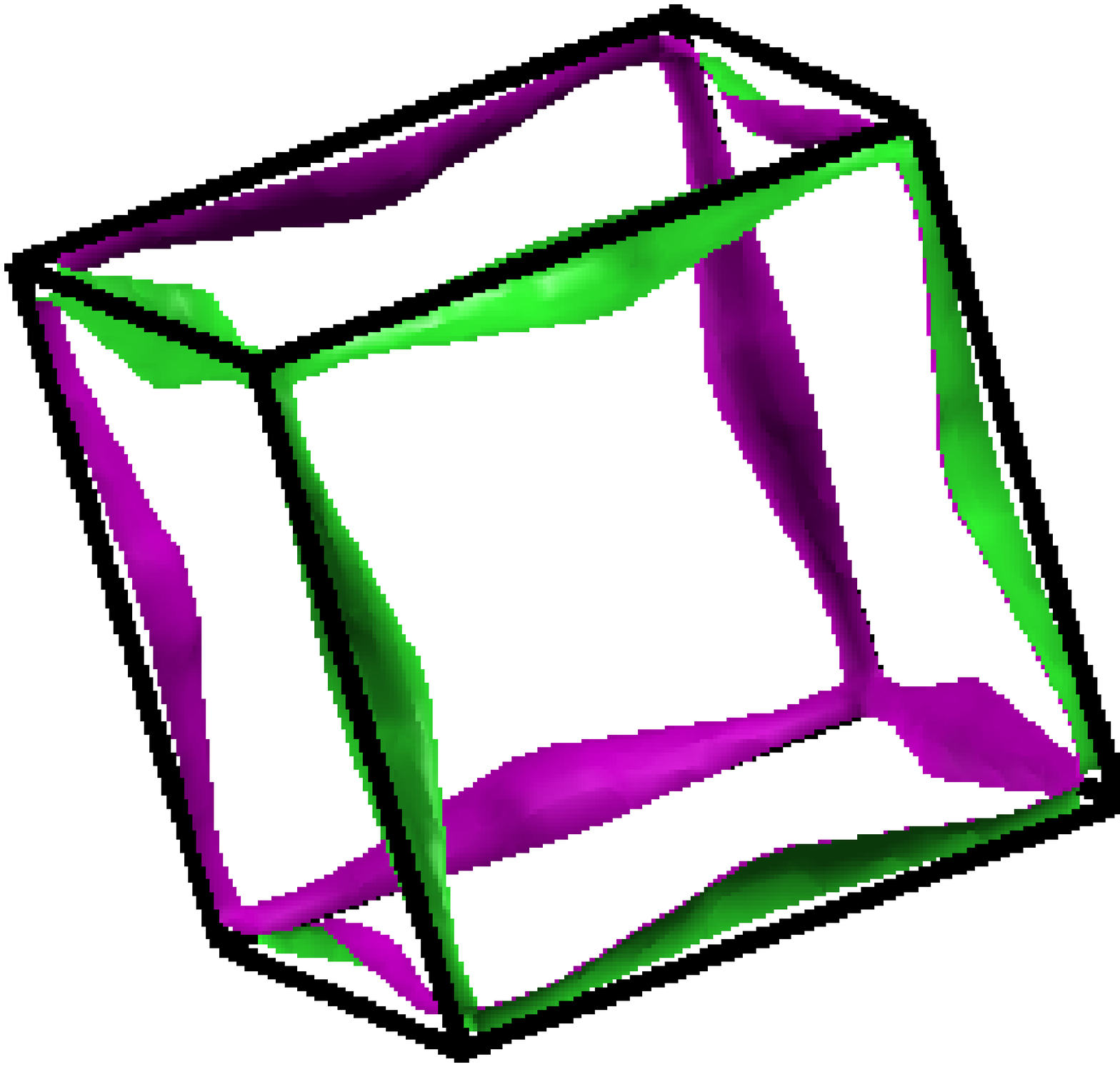}}
\subfigure[]{\includegraphics[width=35mm,height=35mm]{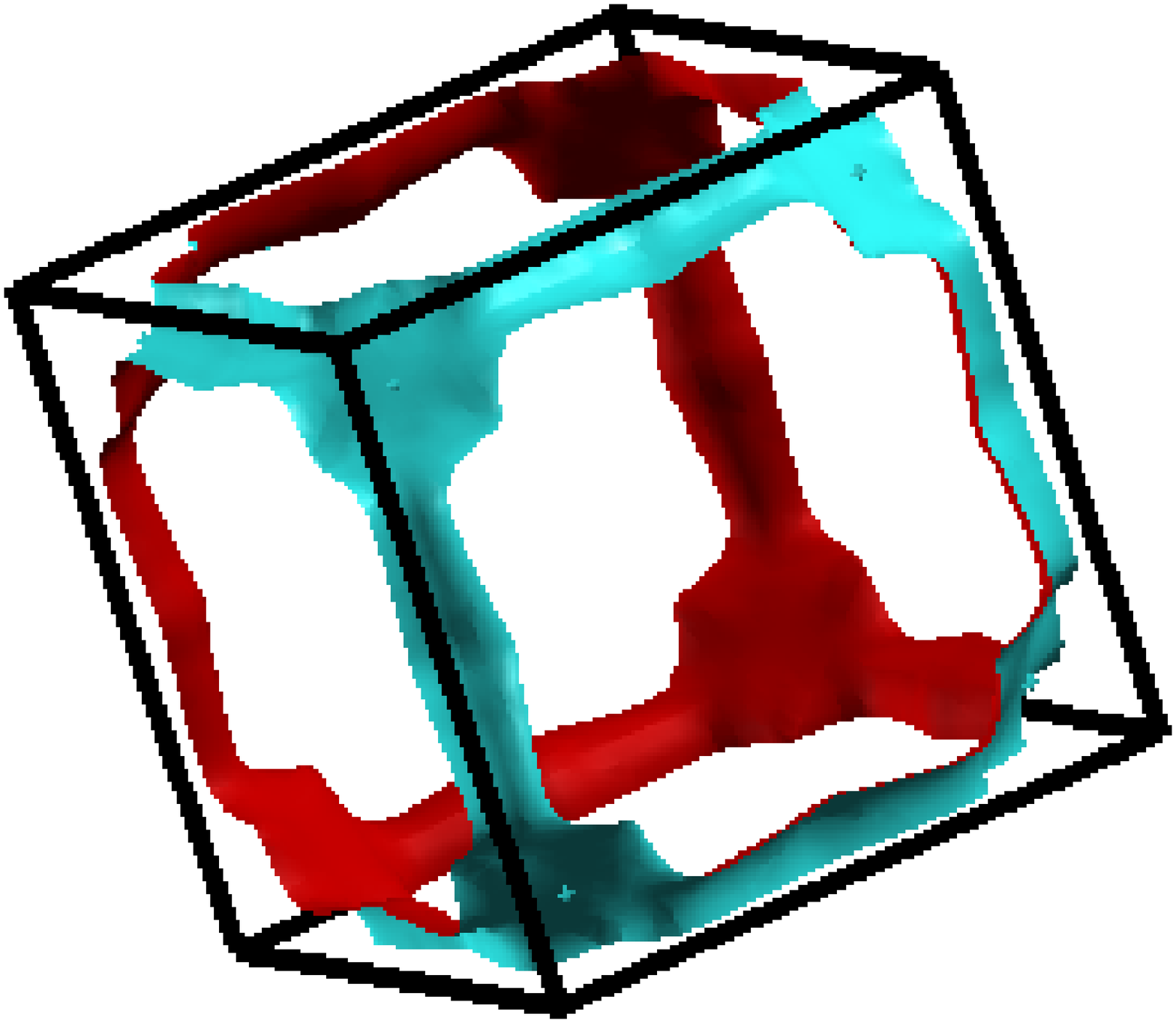}}\\
\subfigure[]{\includegraphics[width=35mm,height=35mm]{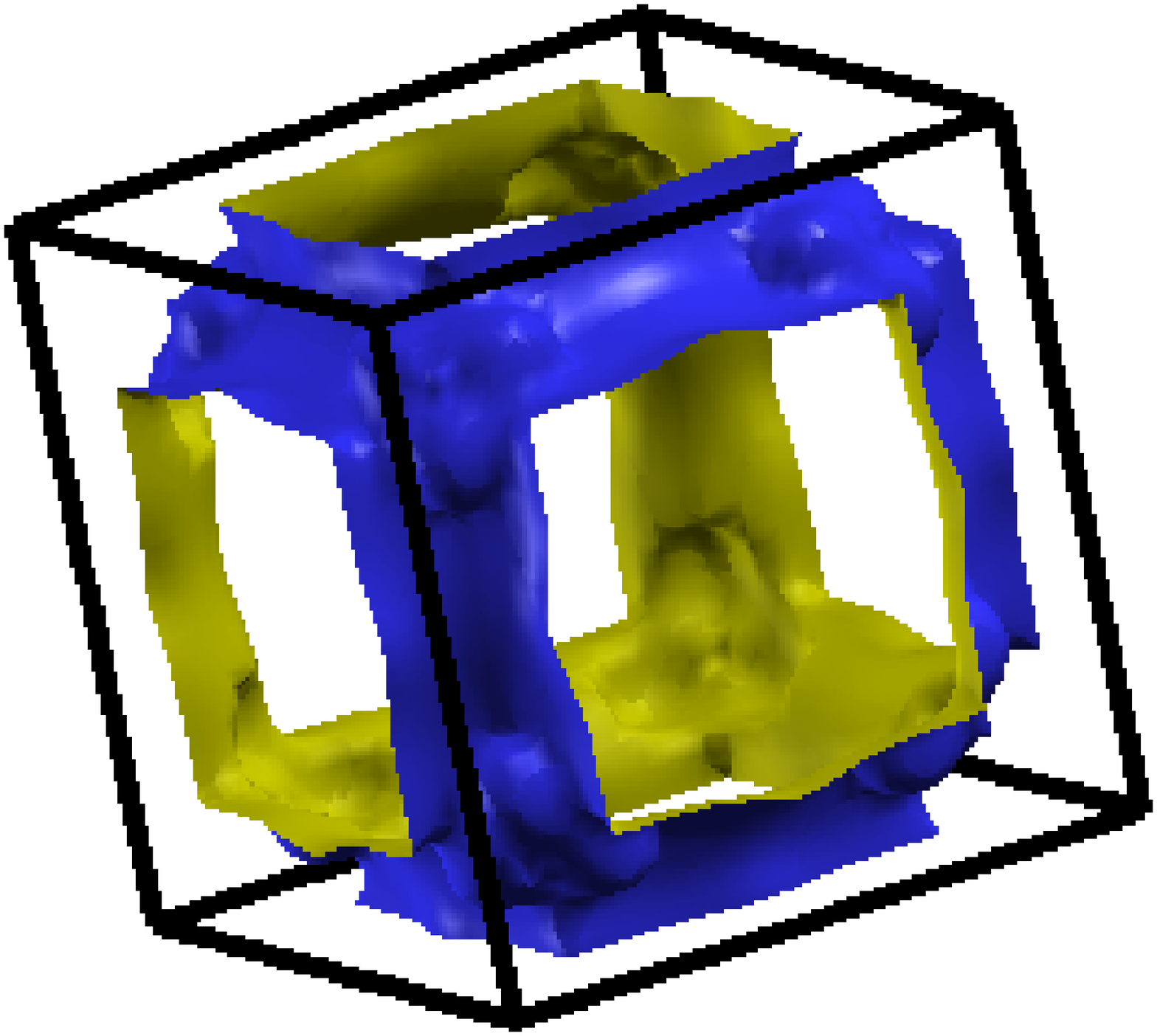}}
\subfigure[]{\includegraphics[width=35mm,height=35mm]{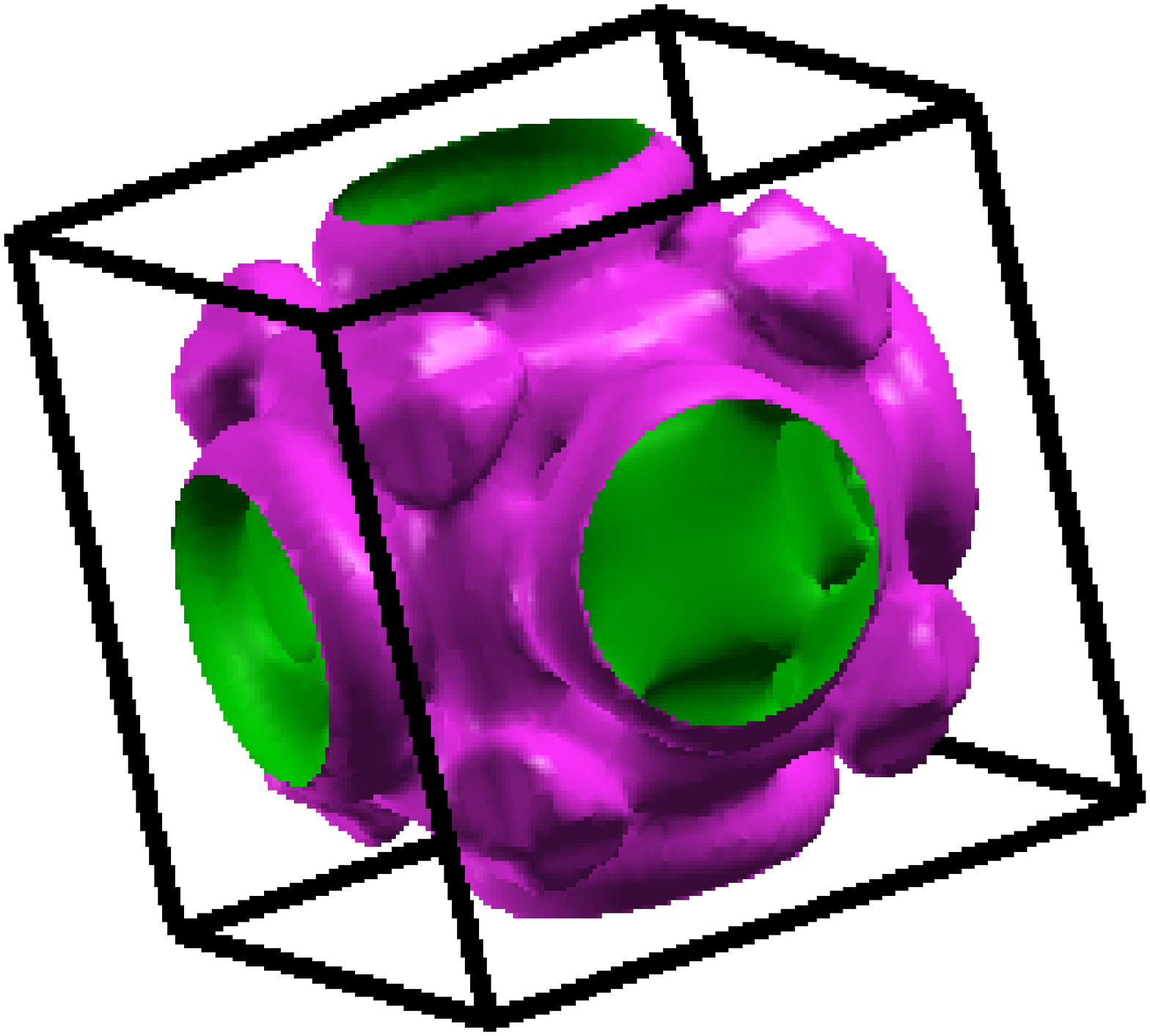}}
\subfigure[]{\includegraphics[width=35mm,height=35mm]{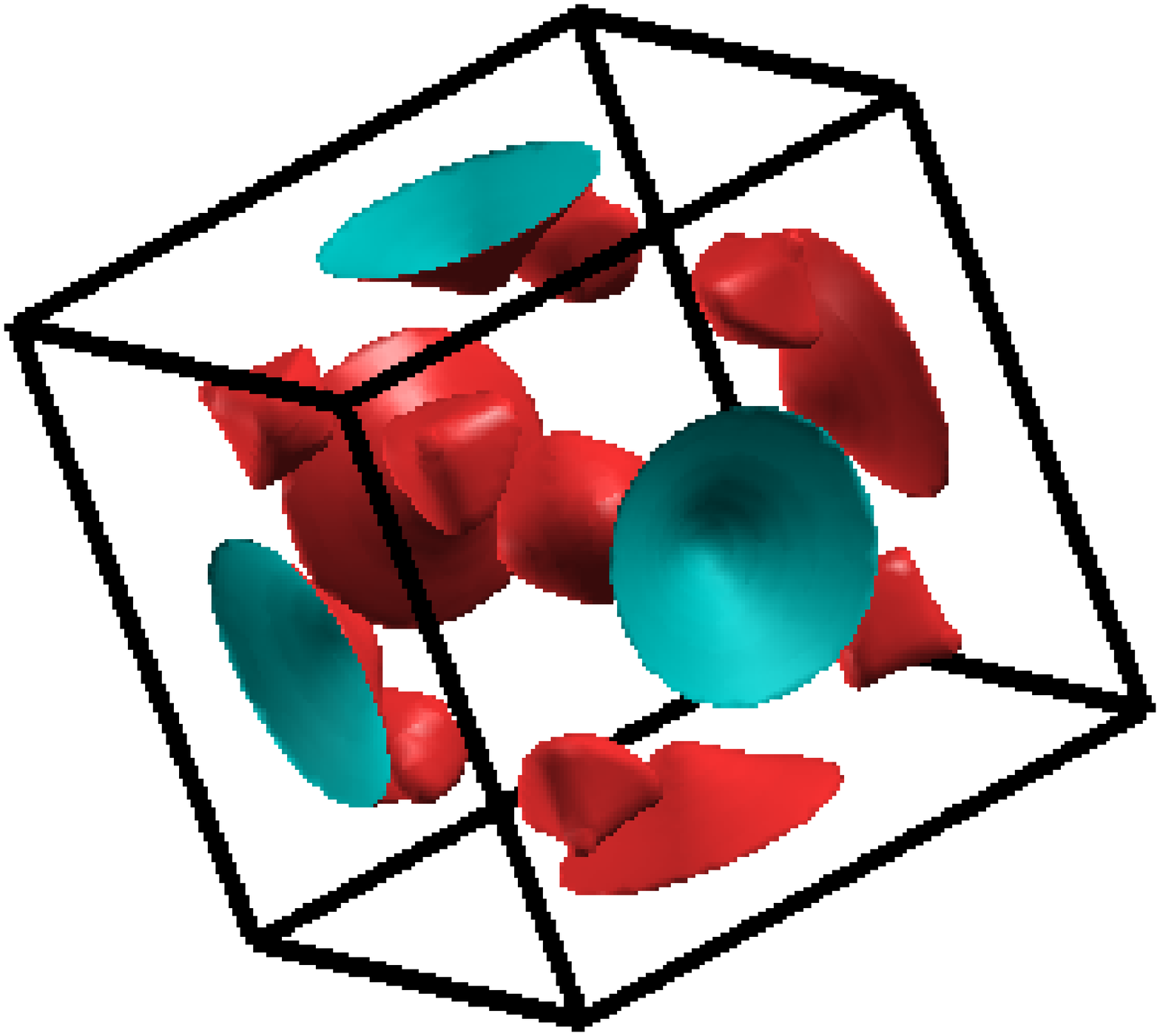}}\\
\caption{Fermi surface for Nb$_3$In at ambient (a)band no. 49, (b)band no. 50, (c)band no. 51, (d)band no. 52, (e)band no. 53 and  (f)band no. 54. The first four FS are having hole nature and remaining two are having electronic nature. The first FS having pockets at M points and BZ corners. The next three FS are having ribbon like sheets along BZ edges. The last two FS are having sheets in middle of the BZ faces with a pocket at $\Gamma$ point.}
\end{center}
\end{figure*}

\begin{figure*}
\begin{center}
\subfigure[]{\includegraphics[width=35mm,height=35mm]{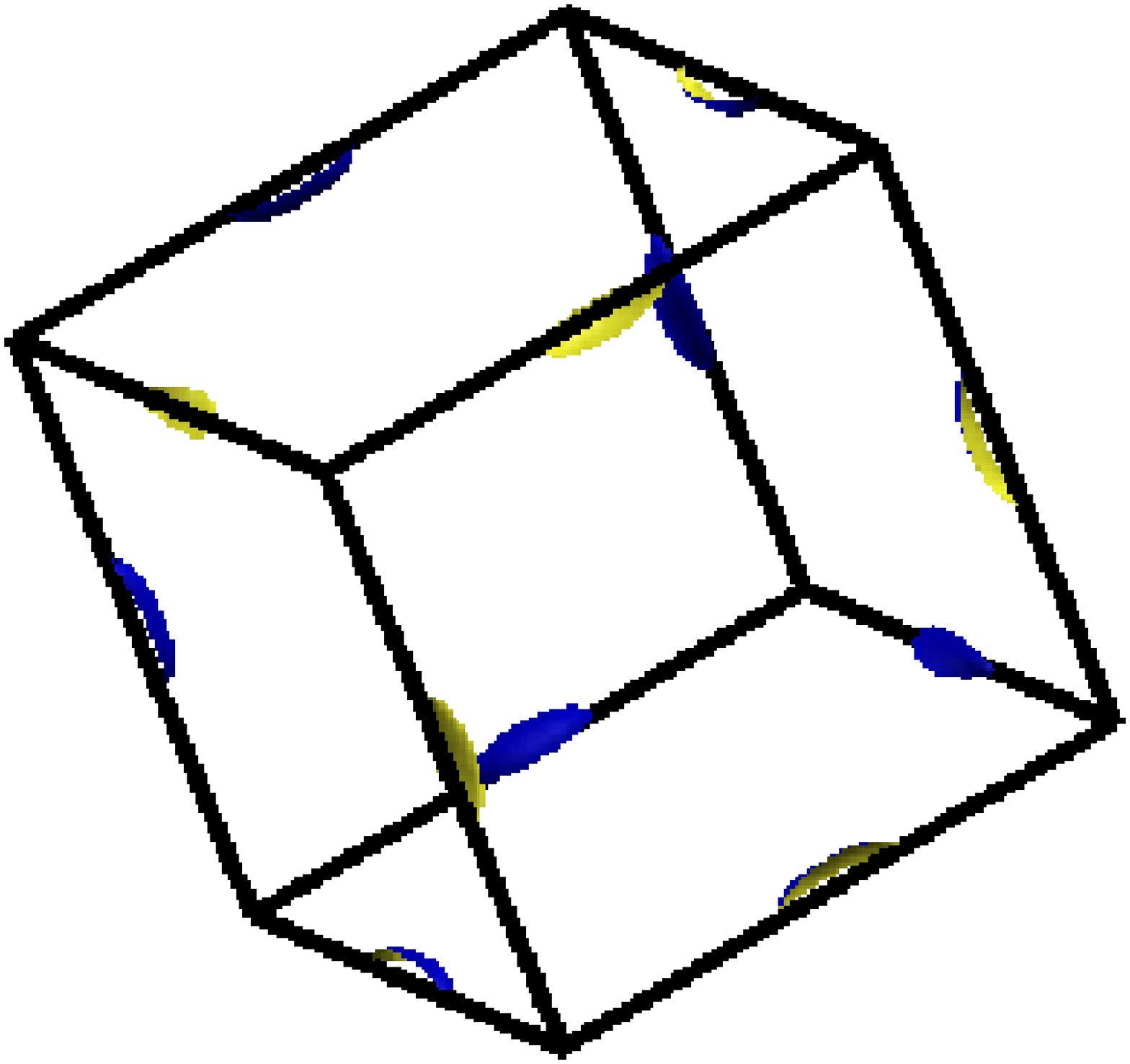}}
\subfigure[]{\includegraphics[width=35mm,height=35mm]{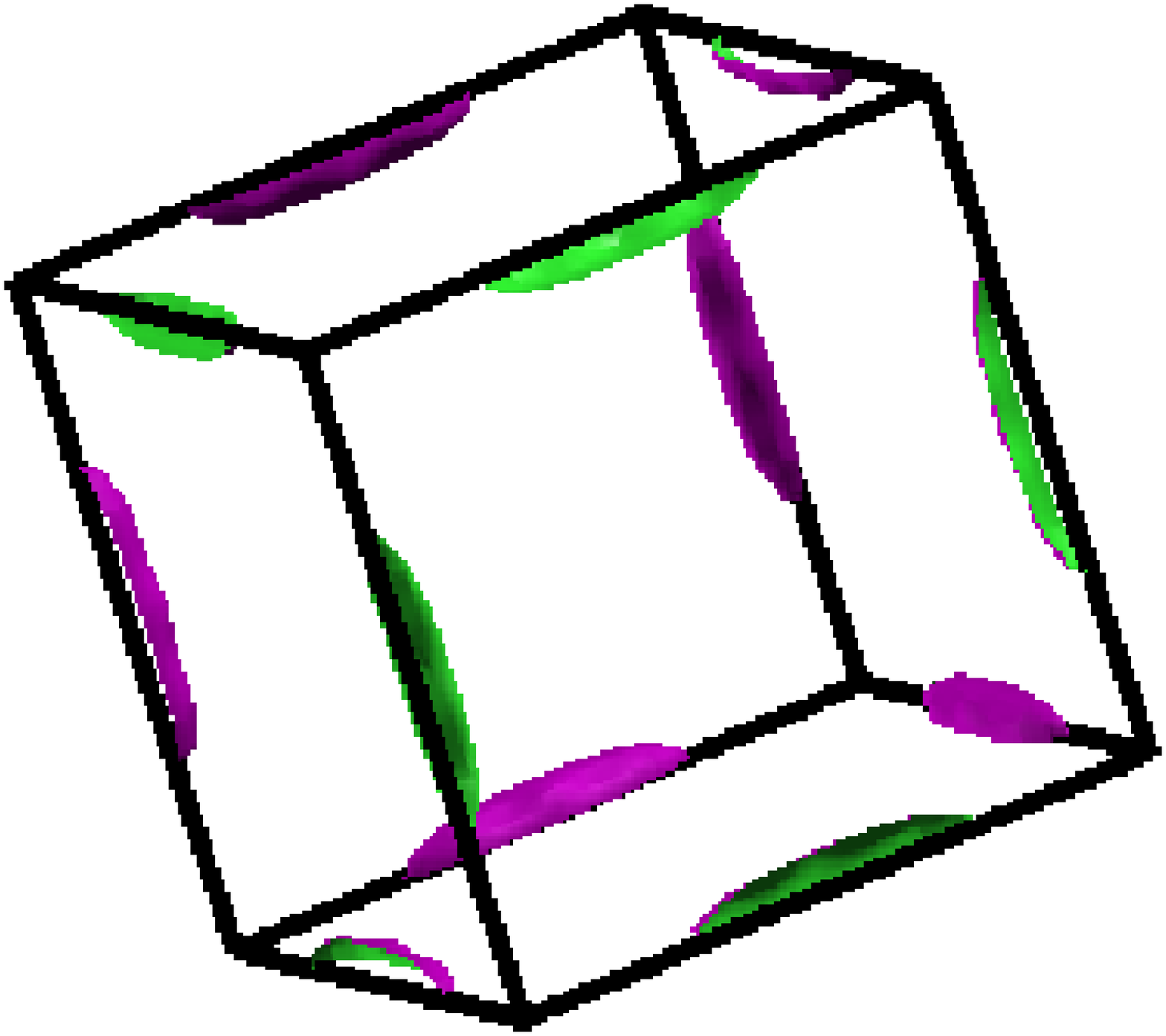}}
\subfigure[]{\includegraphics[width=35mm,height=35mm]{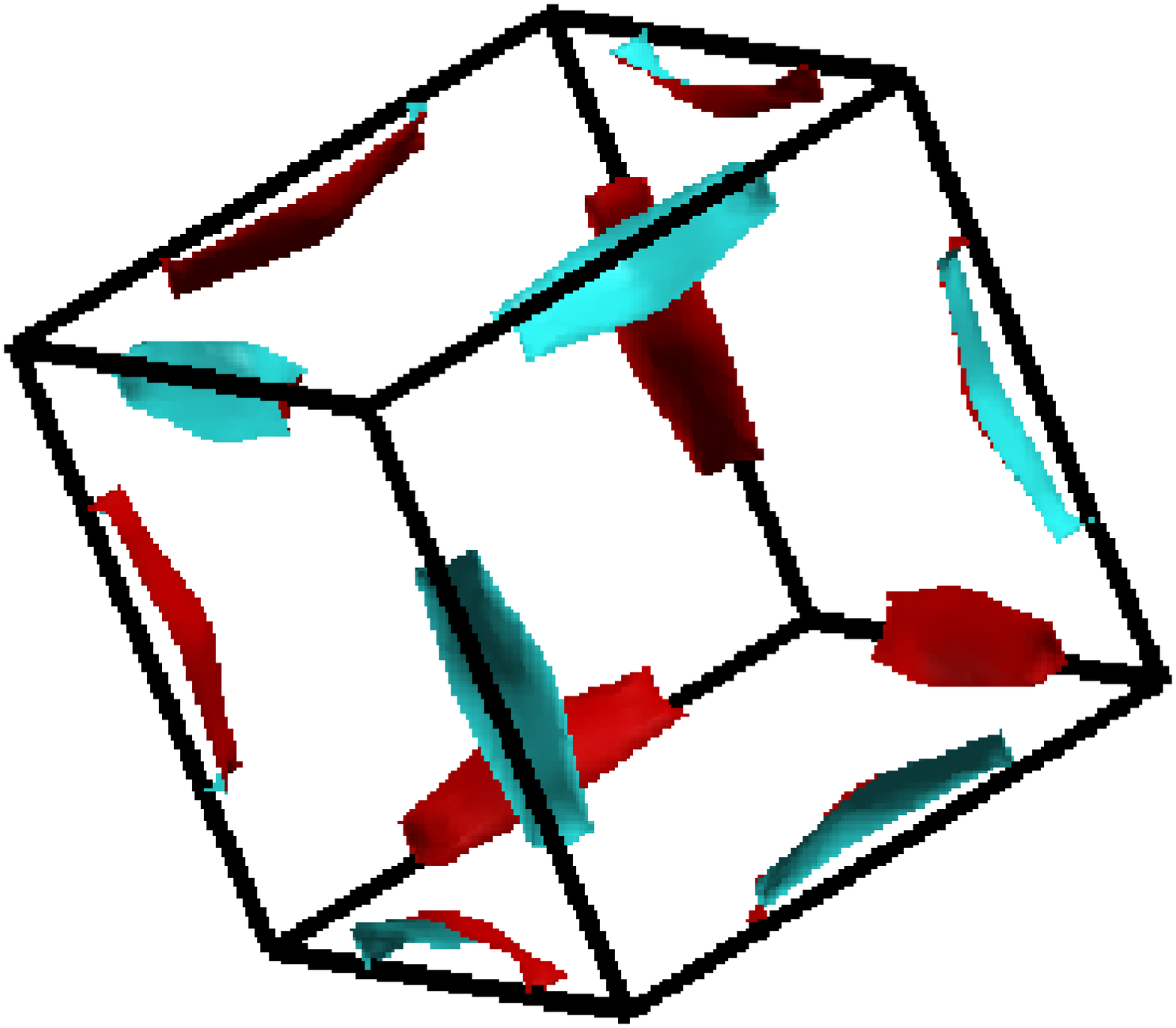}}\\
\subfigure[]{\includegraphics[width=35mm,height=35mm]{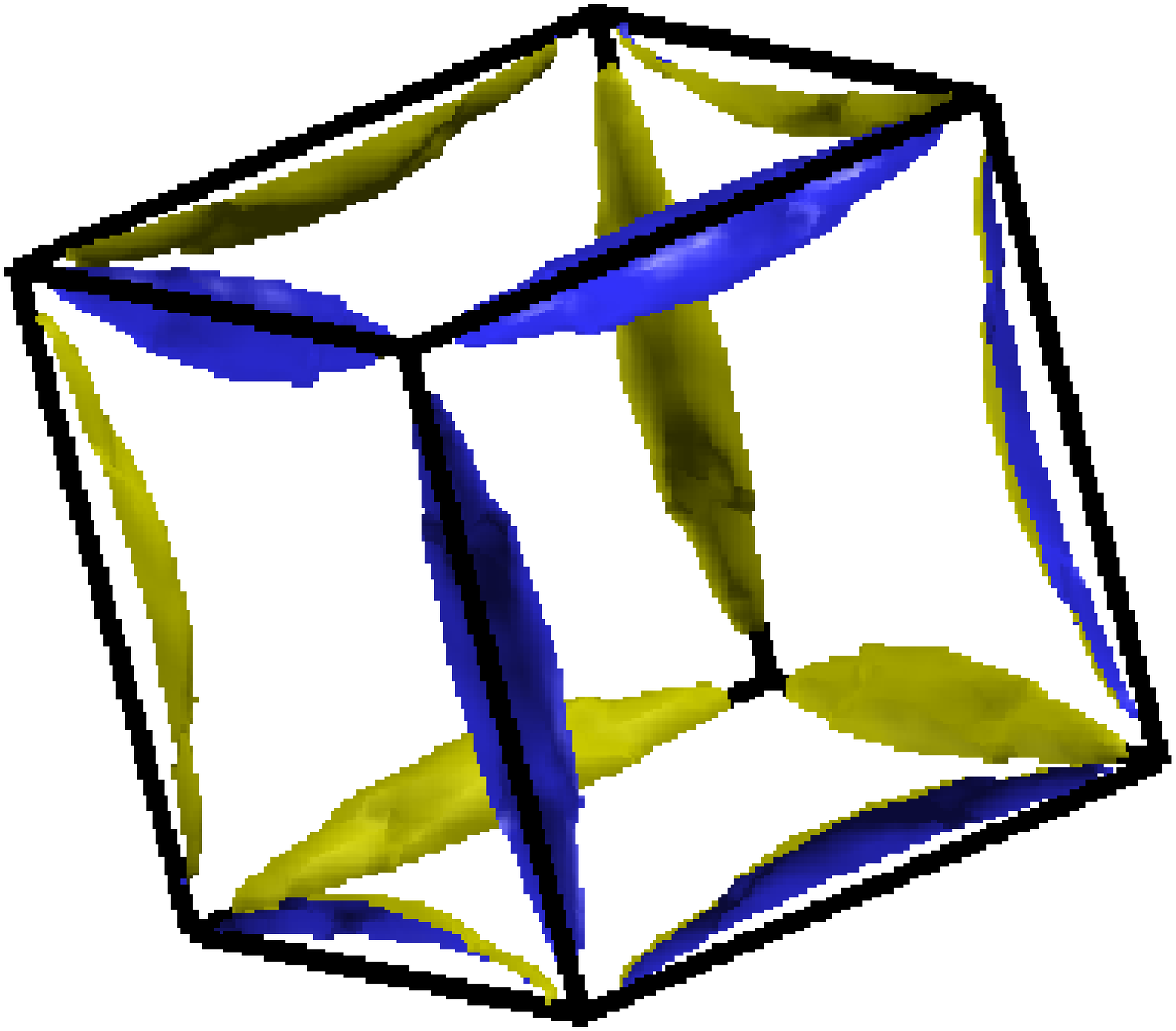}}
\subfigure[]{\includegraphics[width=35mm,height=35mm]{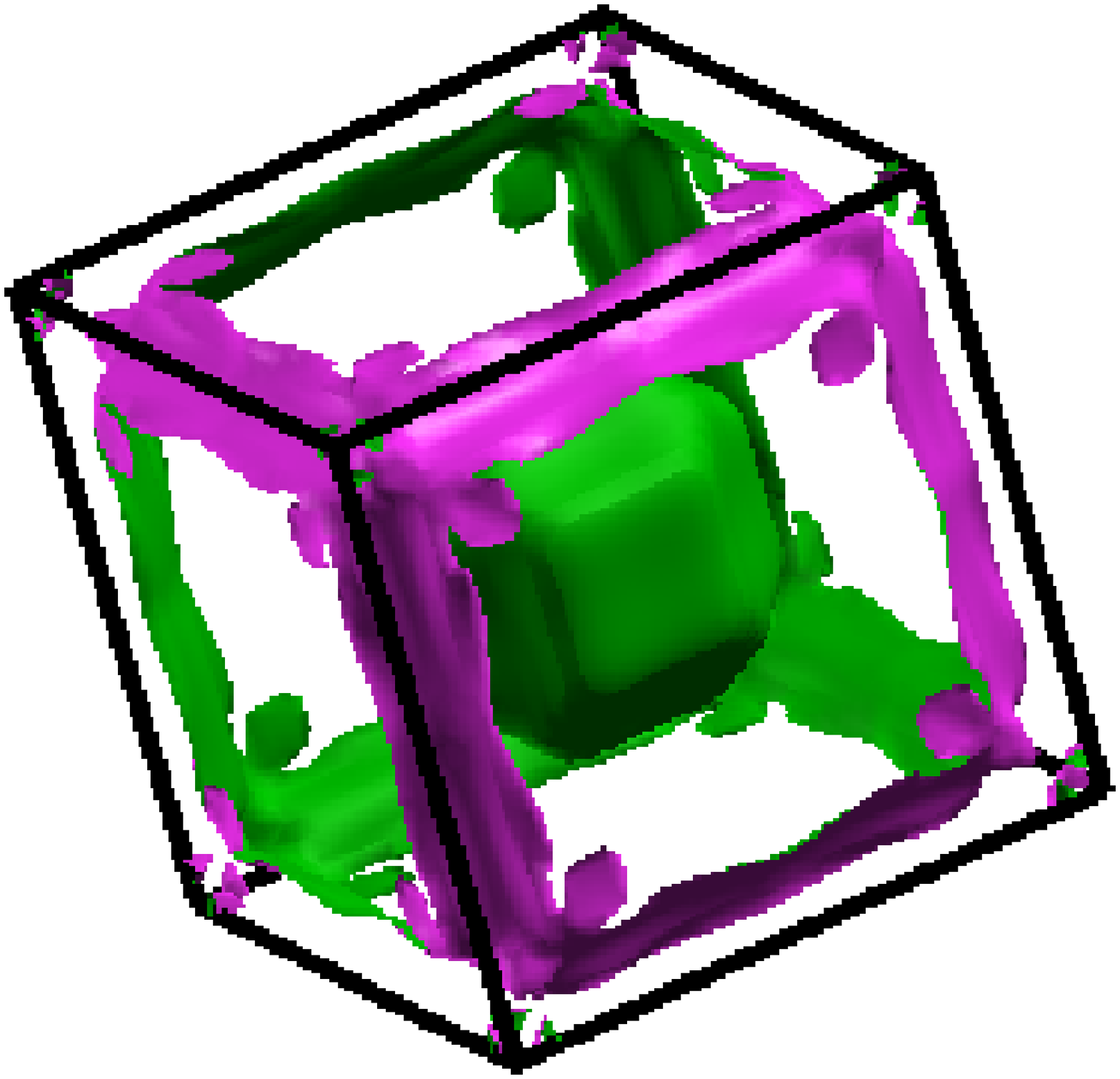}}
\subfigure[]{\includegraphics[width=35mm,height=35mm]{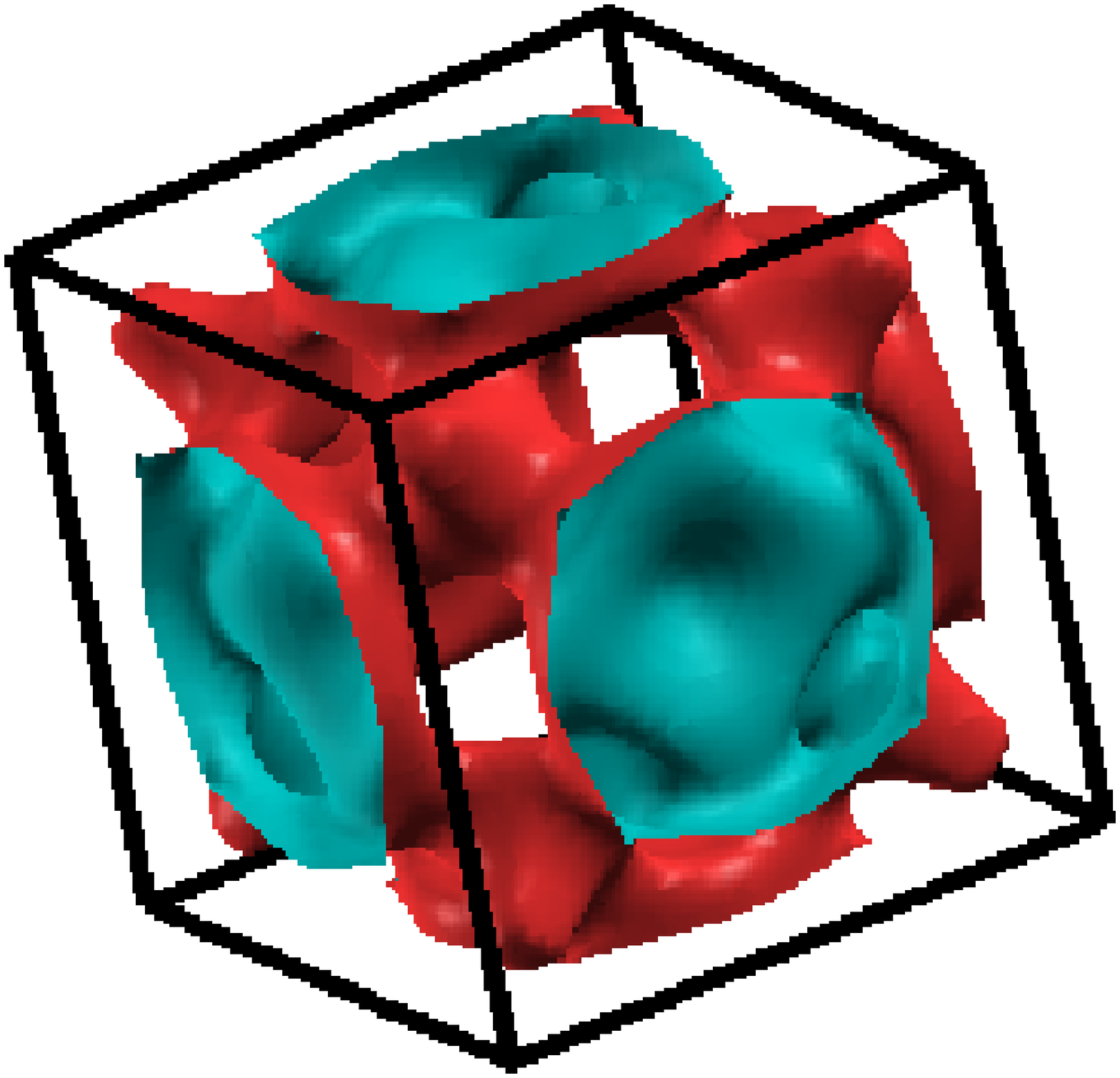}}\\
\caption{Fermi surface for Nb$_3$Ge at ambient (a)band no. 49, (b)band no. 50, (c)band no. 51, (d)band no. 52, (e)band no. 53 and  (f)band no. 54. The first four FS are having hole nature and remaining two are having electronic nature. First four FS are having pockets at M points. Next FS having ribbon like sheets along BZ edges with a pocket at $\Gamma$ point. The last FS having sheets in middle of the BZ faces with a pocket at $\Gamma$ point.}
\end{center}
\end{figure*}

\begin{figure*}
\begin{center}
\subfigure[]{\includegraphics[width=35mm,height=35mm]{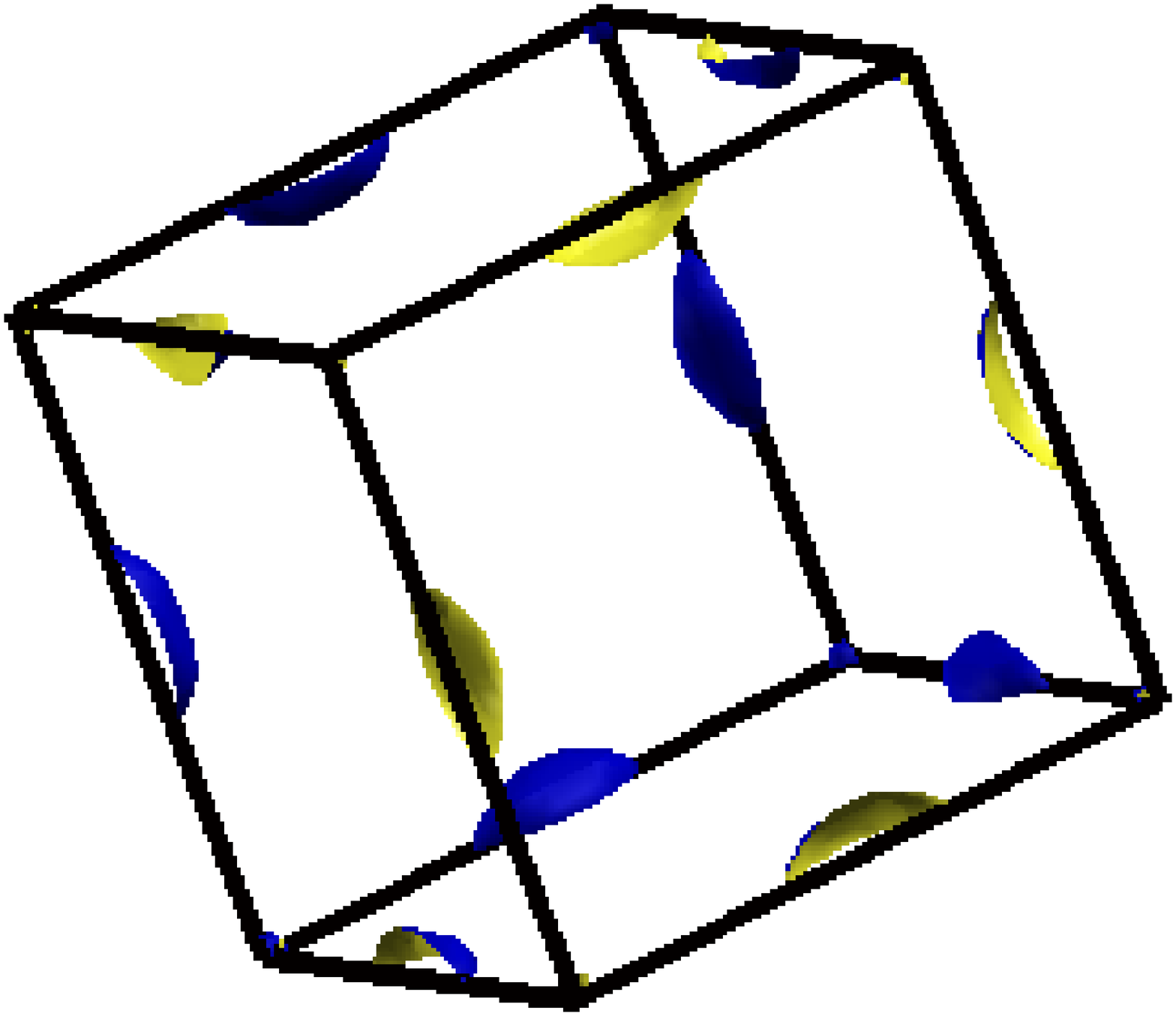}}
\subfigure[]{\includegraphics[width=35mm,height=35mm]{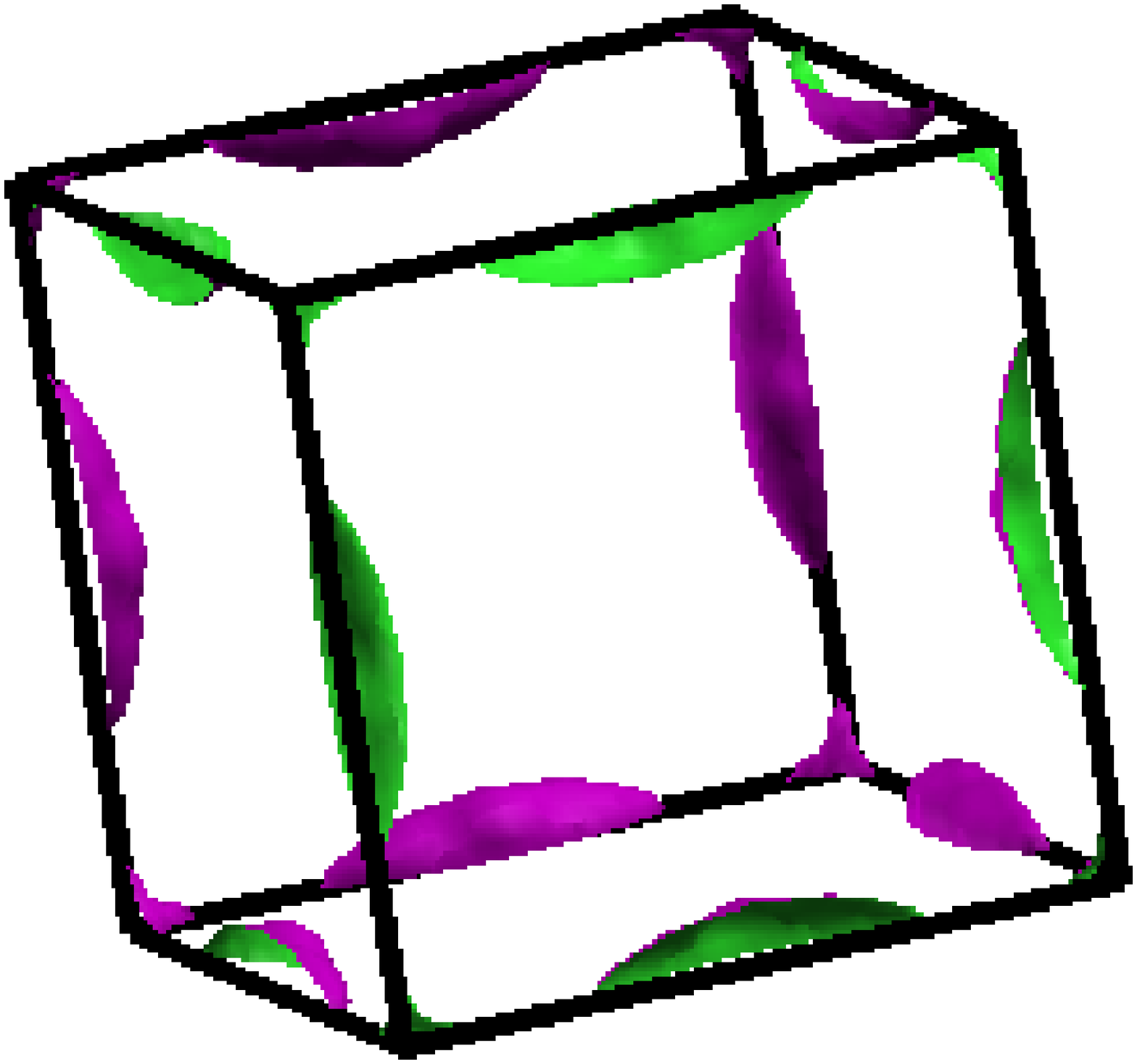}}
\subfigure[]{\includegraphics[width=35mm,height=35mm]{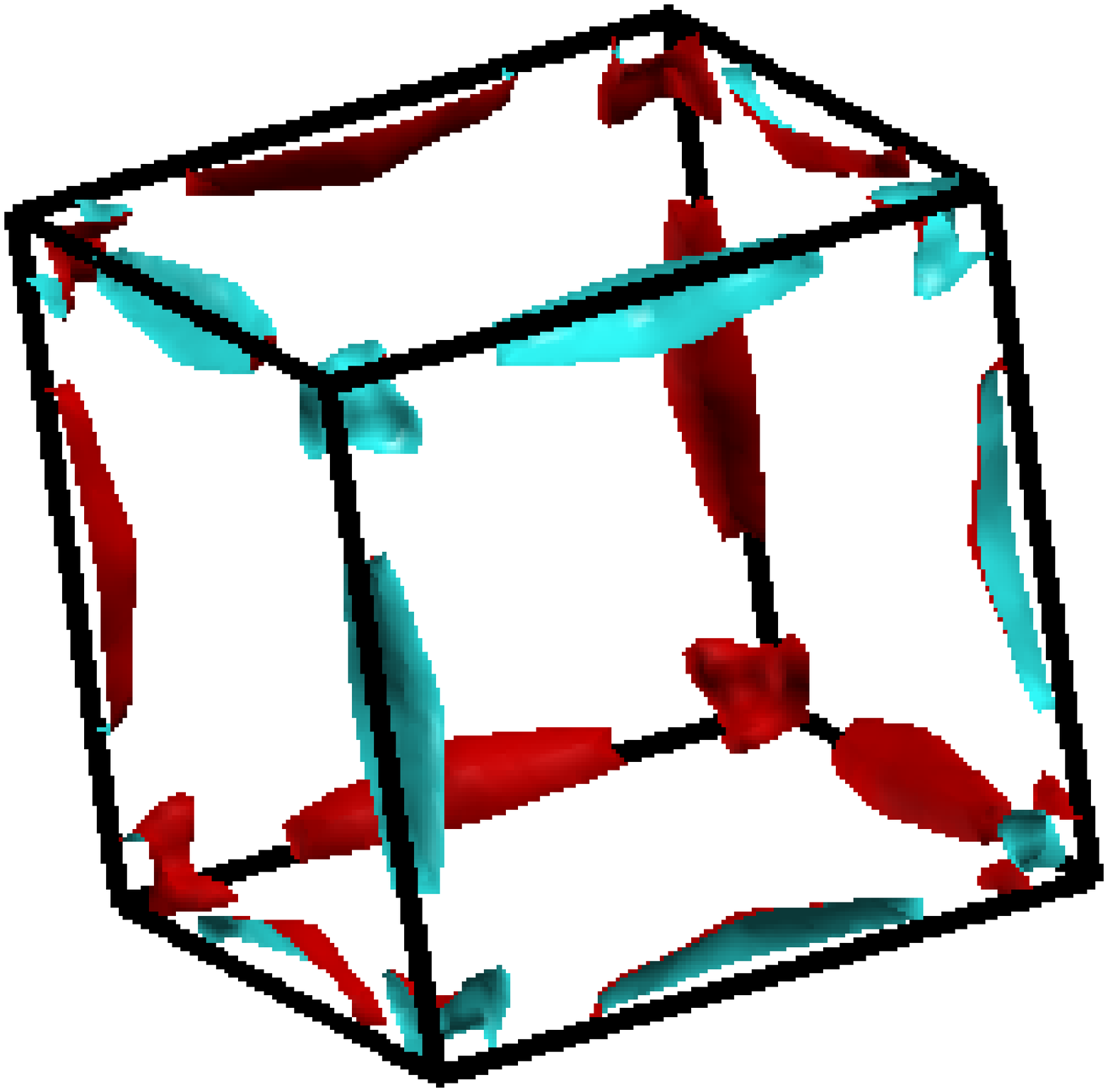}}\\
\subfigure[]{\includegraphics[width=35mm,height=35mm]{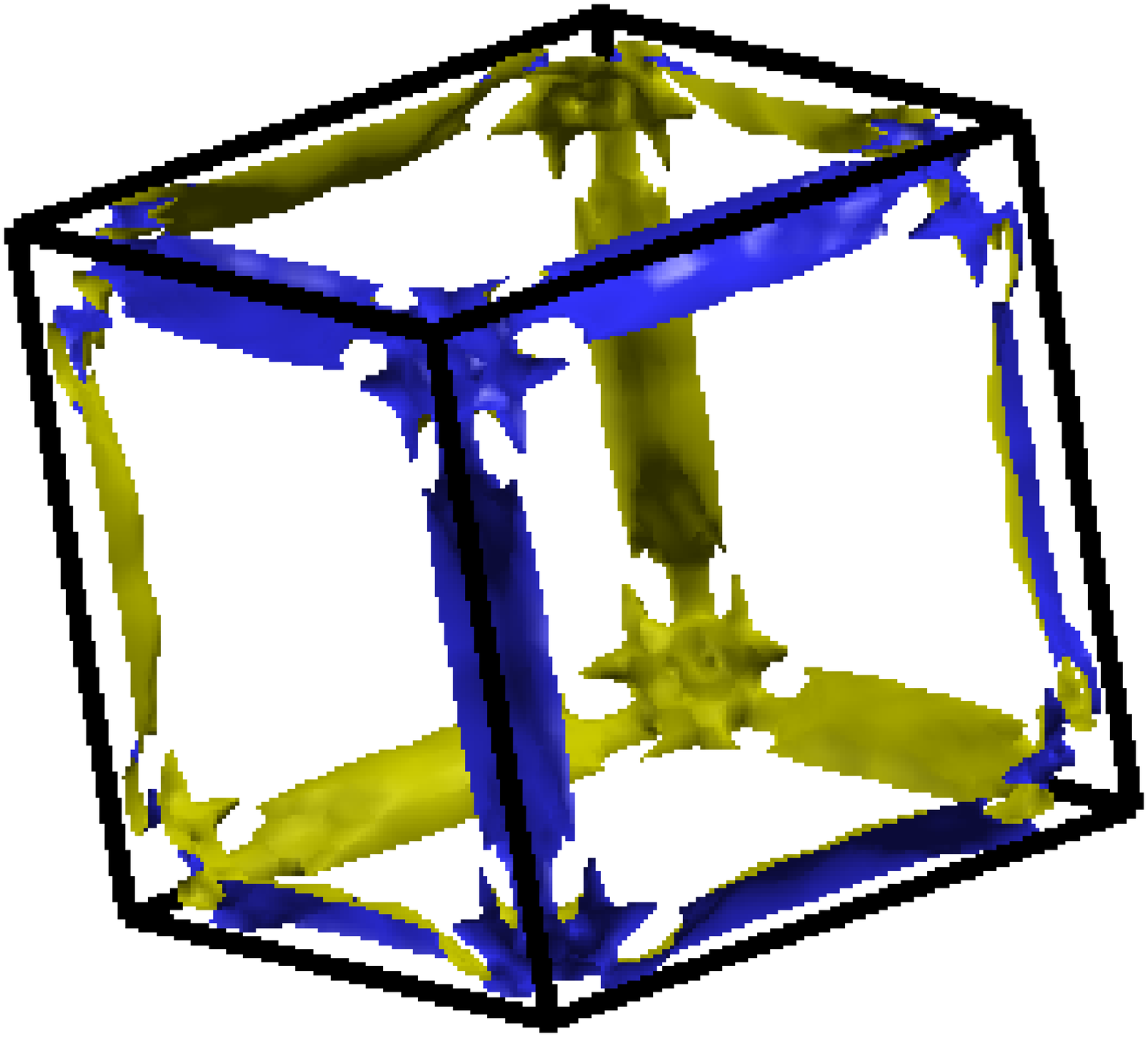}}
\subfigure[]{\includegraphics[width=35mm,height=35mm]{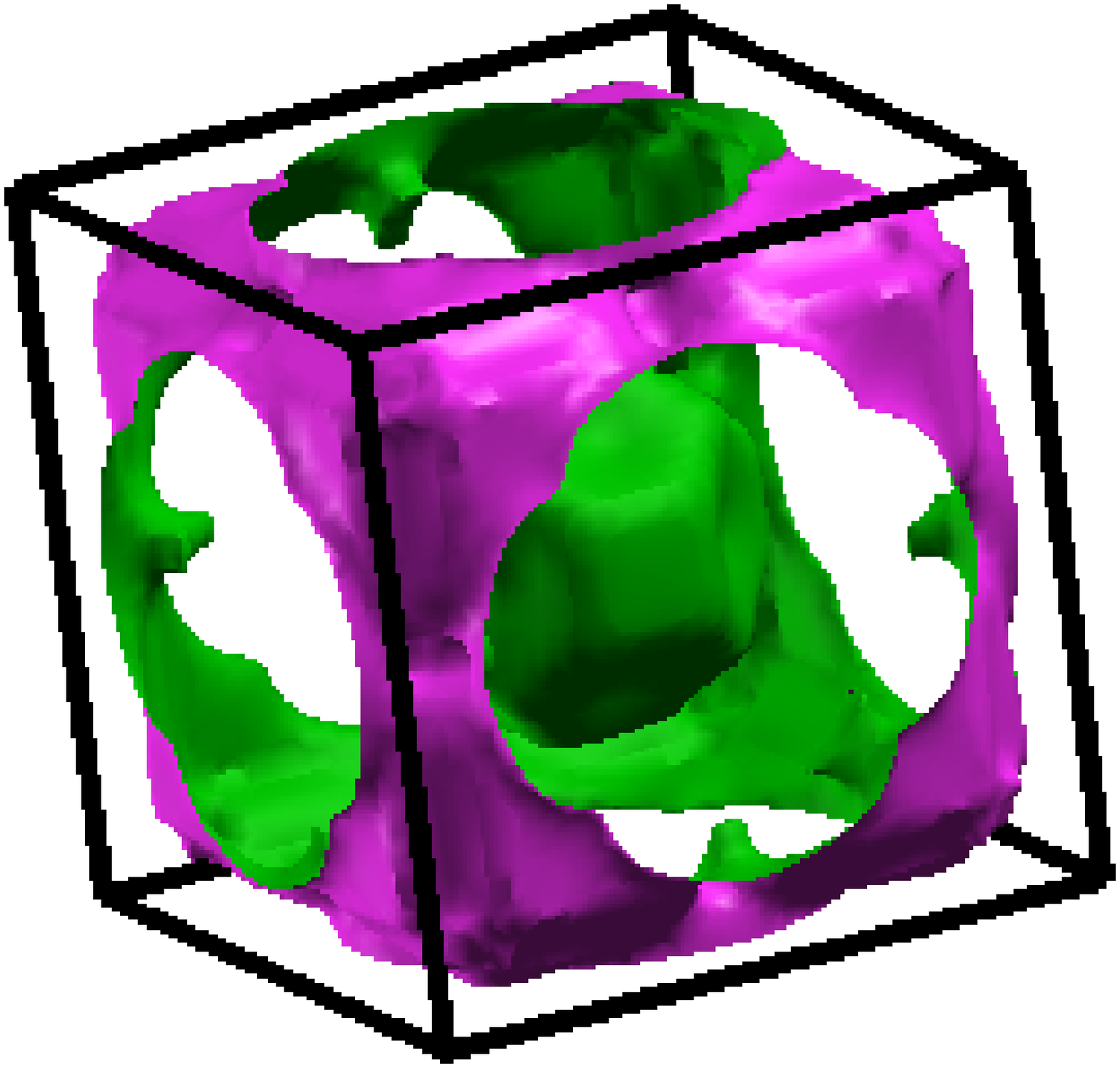}}
\subfigure[]{\includegraphics[width=35mm,height=35mm]{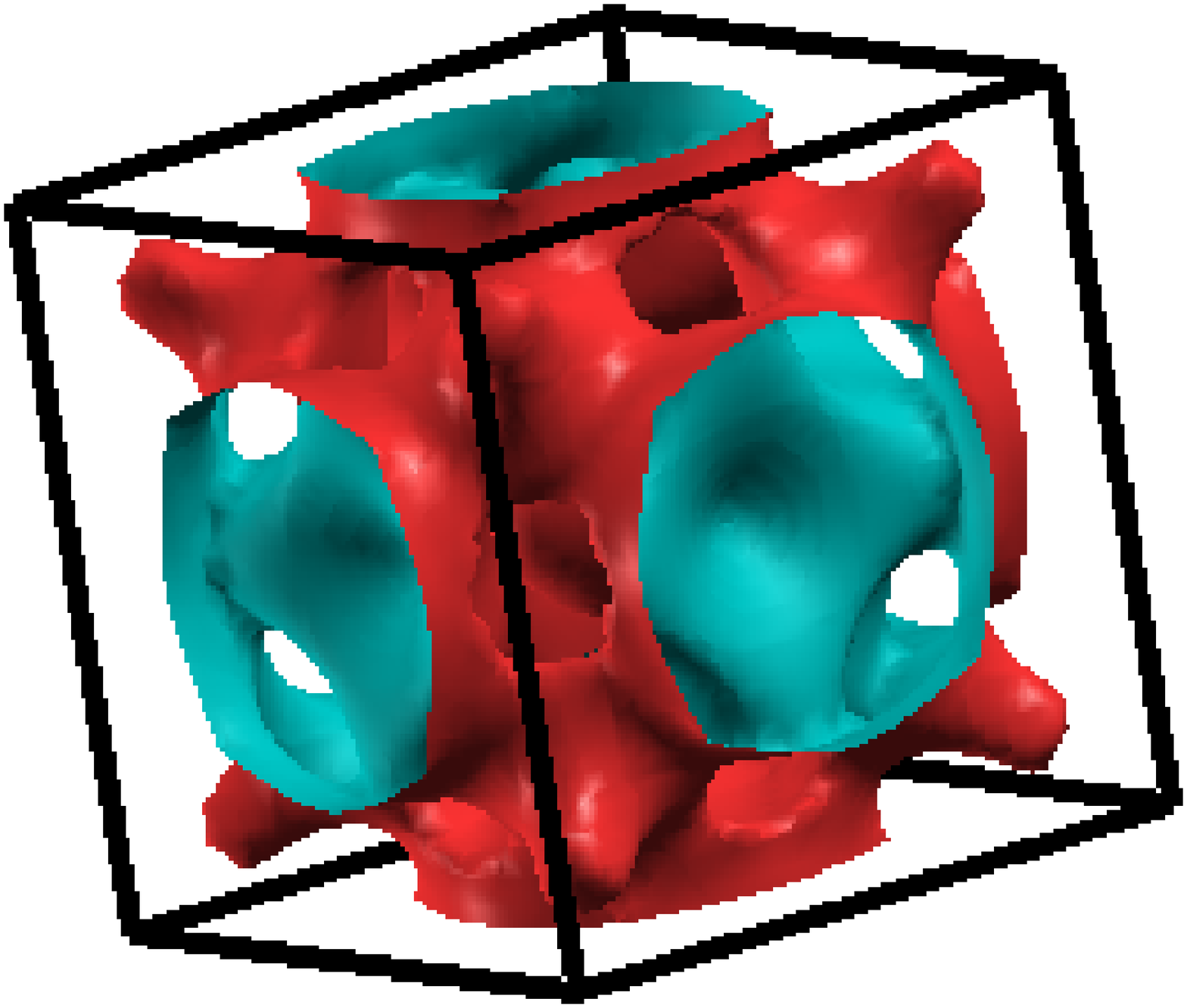}}\\
\caption{Fermi surface for Nb$_3$Sn at ambient (a)band no. 49, (b)band no. 50, (c)band no. 51, (d)band no. 52, (e)band no. 53 and  (f)band no. 54. The first four FS are having hole nature and remaining two are having electronic nature. First three FS are having pockets at M points and corners of the BZ. Next FS having ribbon like sheets along BZ edges. The next FS having sheets along BZ edges with a pocket at $\Gamma$ point. The last FS having sheets in middle of the BZ faces with a pocket at $\Gamma$ point.}
\end{center}
\end{figure*}


\begin{figure*}
\begin{center}
\subfigure[]{\includegraphics[width=60mm,height=65mm]{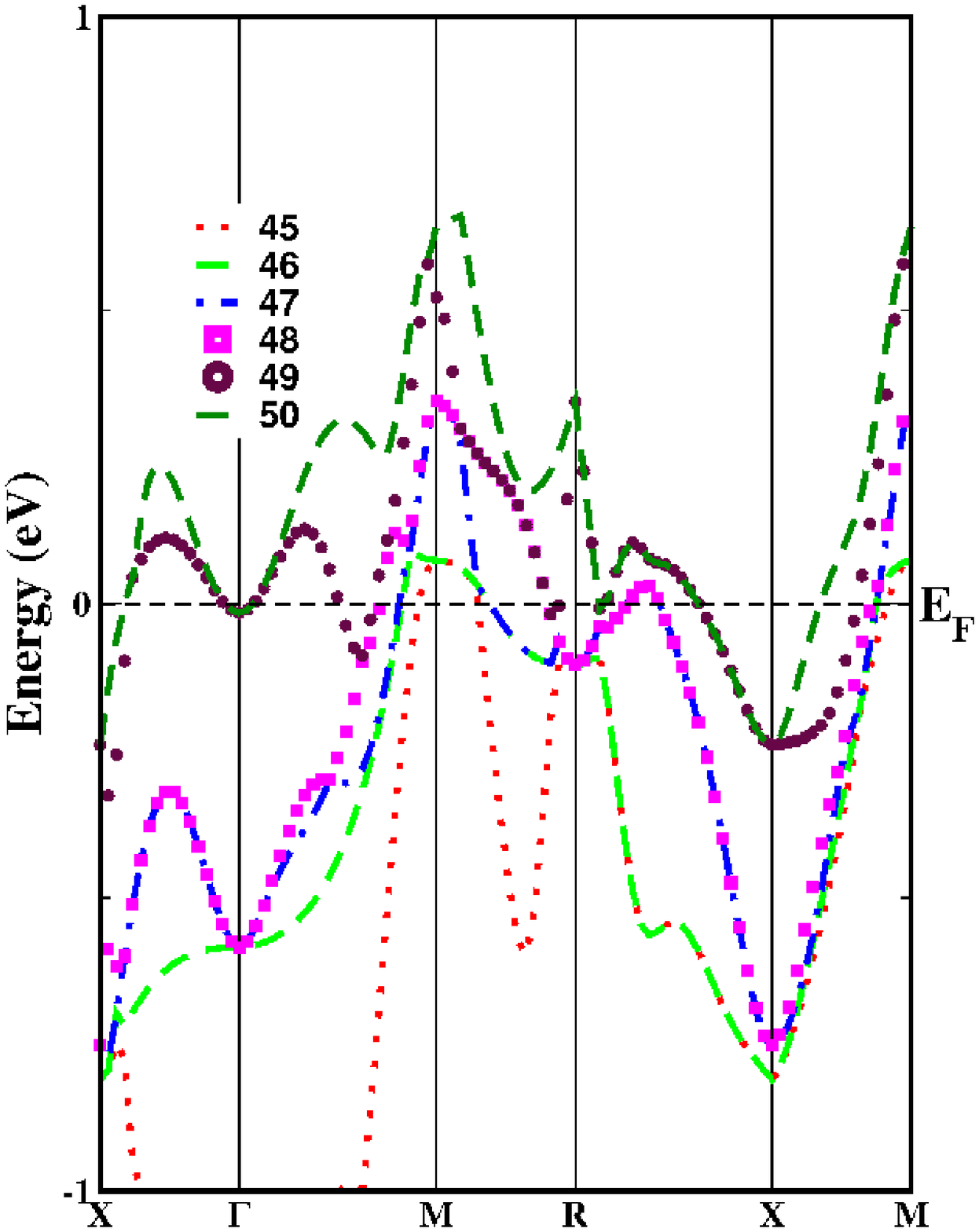}}
\subfigure[]{\includegraphics[width=60mm,height=65mm]{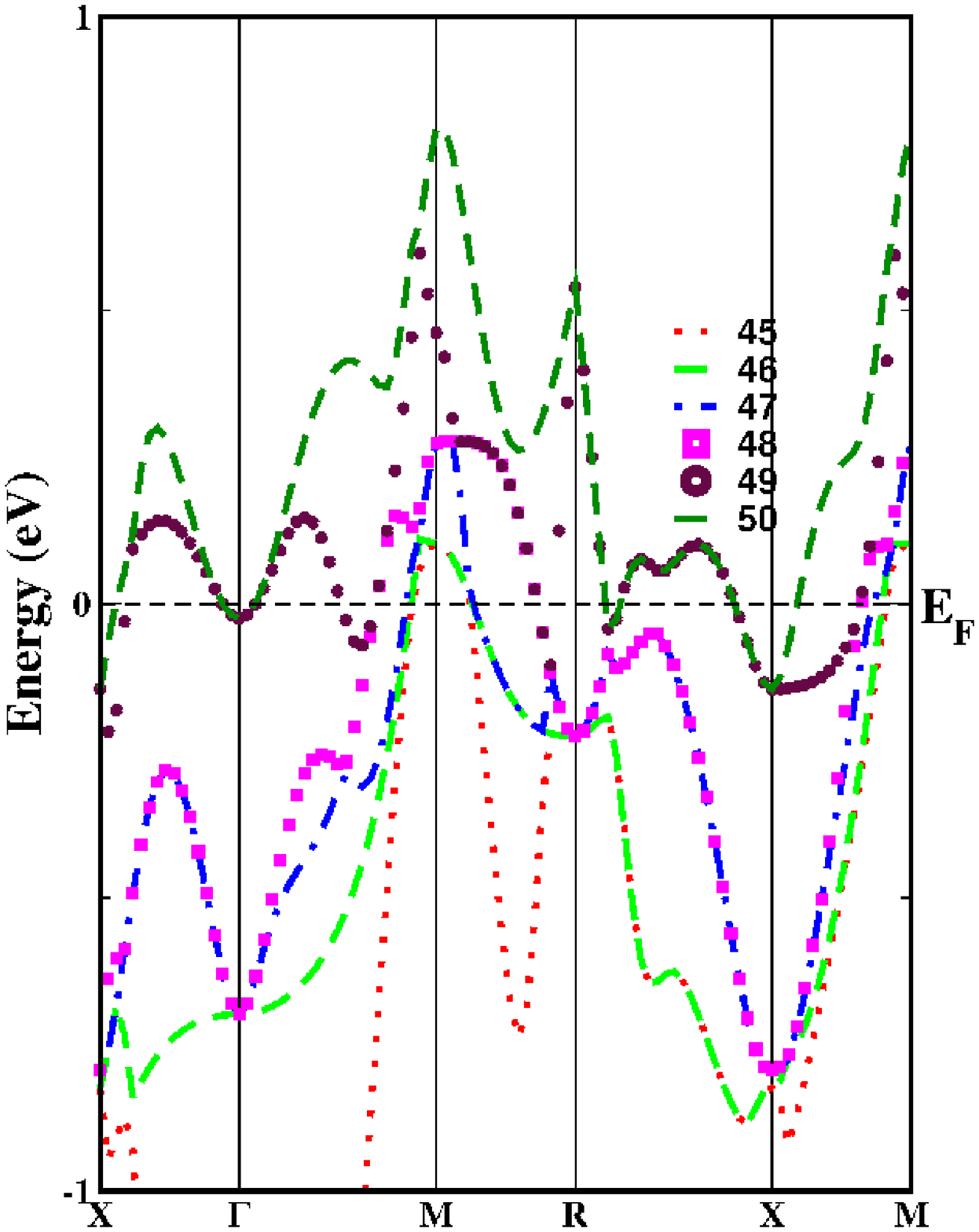}}\\
\subfigure[]{\includegraphics[width=35mm,height=35mm]{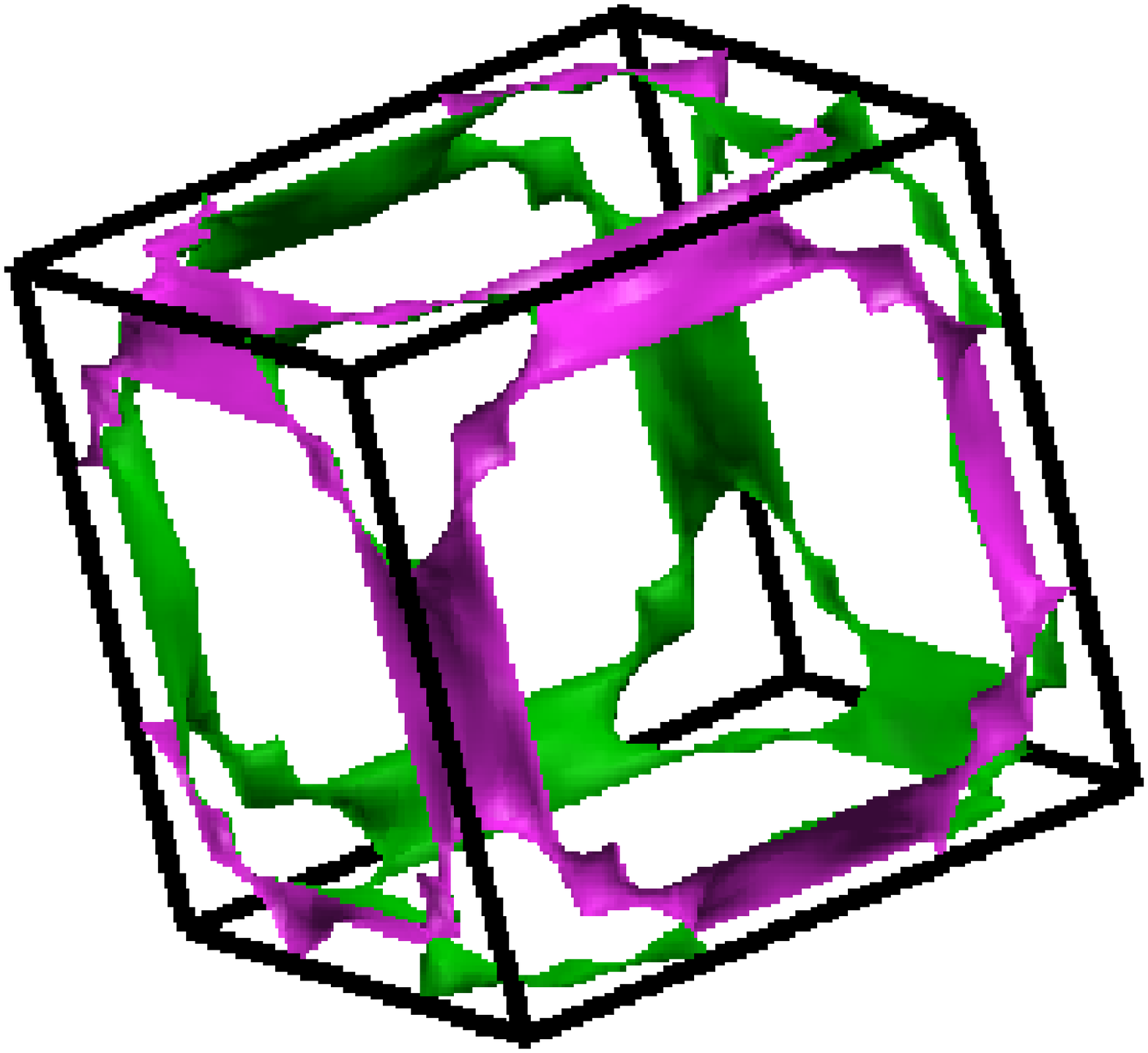}}
\subfigure[]{\includegraphics[width=35mm,height=35mm]{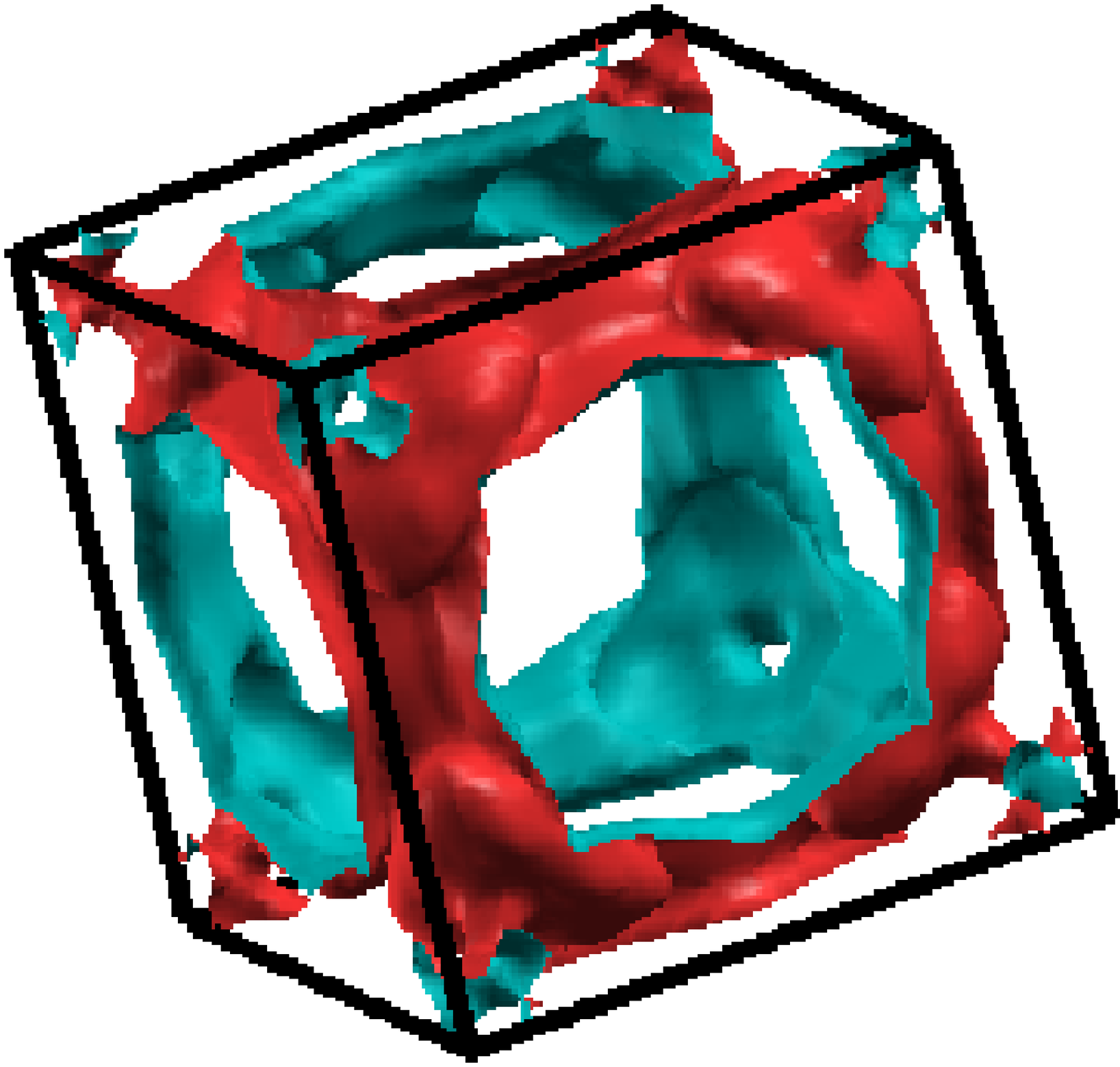}}
\subfigure[]{\includegraphics[width=35mm,height=35mm]{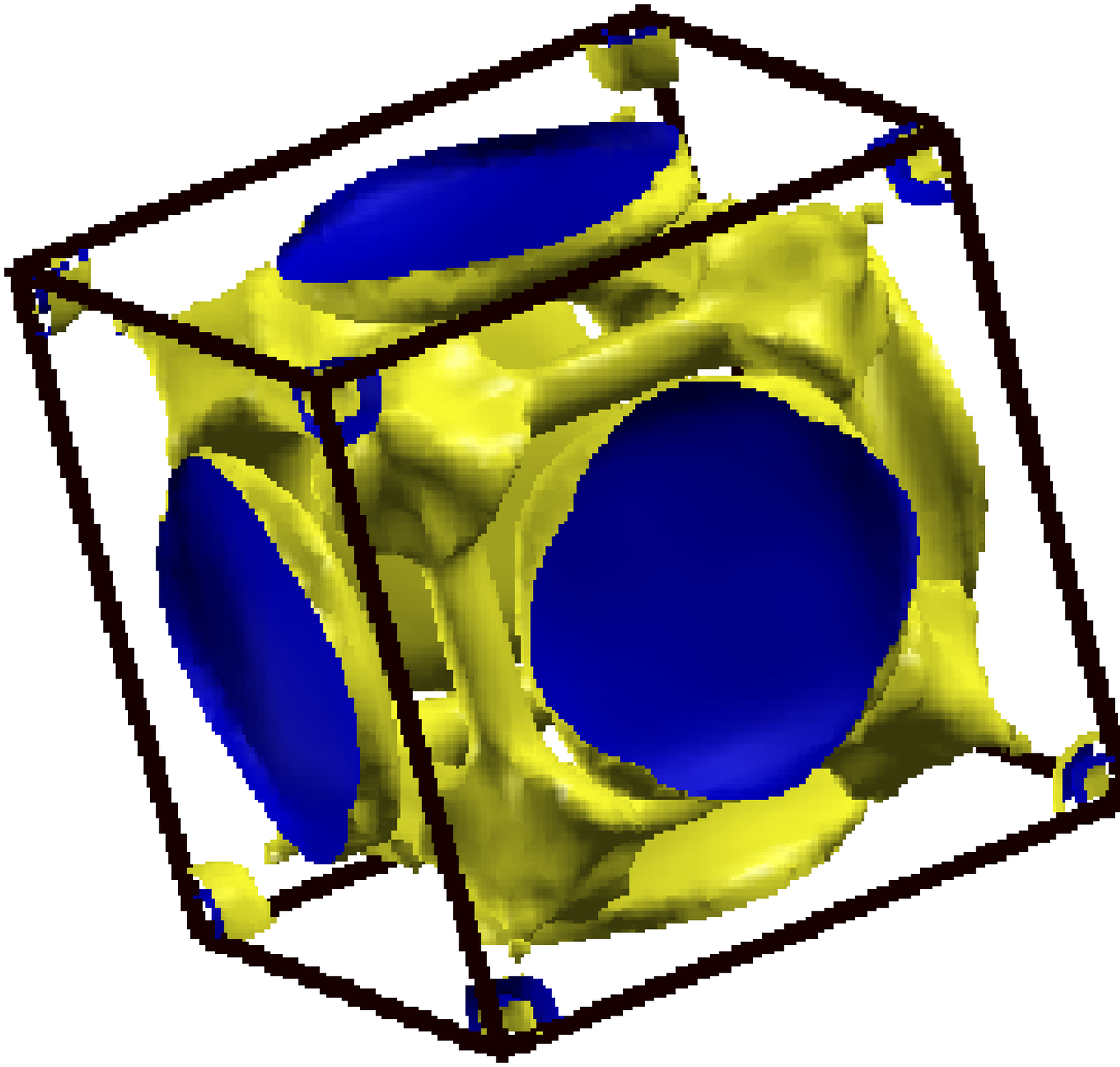}}\\
\subfigure[]{\includegraphics[width=35mm,height=35mm]{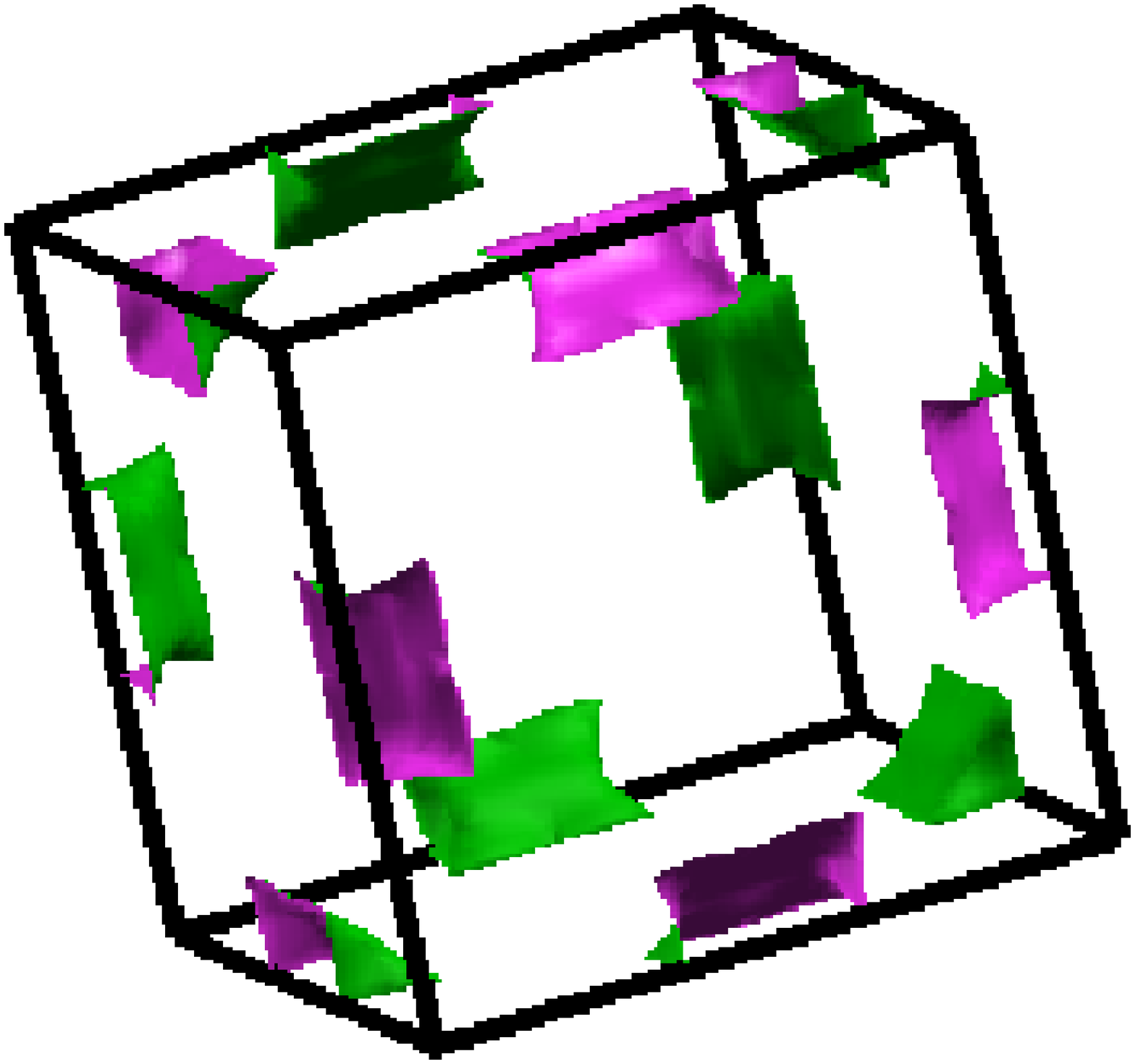}}
\subfigure[]{\includegraphics[width=35mm,height=35mm]{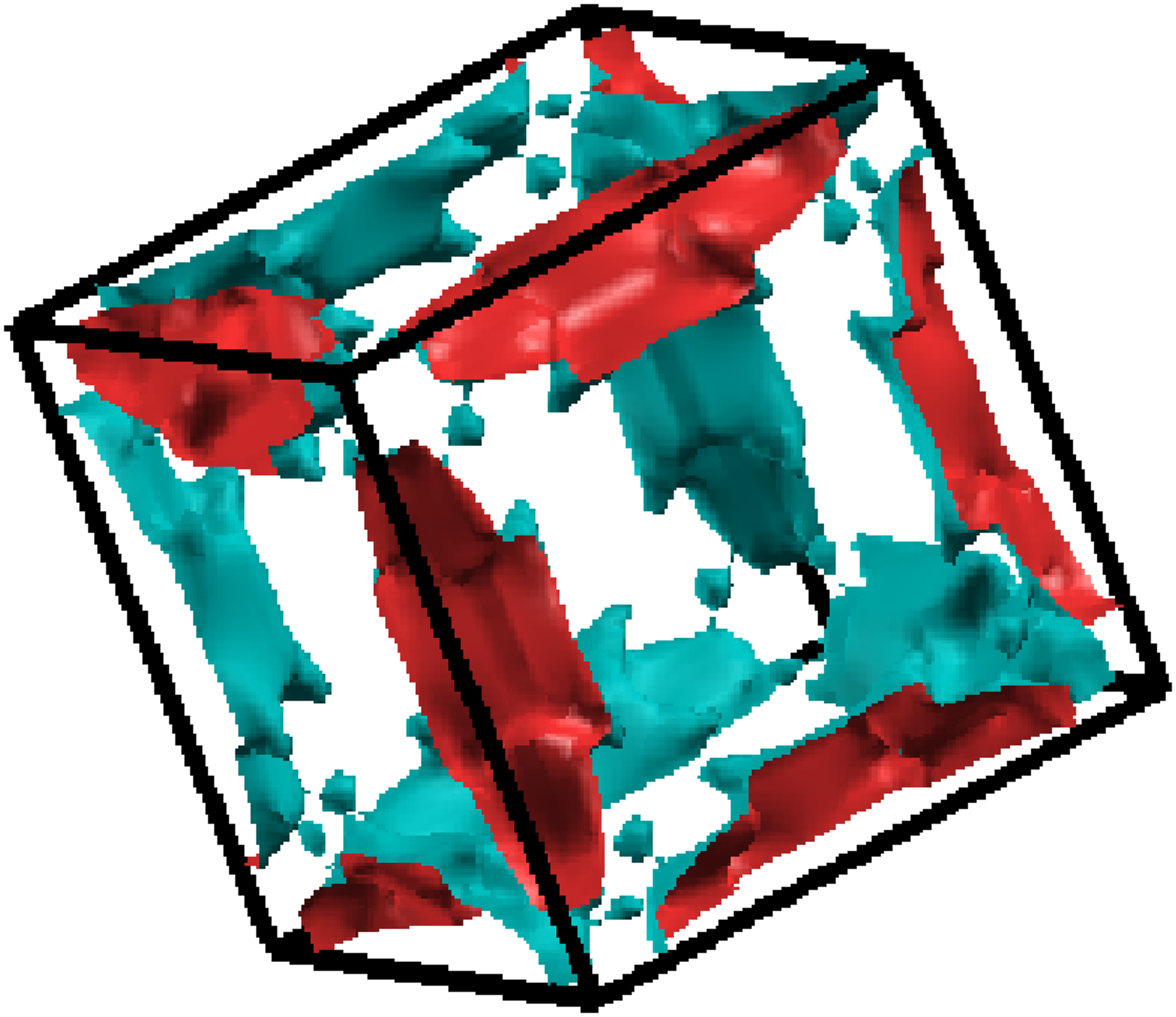}}
\subfigure[]{\includegraphics[width=35mm,height=35mm]{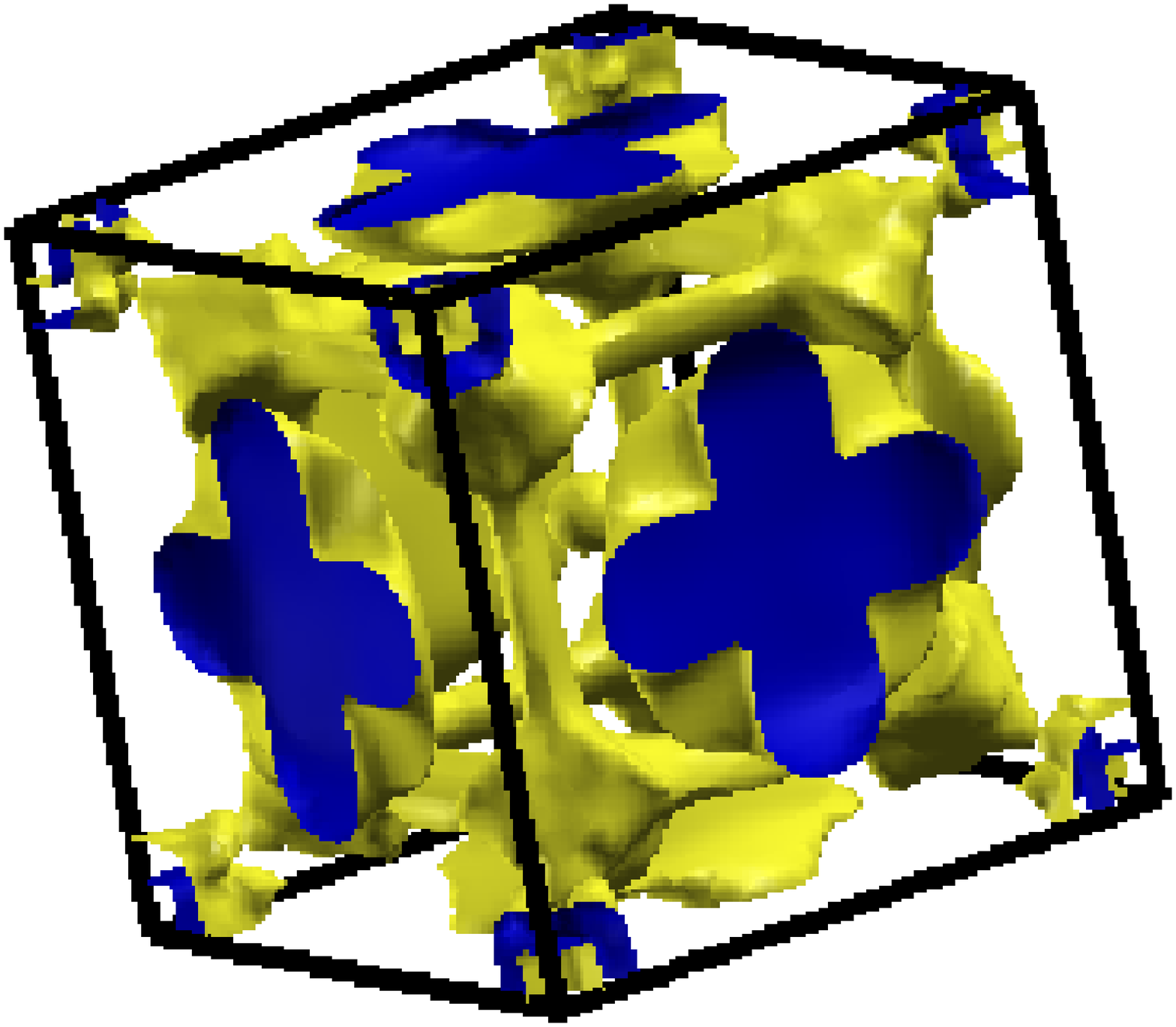}}\\
\caption{Band structure of Nb$_3$Al at (a)V/V$_0$ = 1.00, (b)V/V$_0$ = 0.90 (pressure of 21.5 GPa) and FS for which change in FS is observed at ambient (c), (d), (e)for band no. 47, 48 and 49 and at V/V$_0$ = 0.90 (f), (g) and (h) where change in FS topology is observed.}
\end{center}
\end{figure*}

\begin{figure*}
\begin{center}
\subfigure[]{\includegraphics[width=60mm,height=65mm]{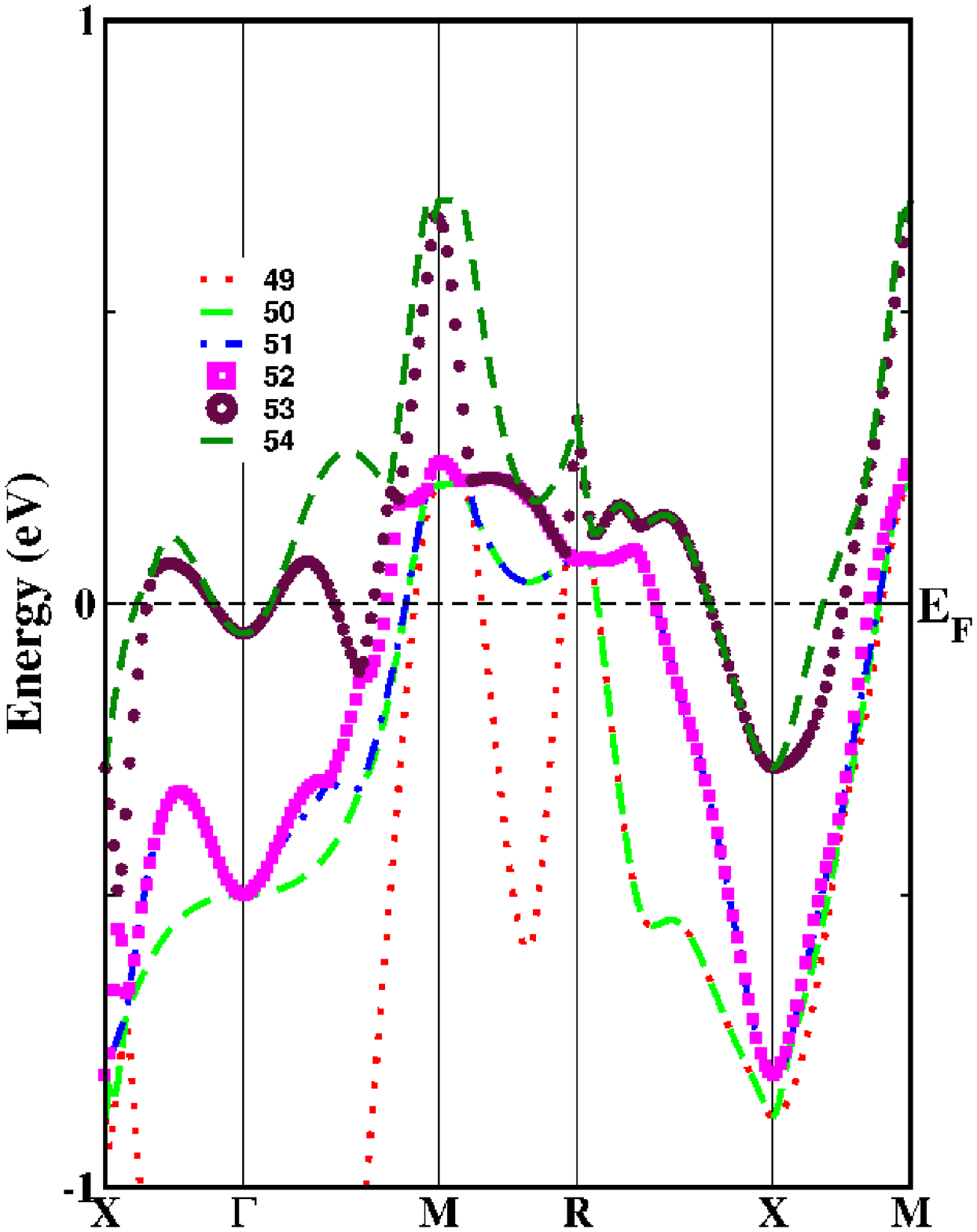}}
\subfigure[]{\includegraphics[width=60mm,height=65mm]{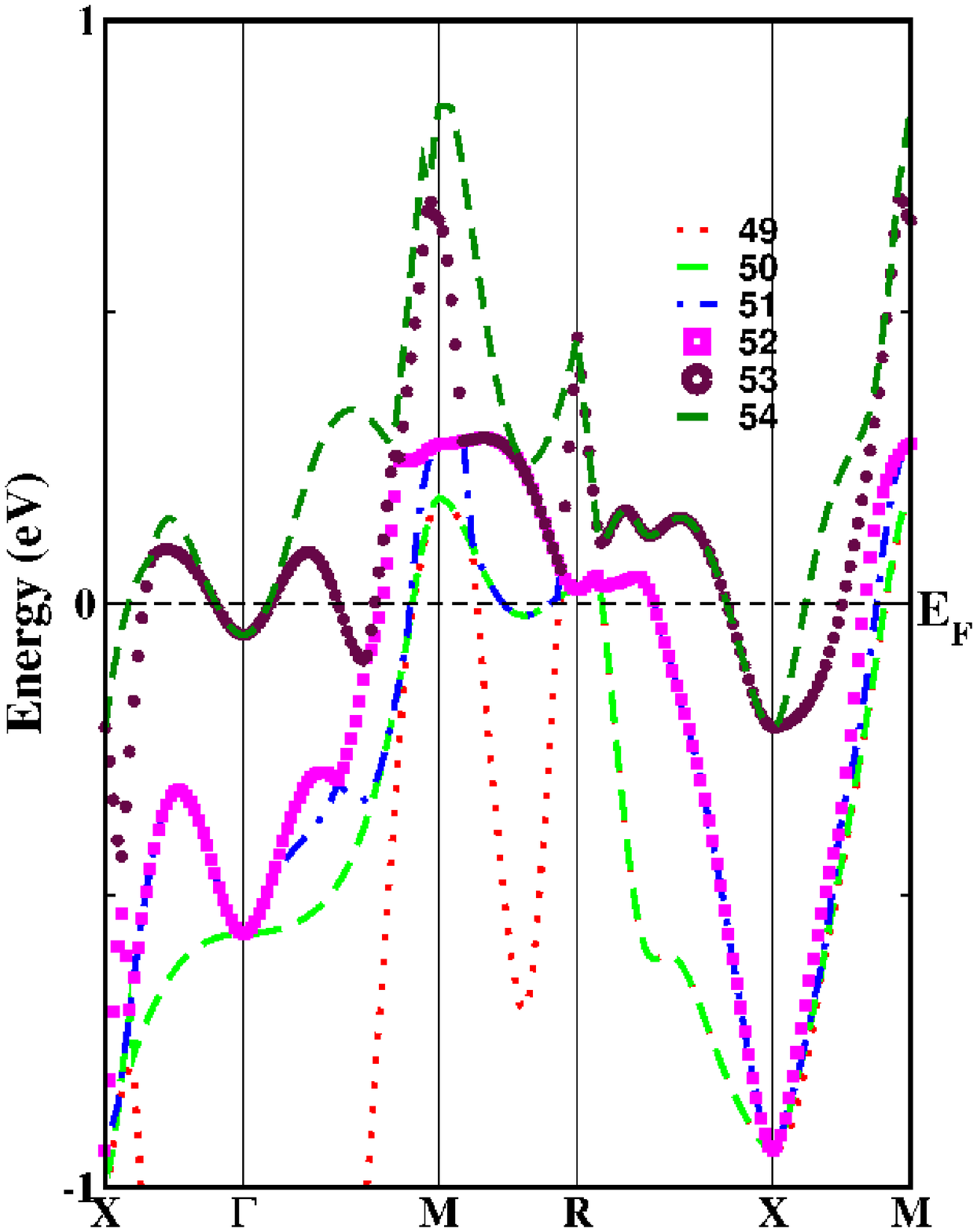}}\\
\subfigure[]{\includegraphics[width=35mm,height=35mm]{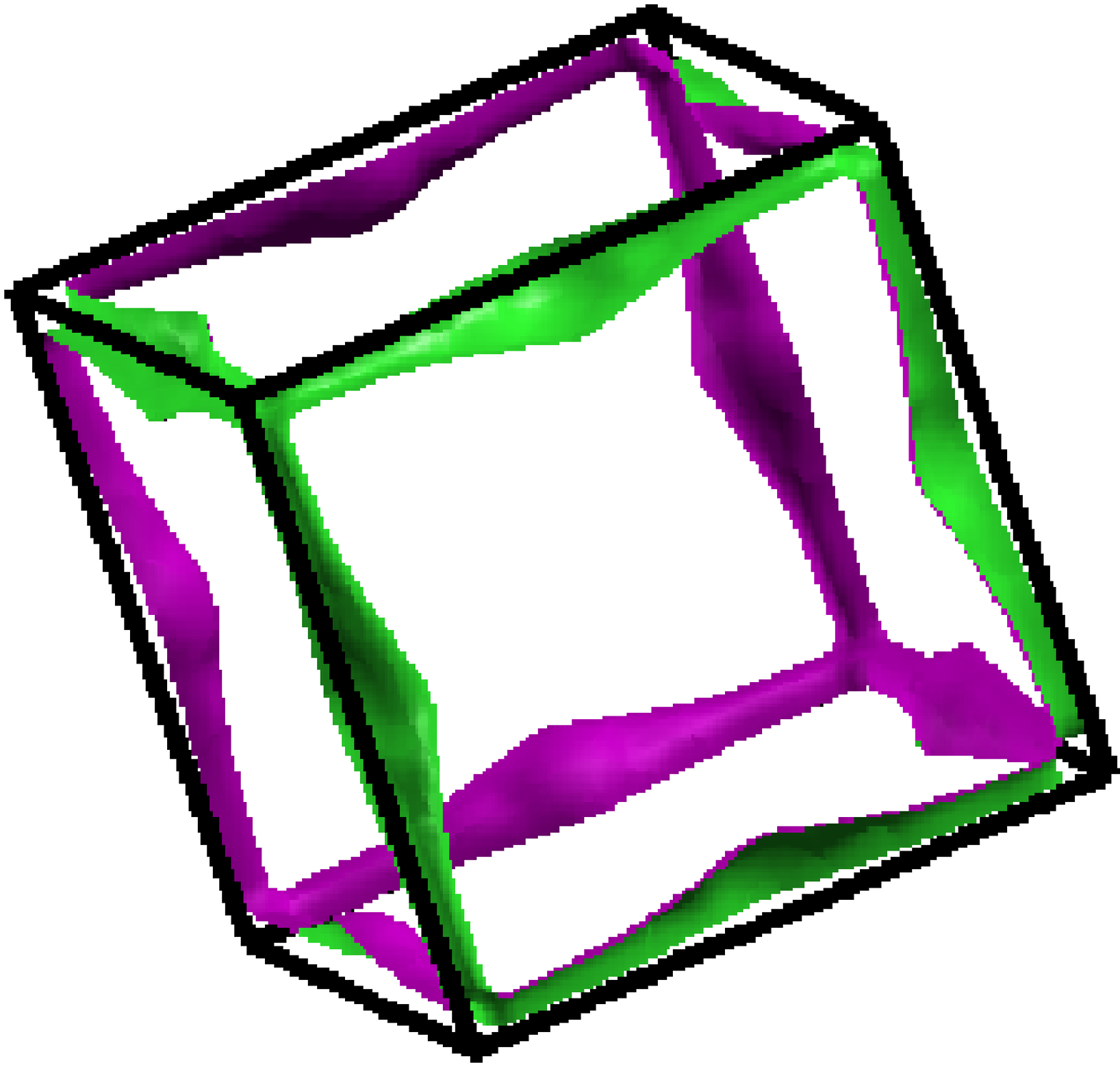}}
\subfigure[]{\includegraphics[width=35mm,height=35mm]{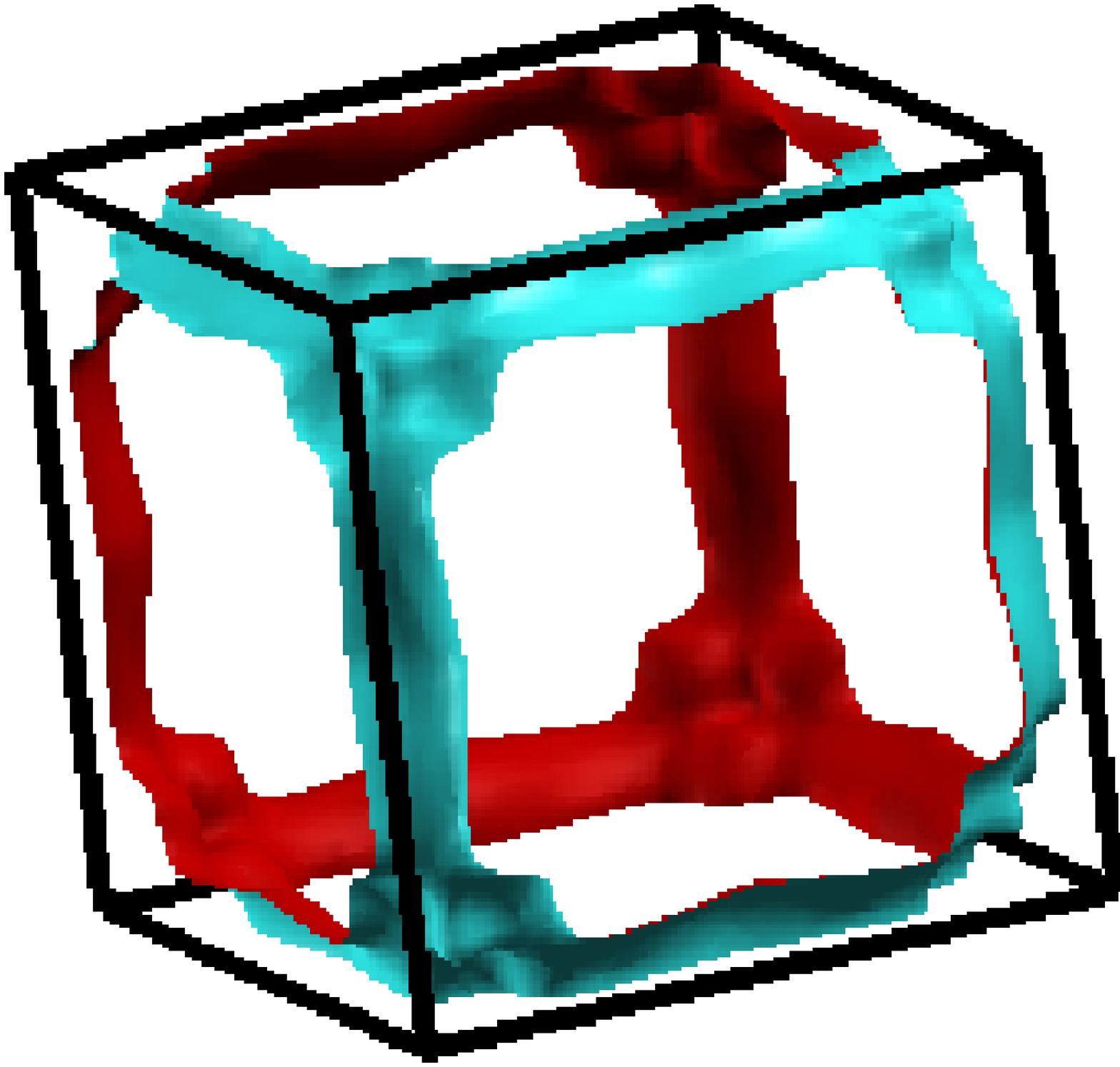}}
\subfigure[]{\includegraphics[width=35mm,height=35mm]{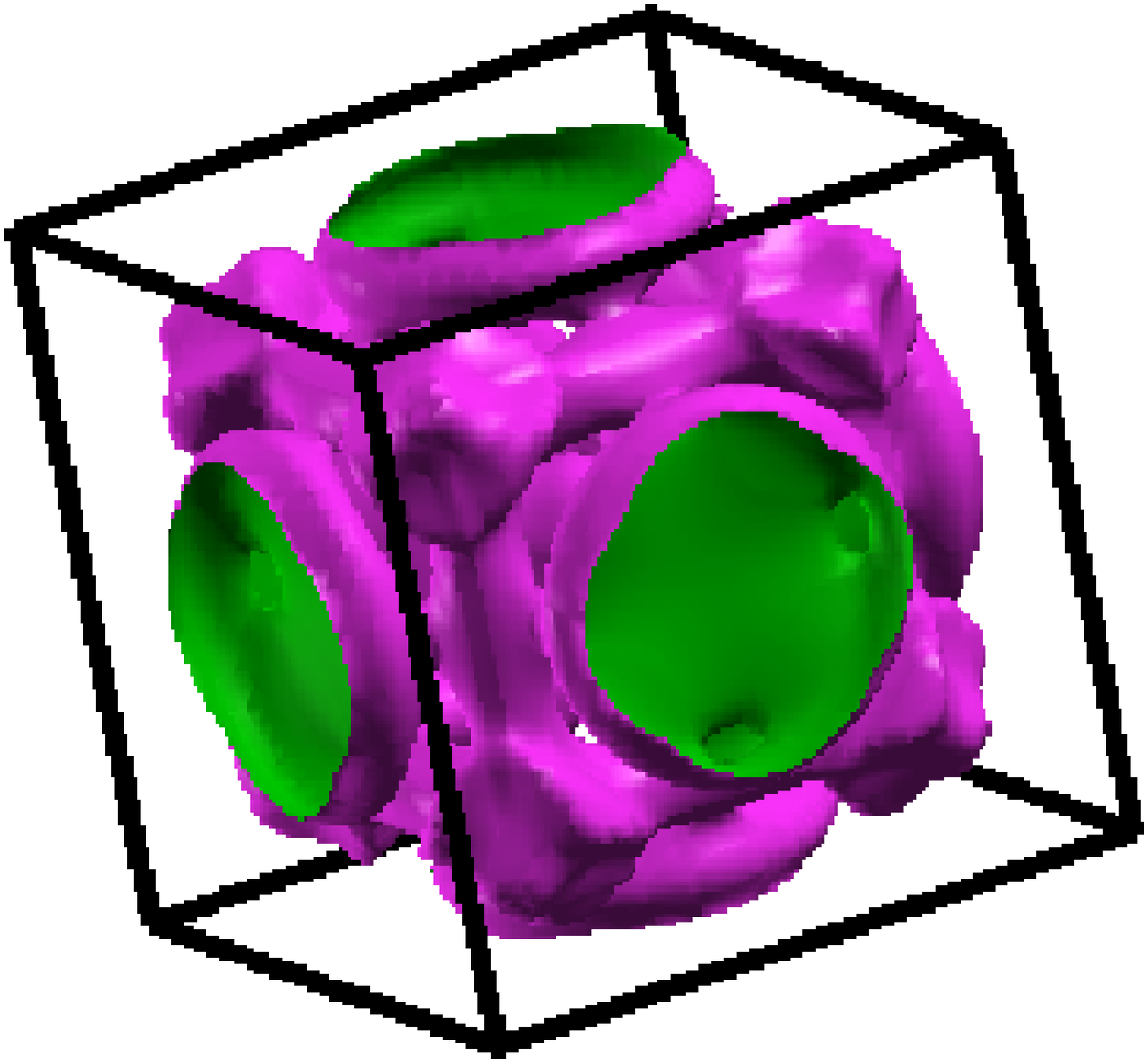}}\\
\subfigure[]{\includegraphics[width=35mm,height=35mm]{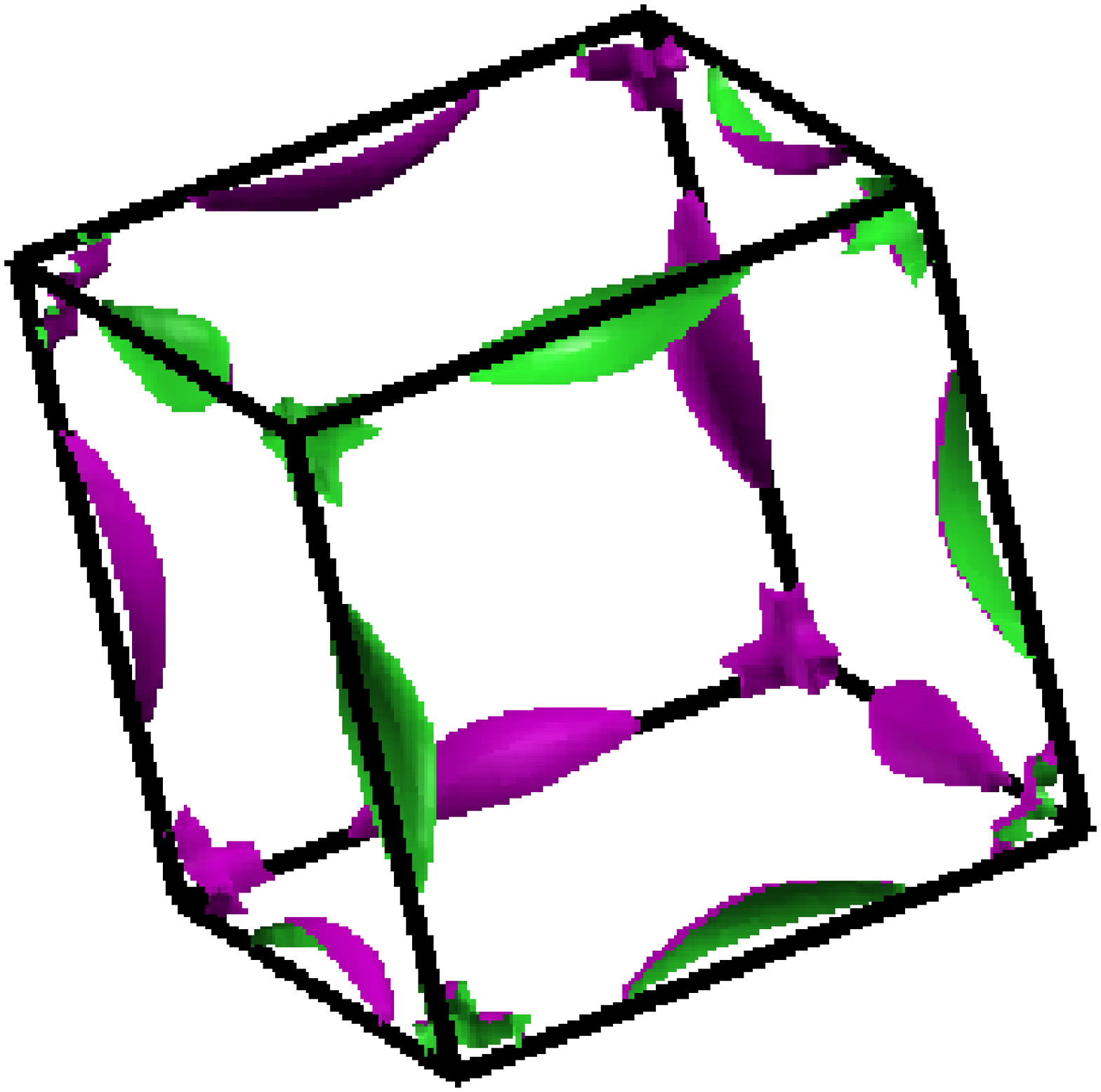}}
\subfigure[]{\includegraphics[width=35mm,height=35mm]{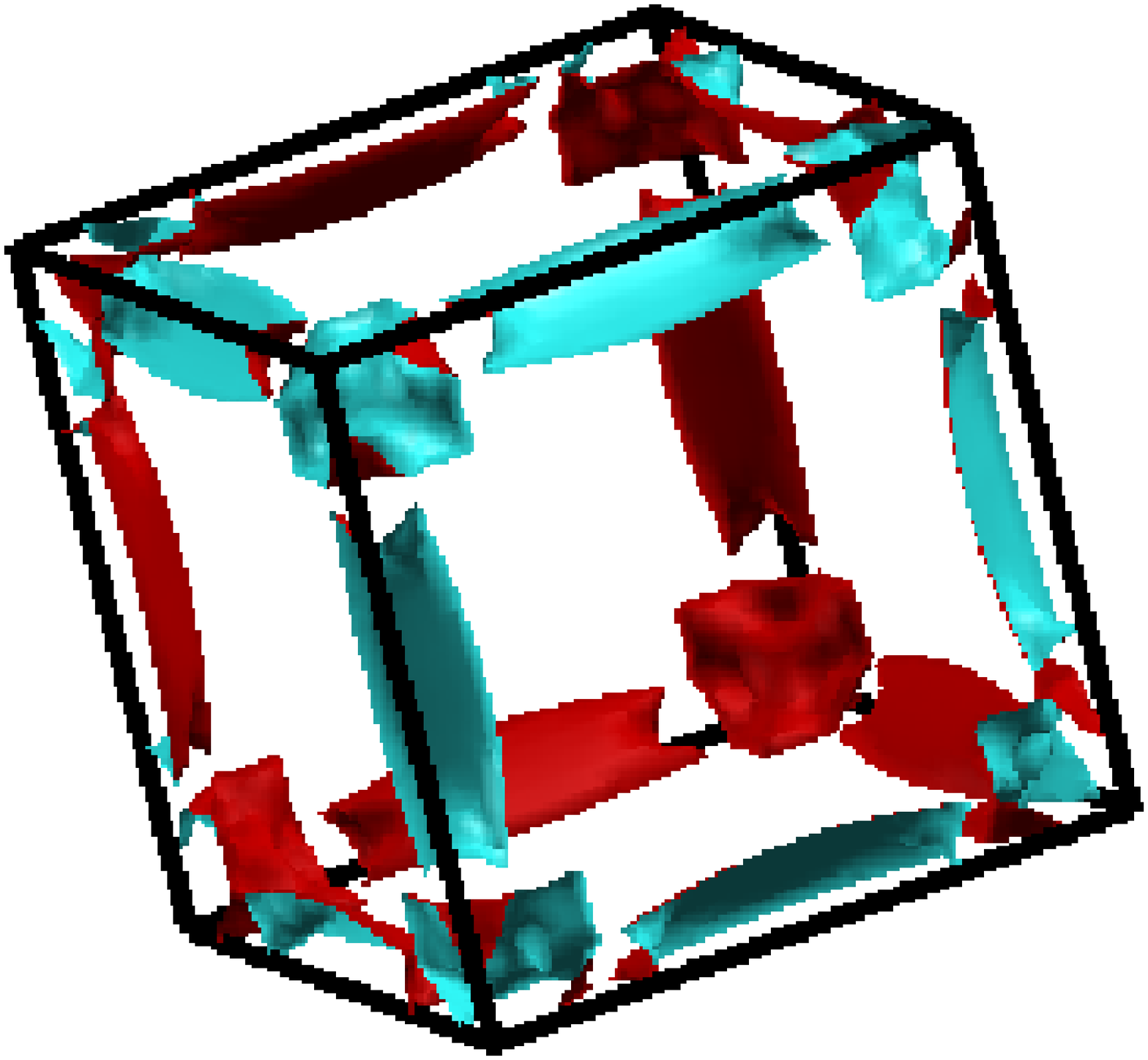}}
\subfigure[]{\includegraphics[width=35mm,height=35mm]{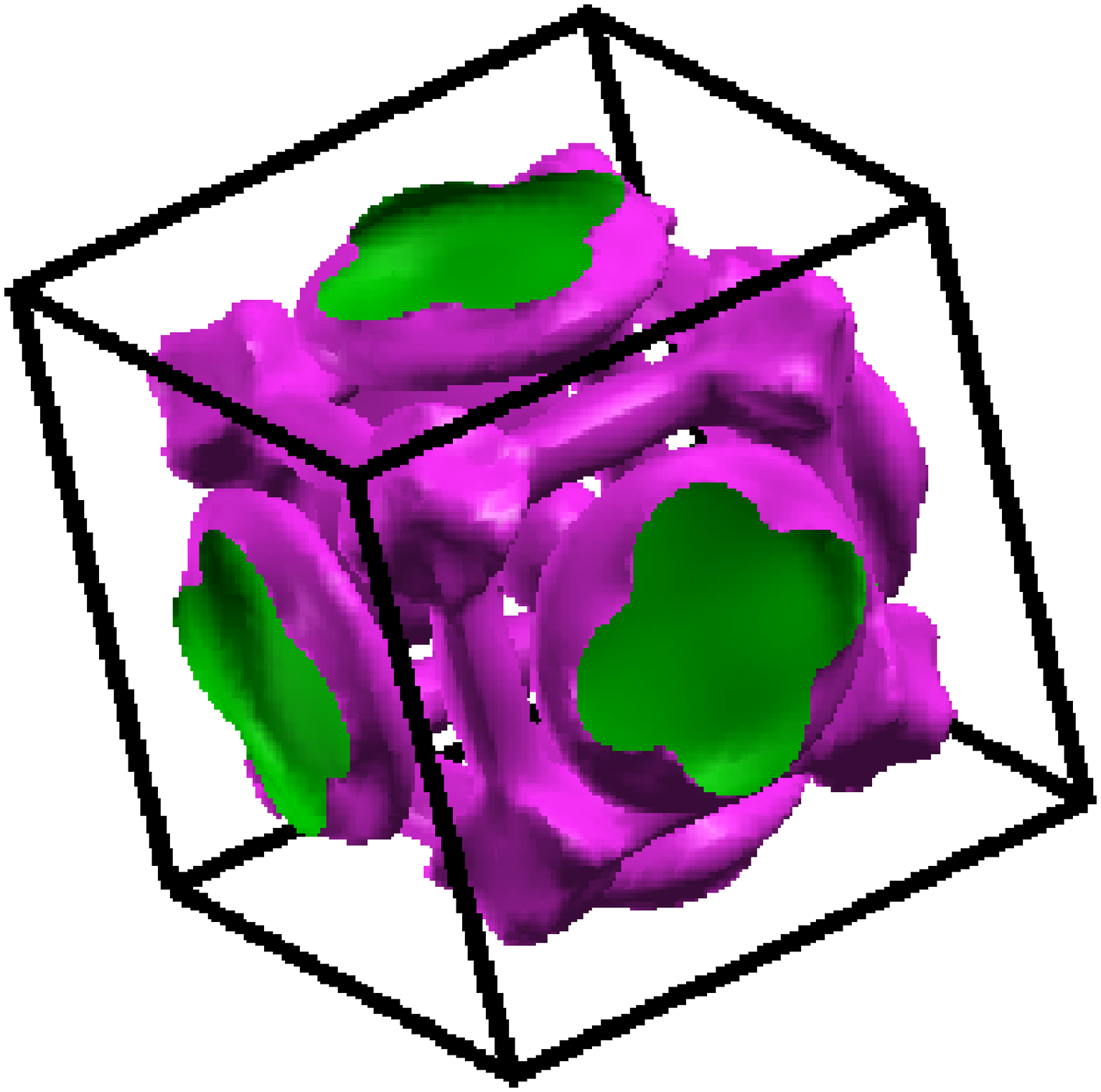}}

\caption{Band structure of Nb$_3$Ga at (a)V/V$_0$ = 1.00, (b)V/V$_0$ = 0.92 (pressure of 17.5 GPa) and FS for which change in FS is observed at ambient (c), (d), (e)for band no. 50, 51 and 53 and at V/V$_0$ = 0.92 (f), (g) and (h) where change in FS topology is observed.}
\end{center}
\end{figure*}

\begin{figure*}
\begin{center}
\subfigure[]{\includegraphics[width=60mm,height=65mm]{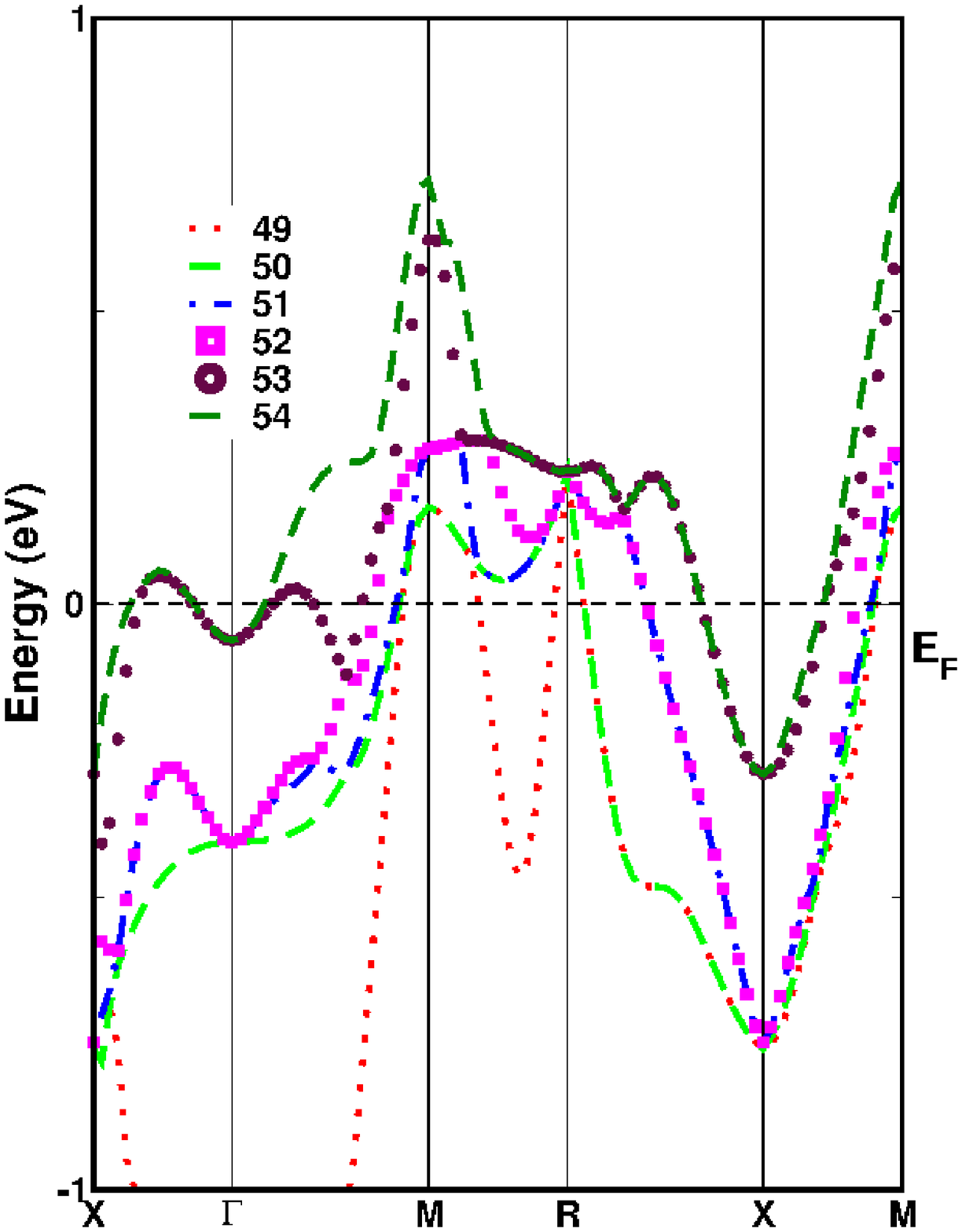}}
\subfigure[]{\includegraphics[width=60mm,height=65mm]{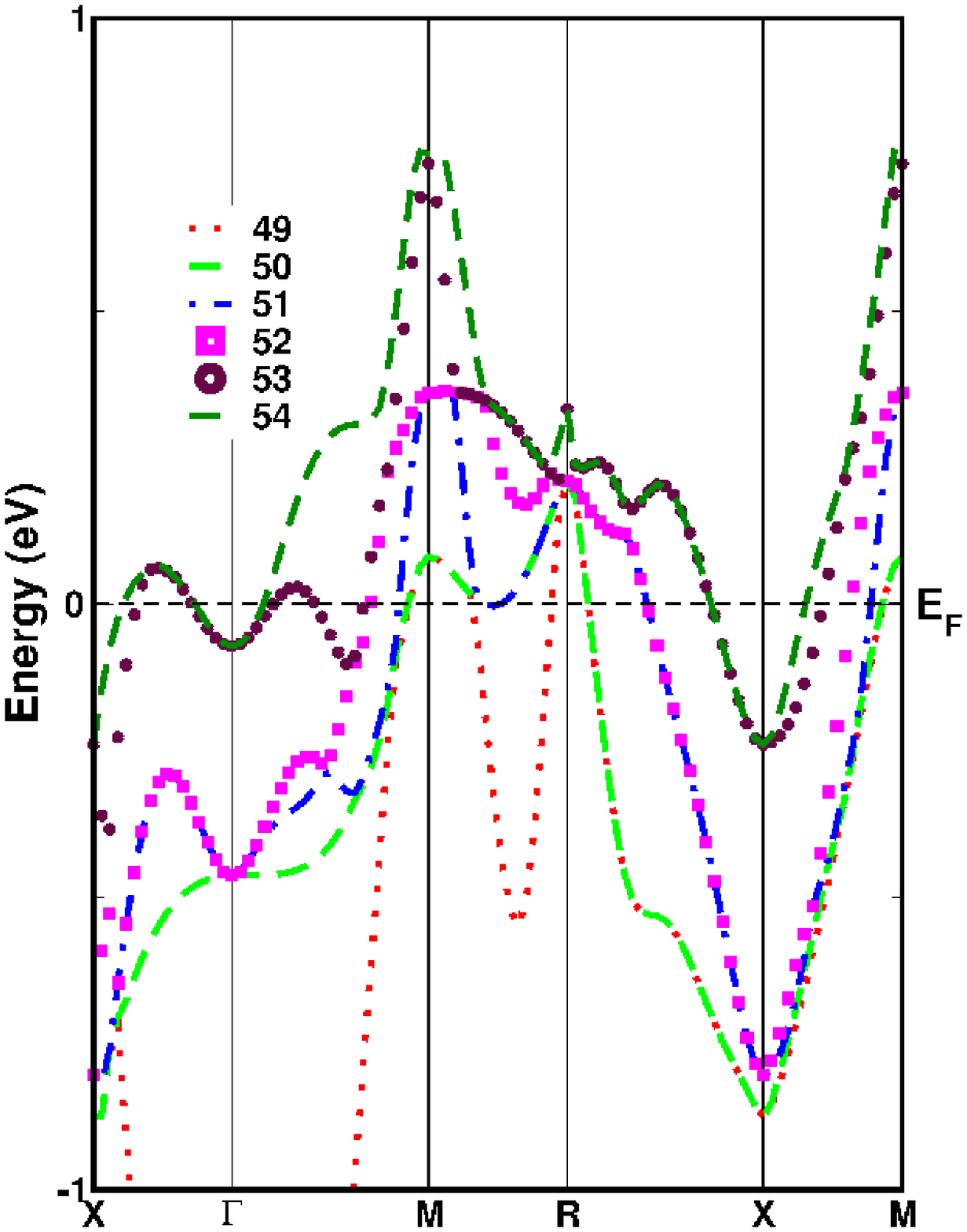}}\\
\subfigure[]{\includegraphics[width=35mm,height=35mm]{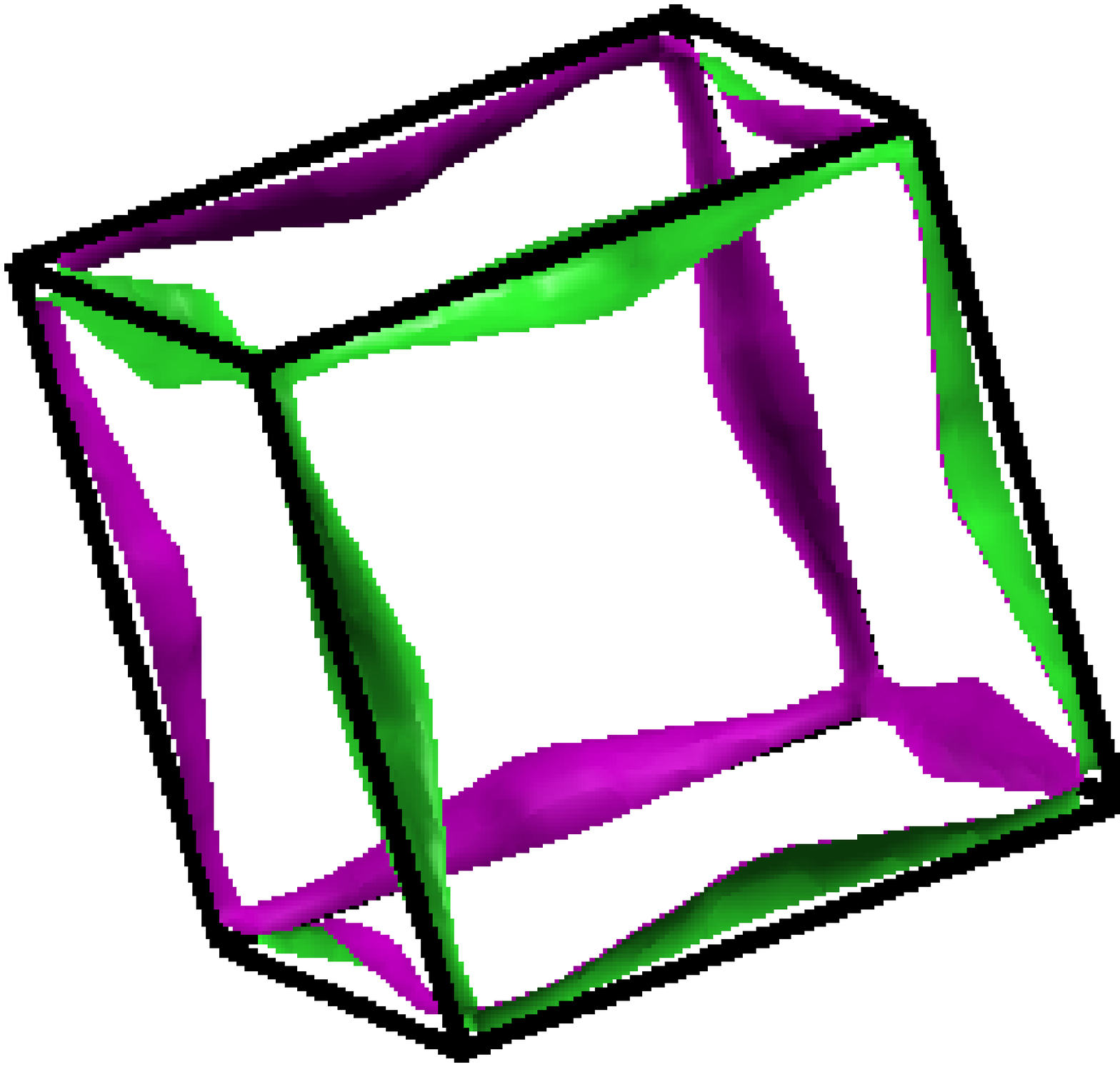}}
\subfigure[]{\includegraphics[width=35mm,height=35mm]{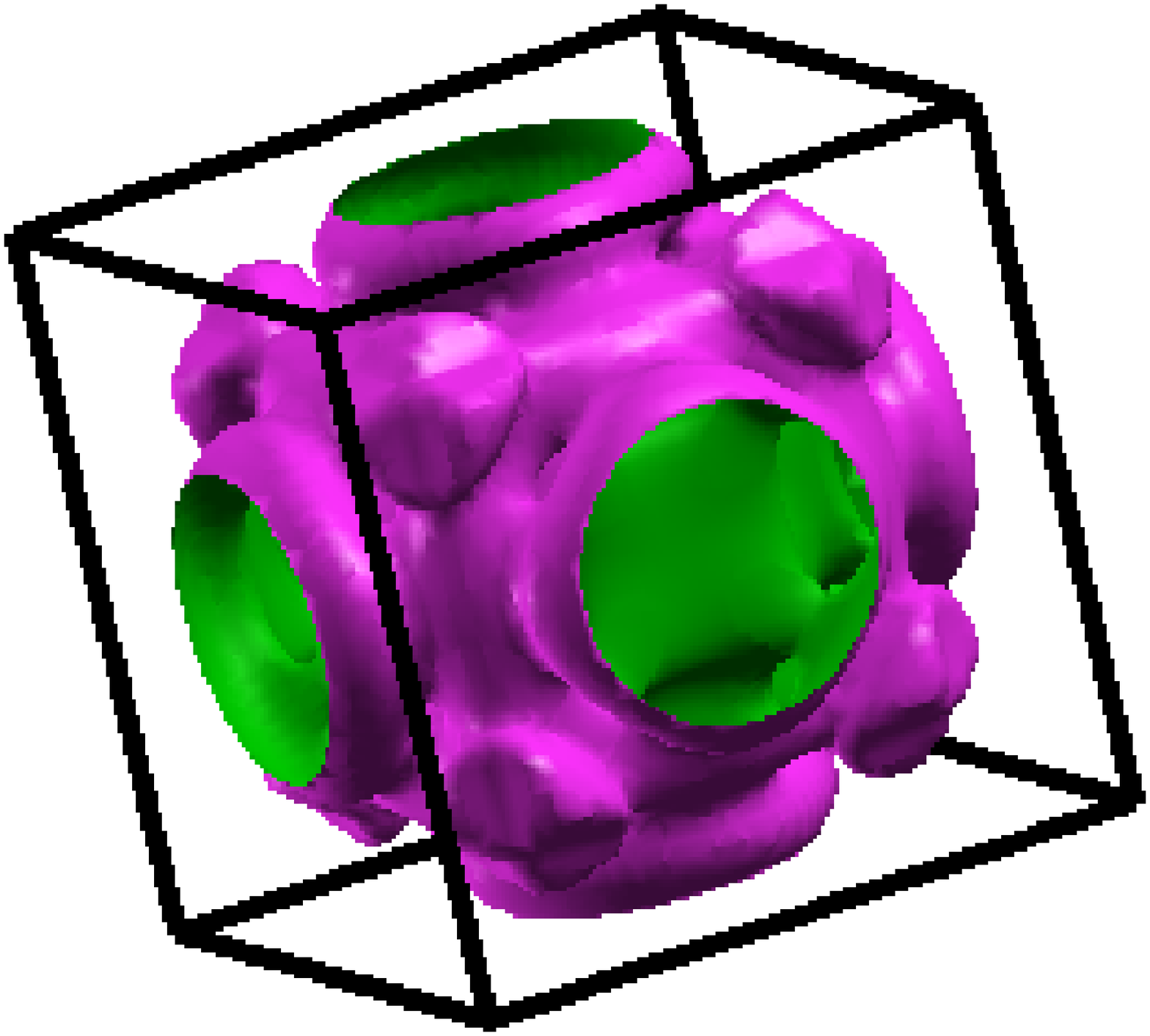}}\\
\subfigure[]{\includegraphics[width=35mm,height=35mm]{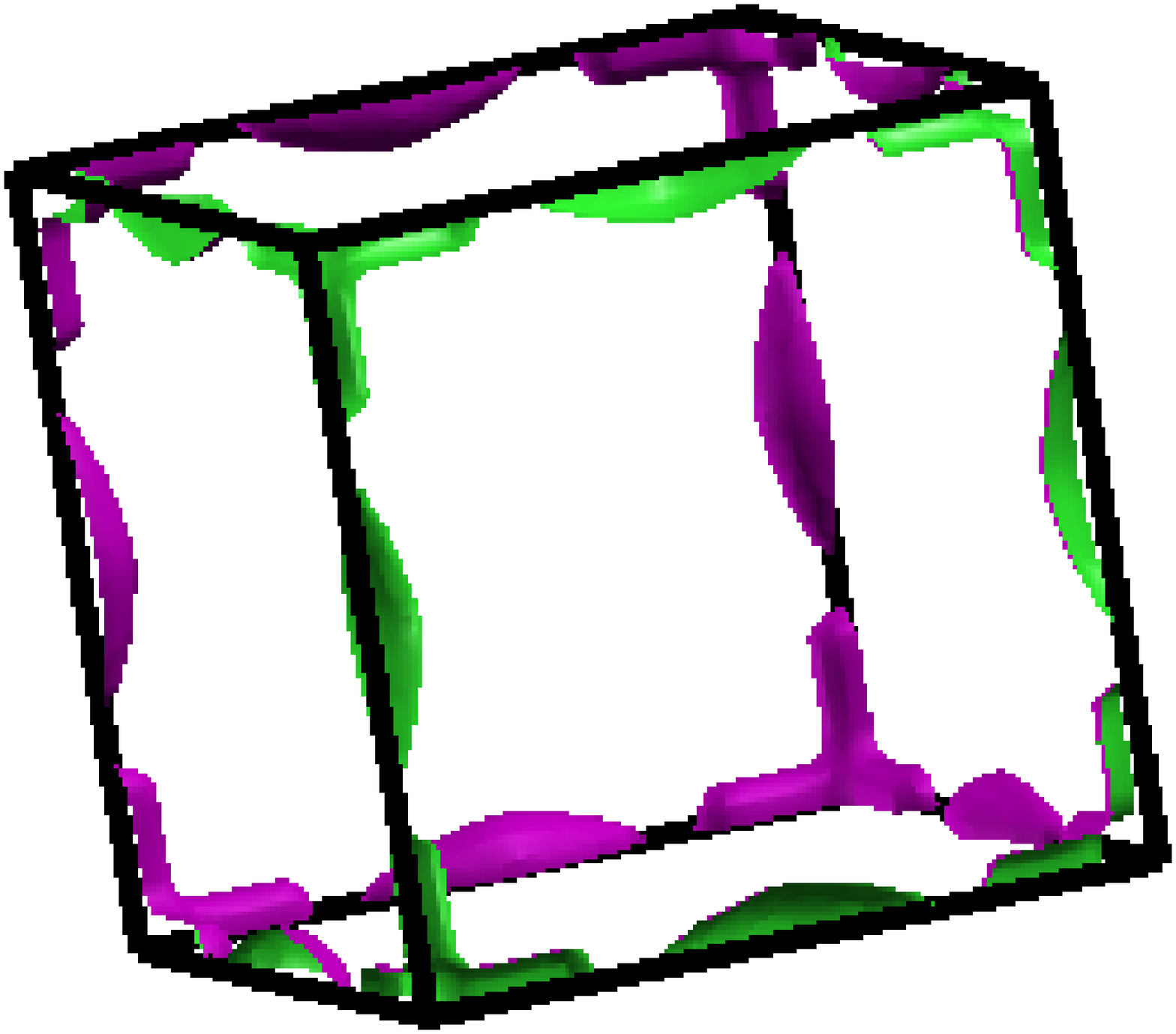}}
\subfigure[]{\includegraphics[width=35mm,height=35mm]{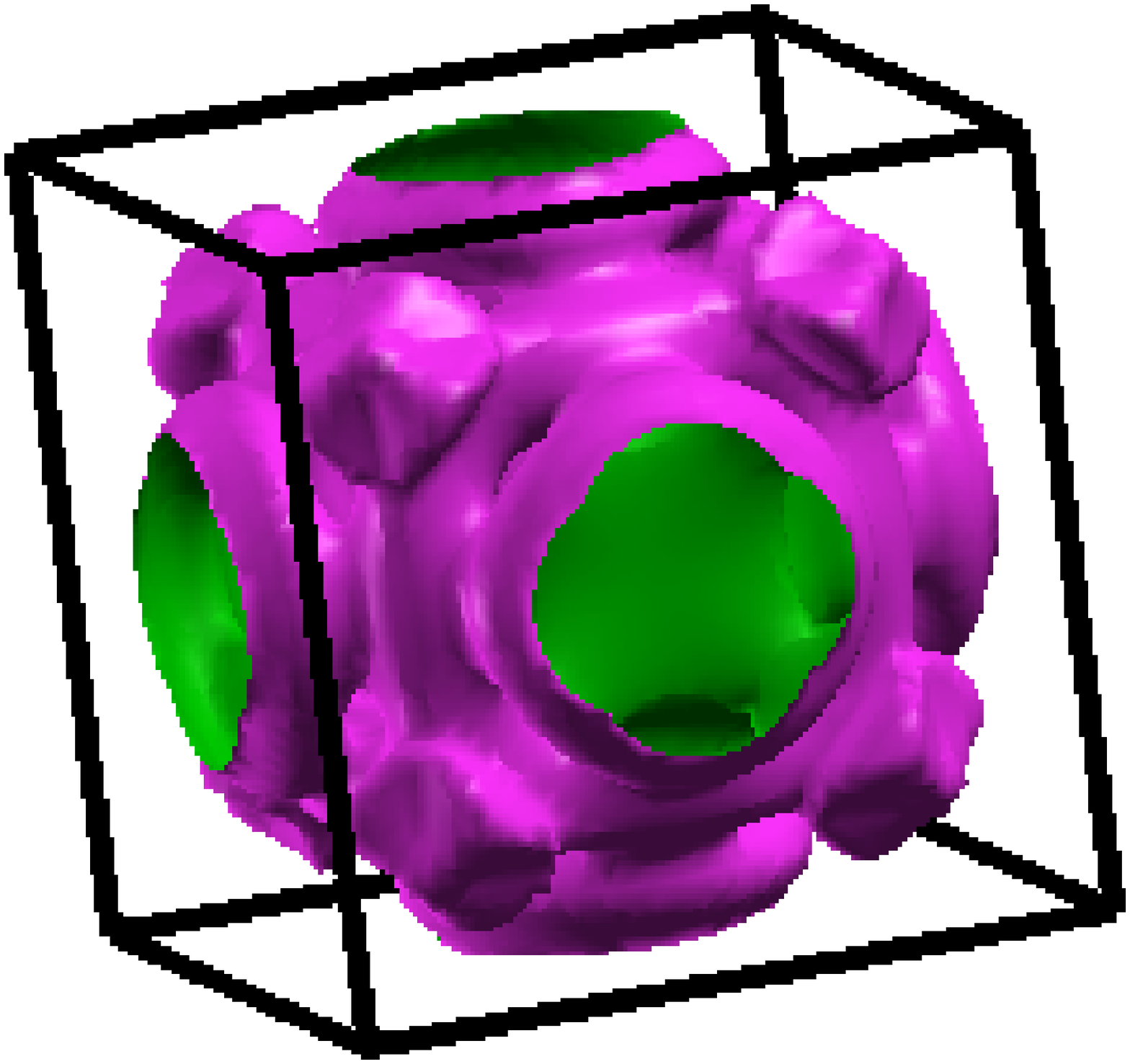}}\\
\caption{Band structure of Nb$_3$In at (a)V/V$_0$ = 1.00, (b)V/V$_0$ = 0.92 (pressure of 21 GPa) and FS for which change in FS is observed at ambient (c), (d) for band no. 50 and 53 and at V/V$_0$ = 0.92 (e) and (f) where change in FS topology is observed.}
\end{center}
\end{figure*}

\begin{figure*}
\begin{center}
\subfigure[]{\includegraphics[width=60mm,height=65mm]{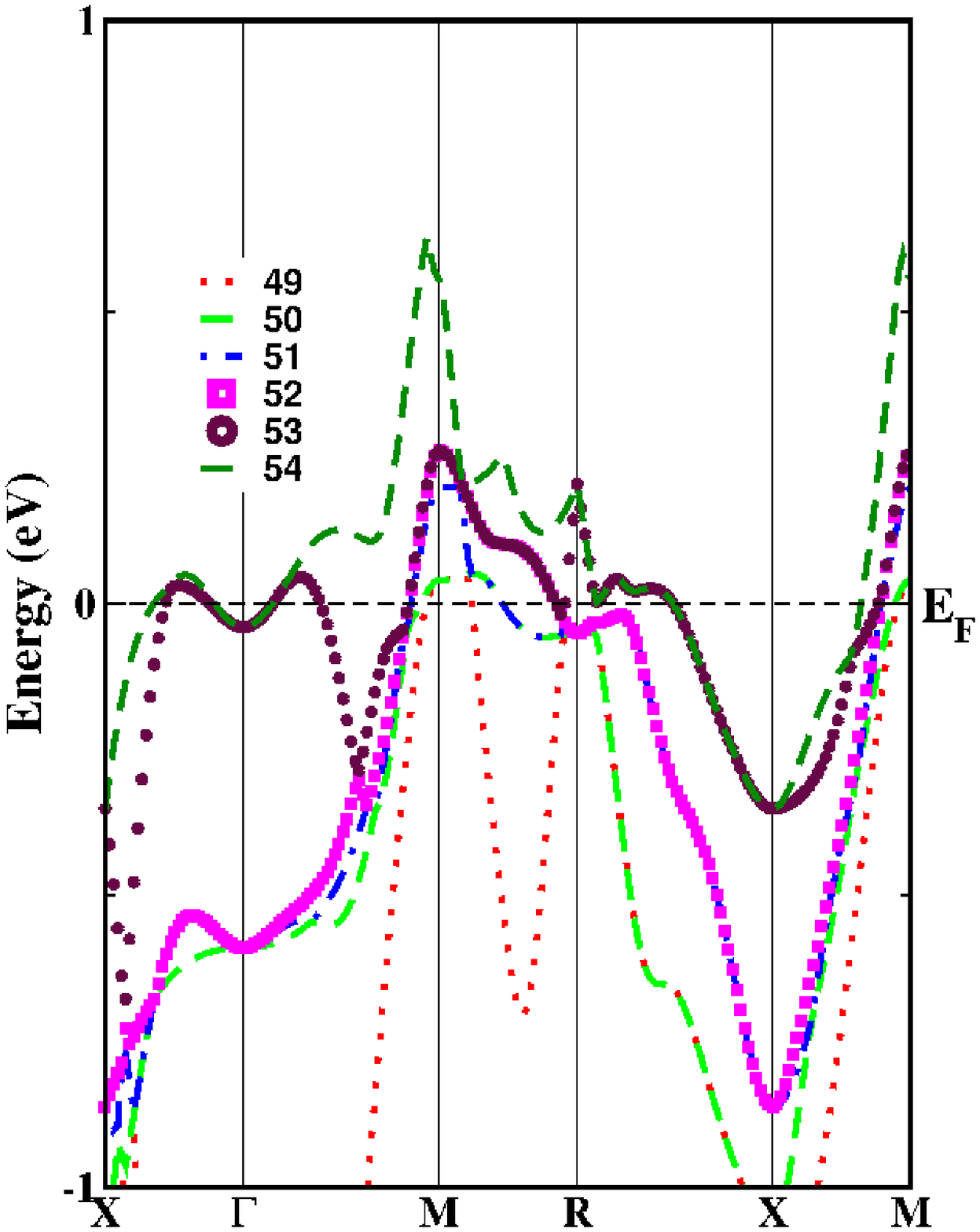}}
\subfigure[]{\includegraphics[width=60mm,height=65mm]{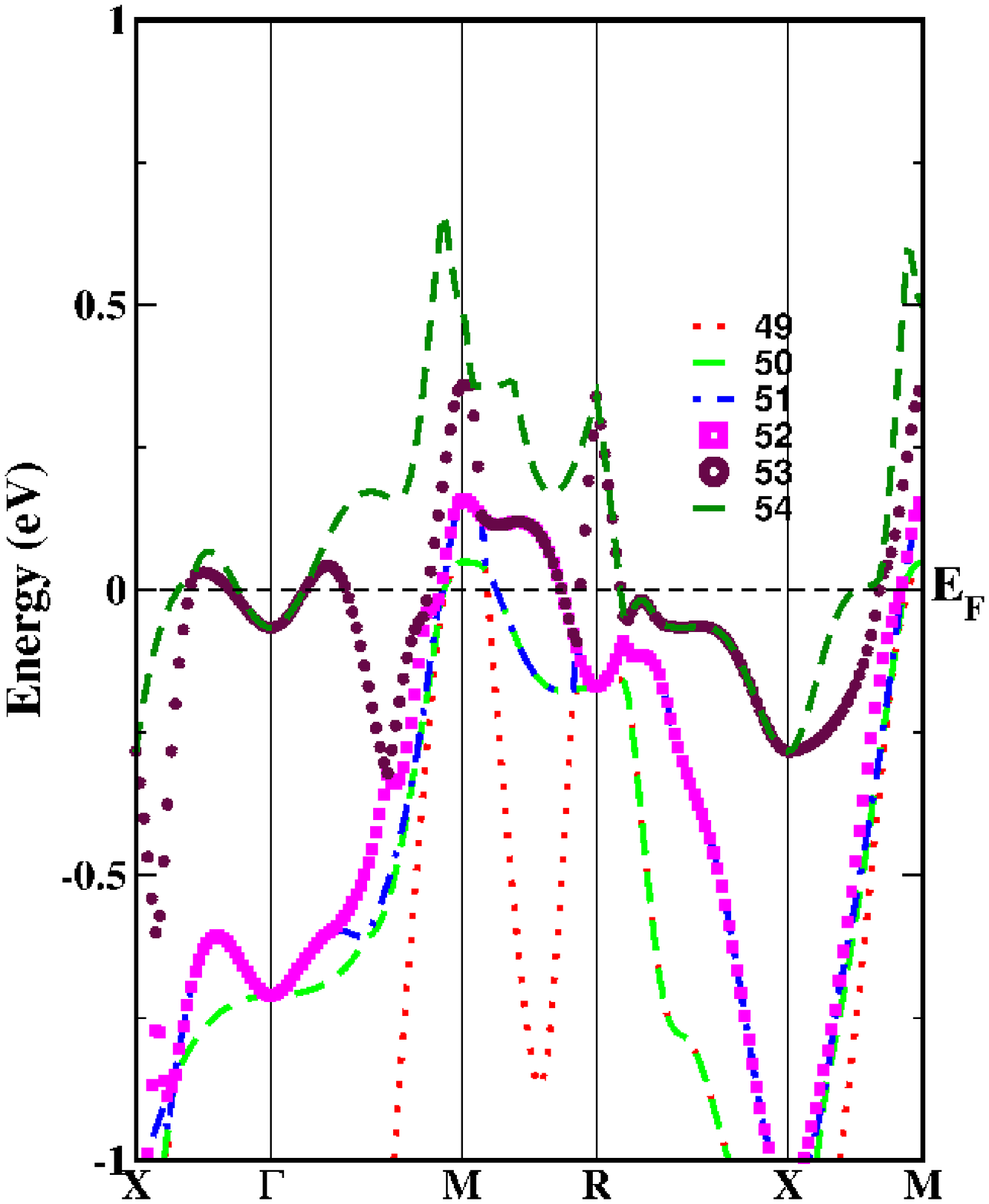}}\\
\subfigure[]{\includegraphics[width=35mm,height=35mm]{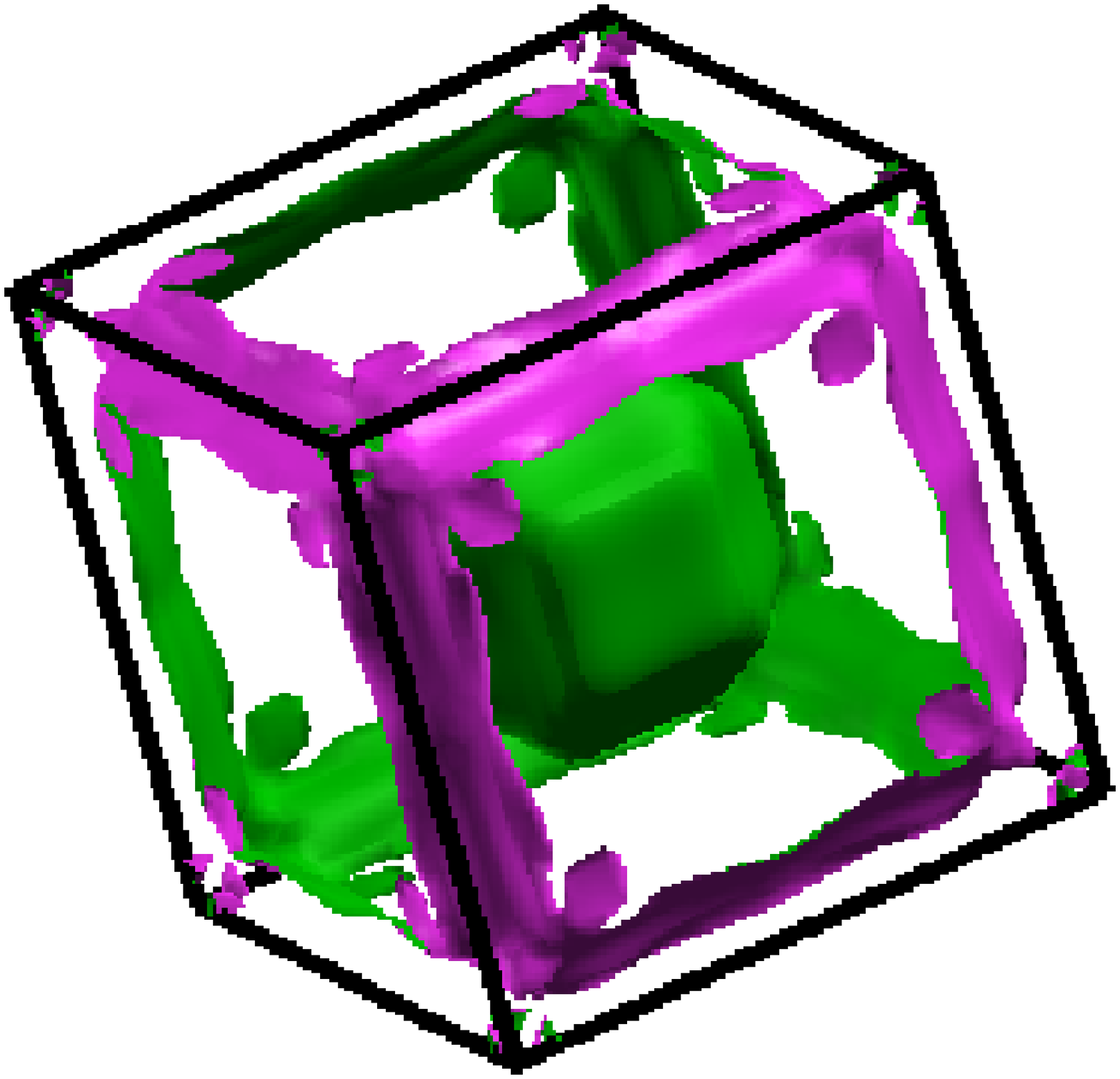}}
\subfigure[]{\includegraphics[width=35mm,height=35mm]{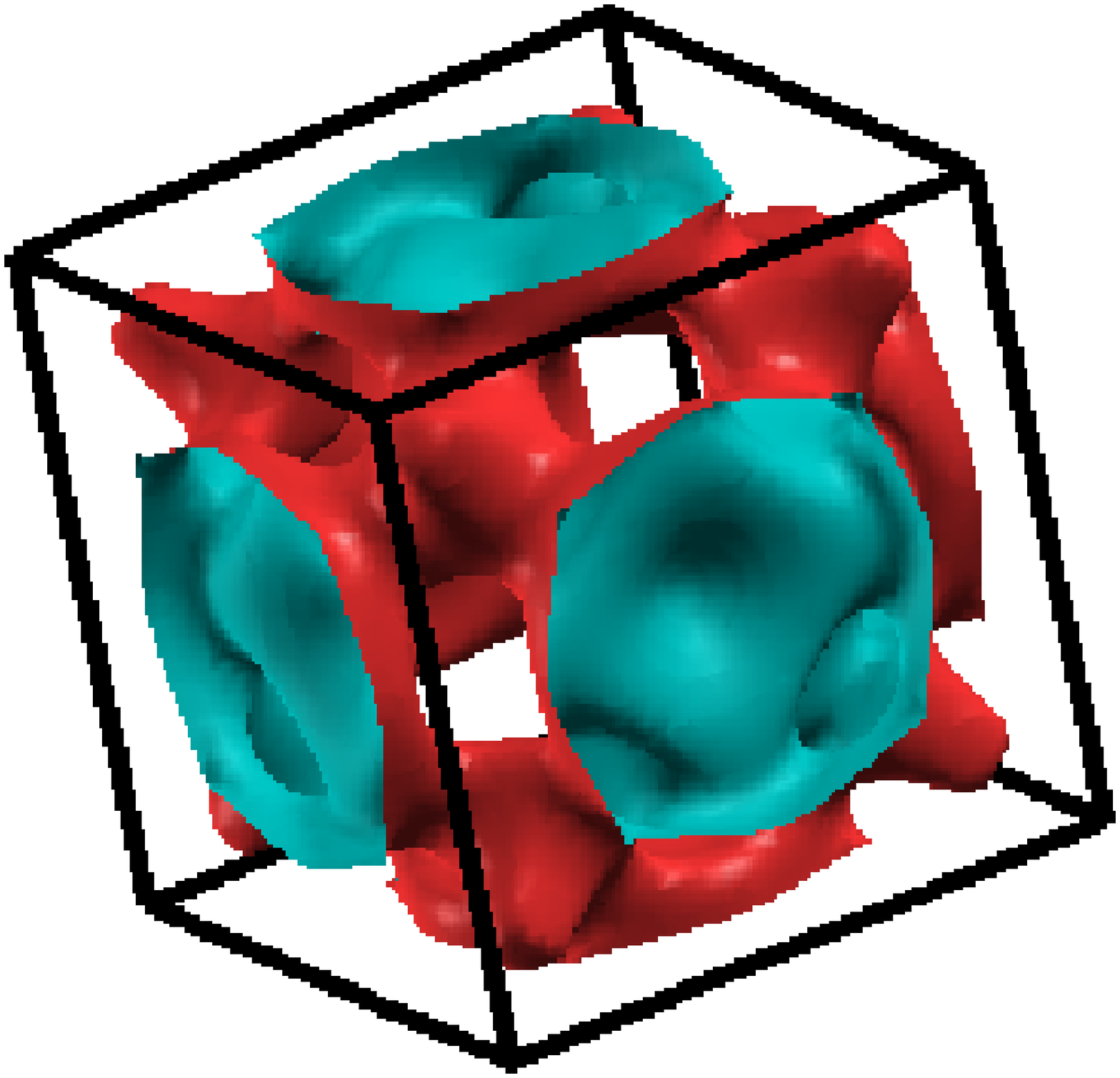}}\\
\subfigure[]{\includegraphics[width=35mm,height=35mm]{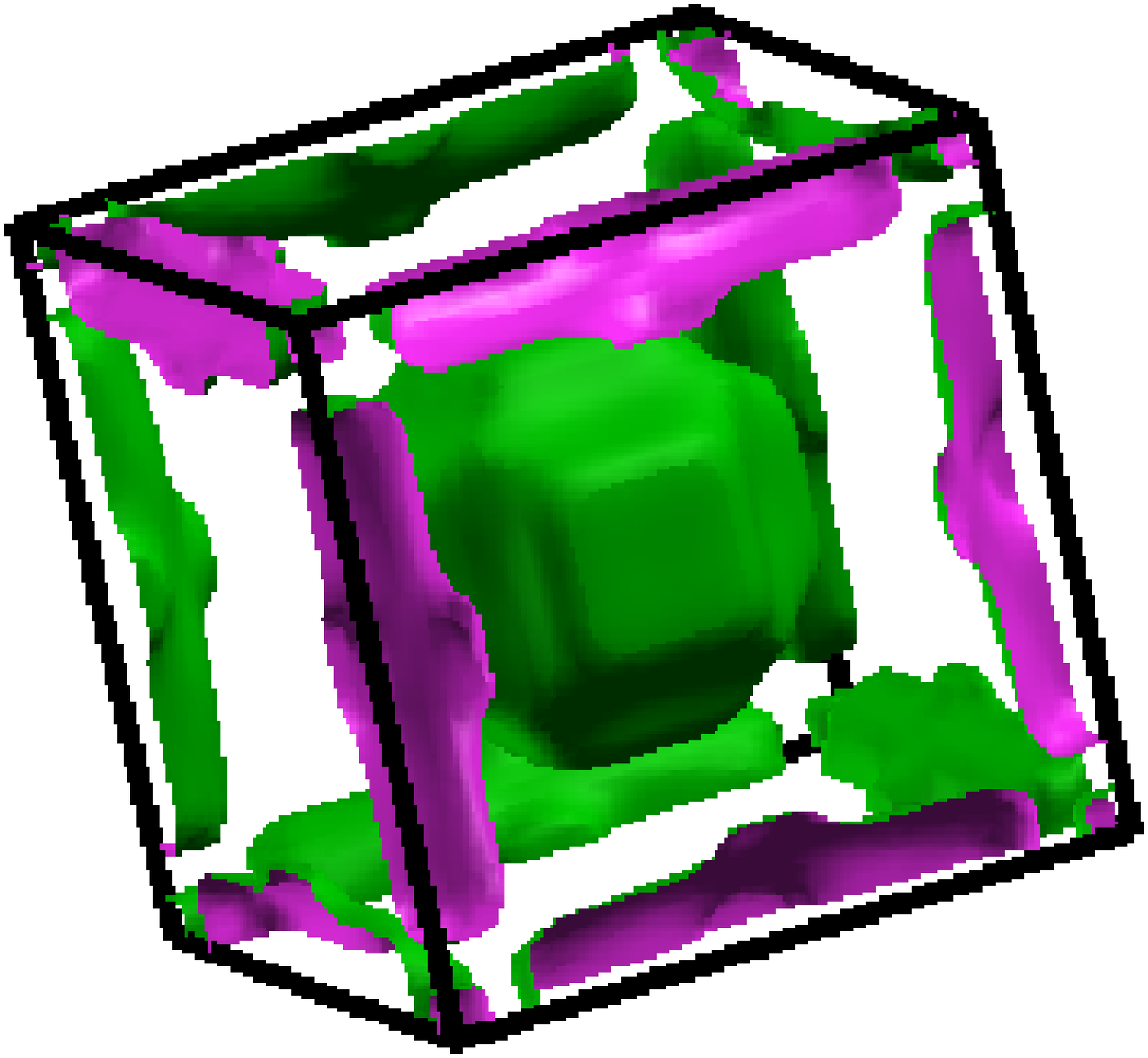}}
\subfigure[]{\includegraphics[width=35mm,height=35mm]{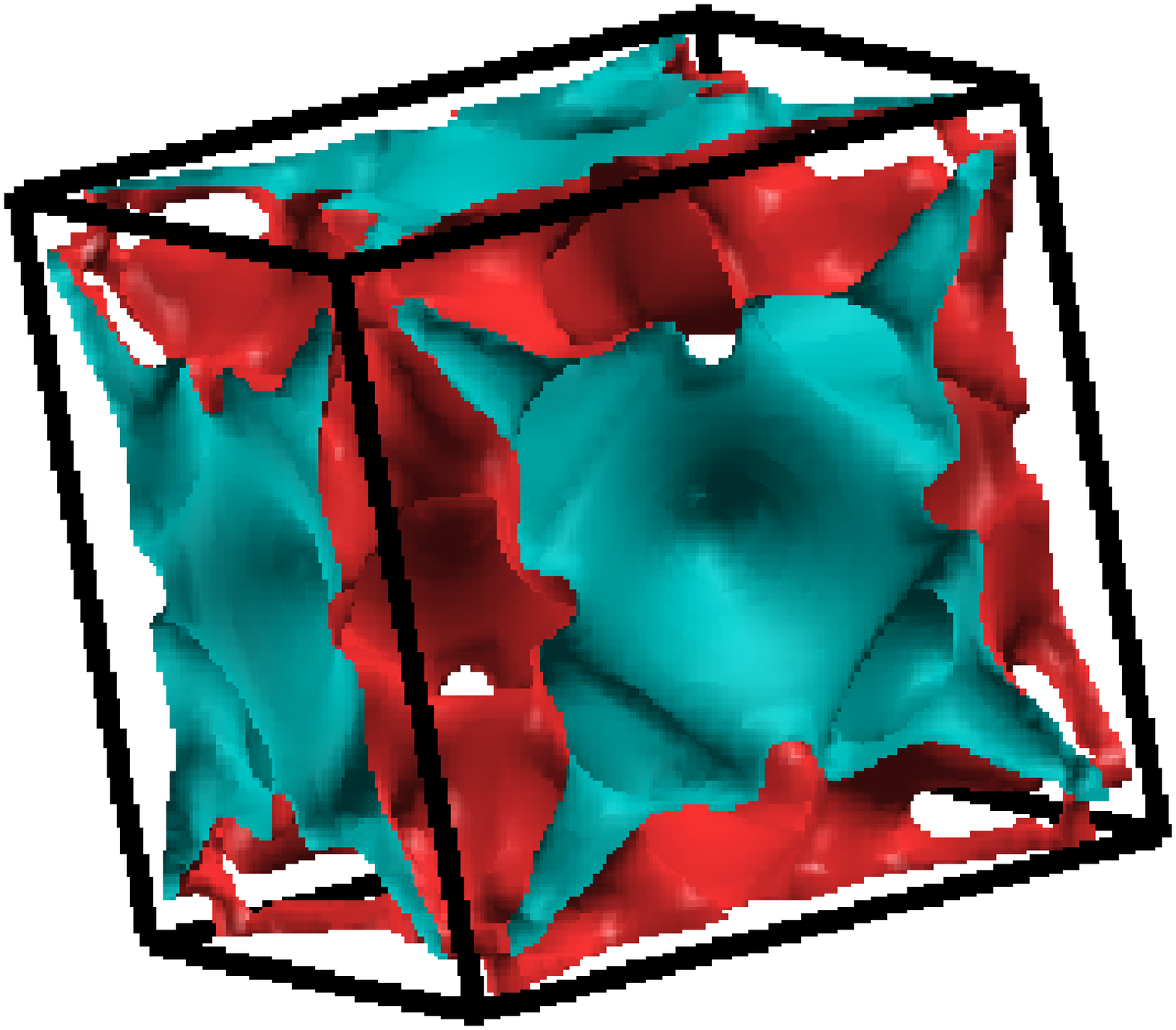}}\\
\caption{Band structure of Nb$_3$Ge at (a)V/V$_0$ = 1.00, (b)V/V$_0$ = 0.90 (pressure of 22.4 GPa) and FS for which change in FS is observed at ambient (c), (d) for band no. 53 and 54 and at V/V$_0$ = 0.90 (e) and (f) where change in FS topology is observed.}
\end{center}
\end{figure*}

\begin{figure*}
\begin{center}
\subfigure[]{\includegraphics[width=60mm,height=65mm]{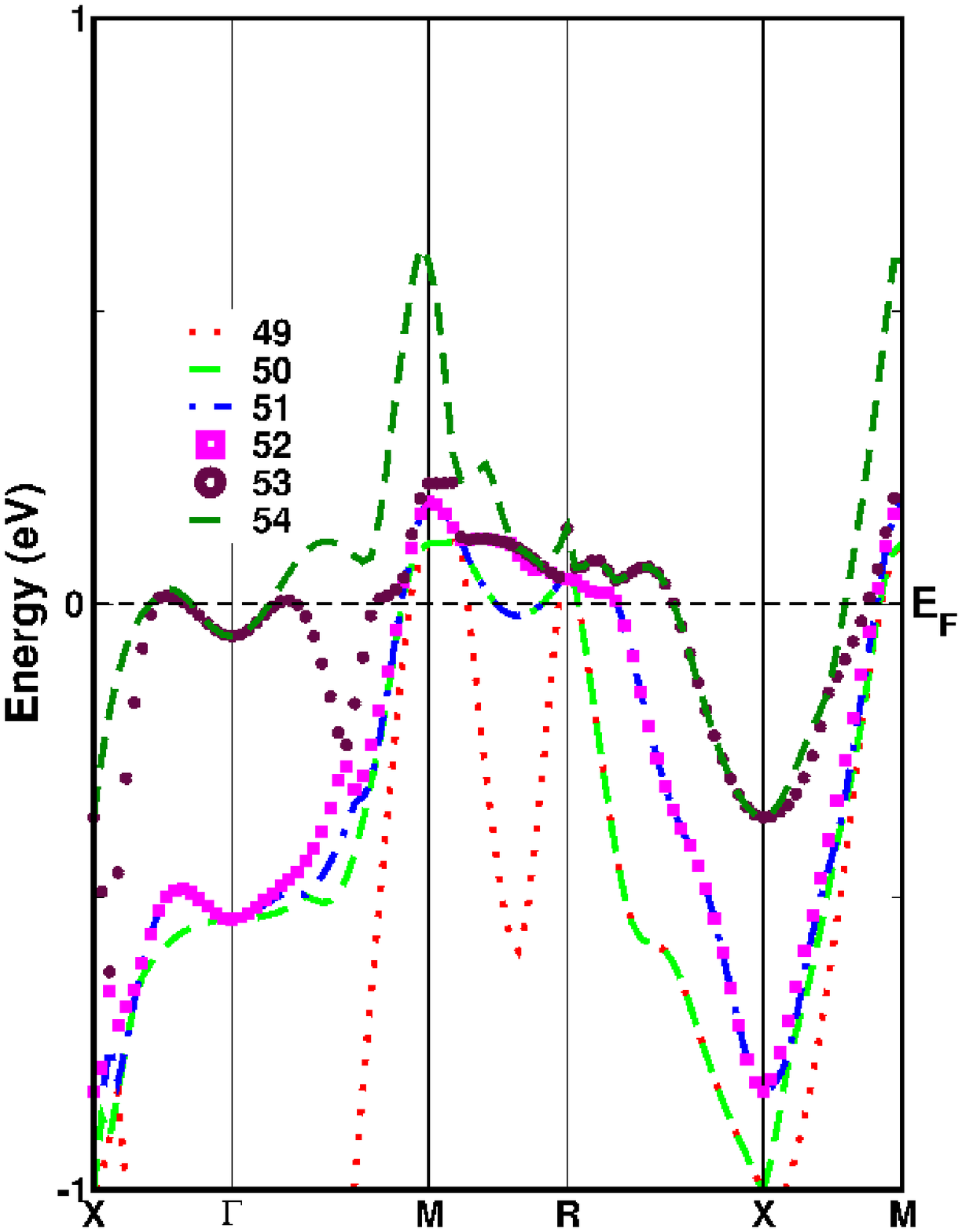}}
\subfigure[]{\includegraphics[width=60mm,height=65mm]{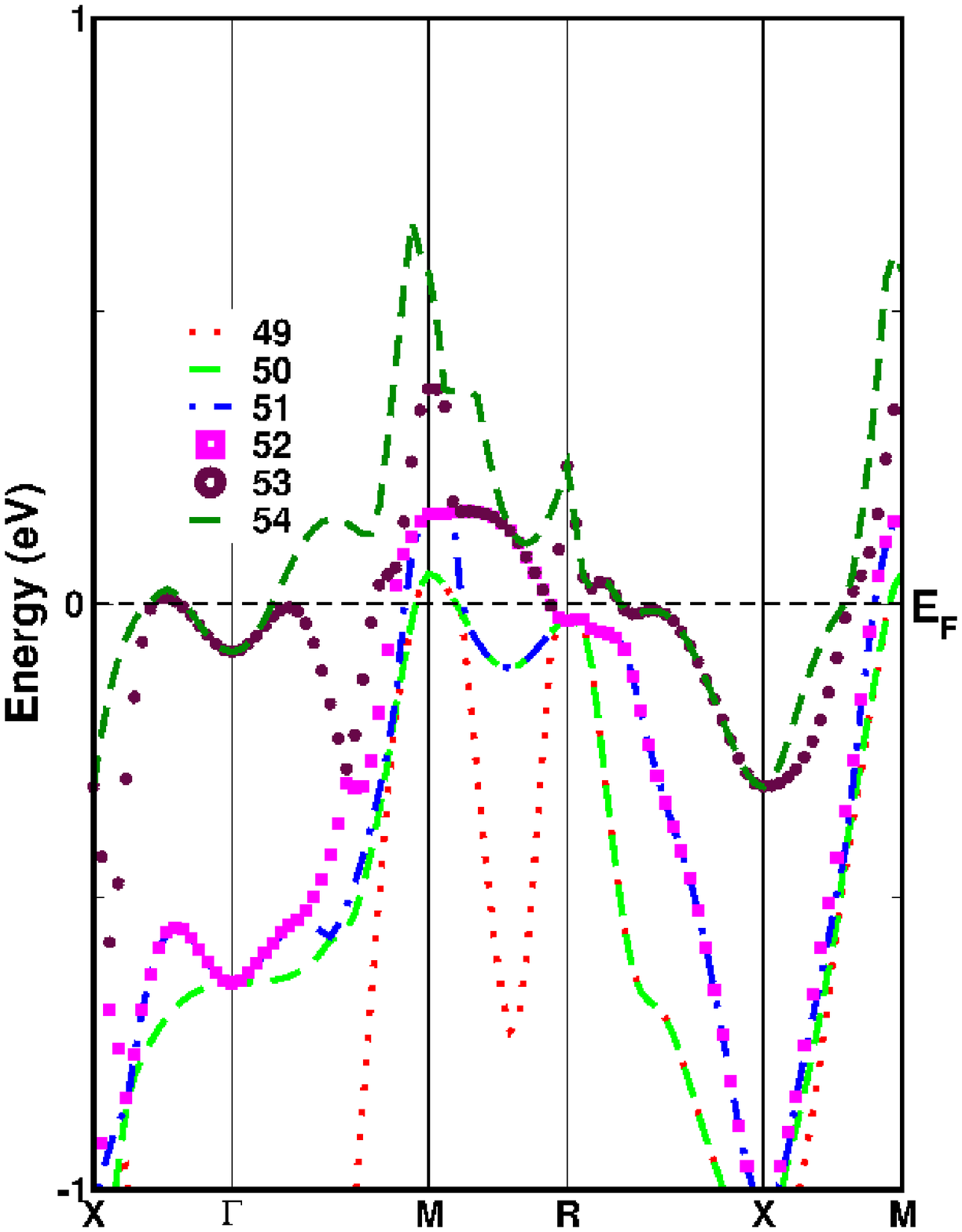}}\\
\subfigure[]{\includegraphics[width=32mm,height=32mm]{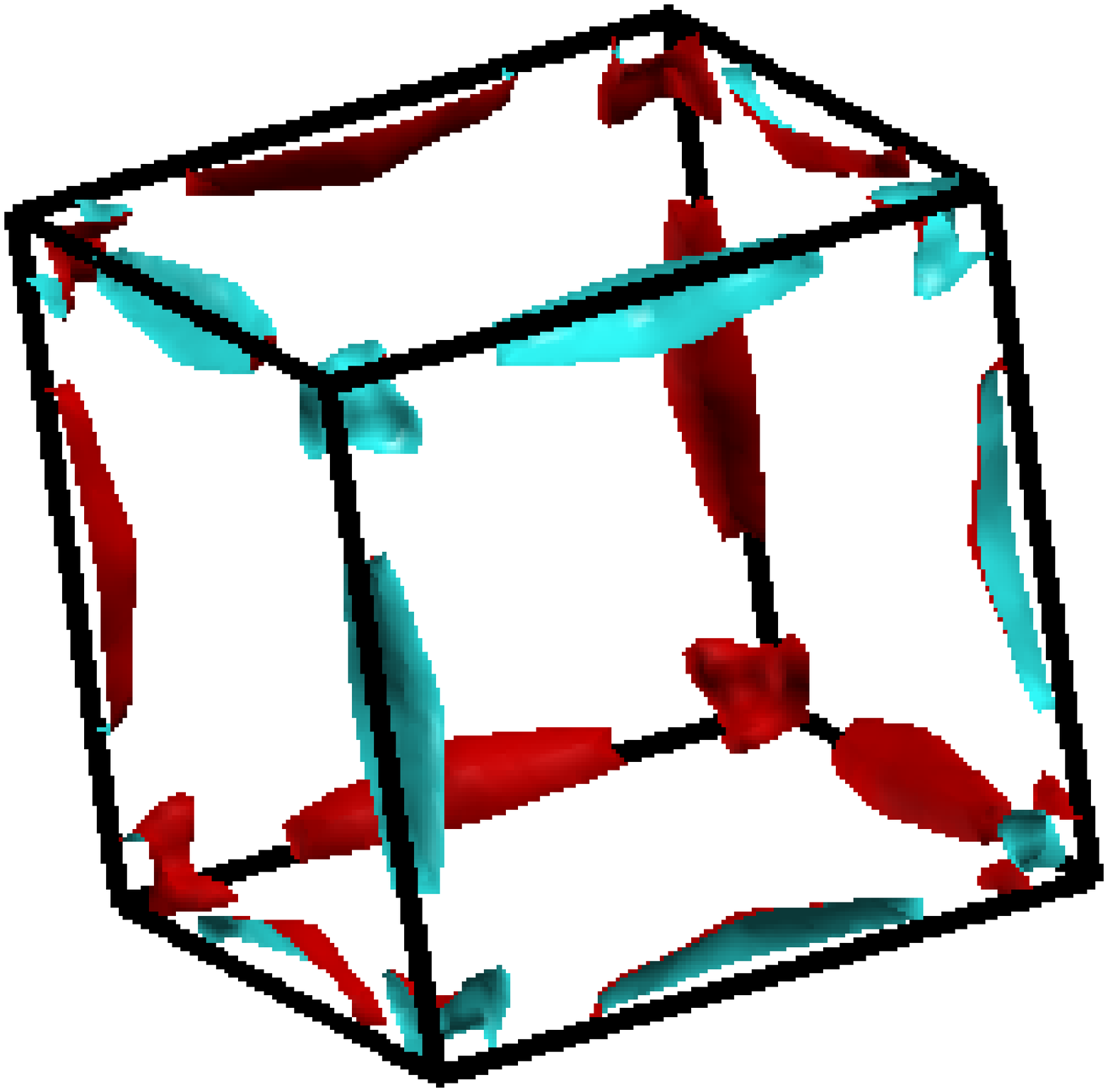}}
\subfigure[]{\includegraphics[width=32mm,height=32mm]{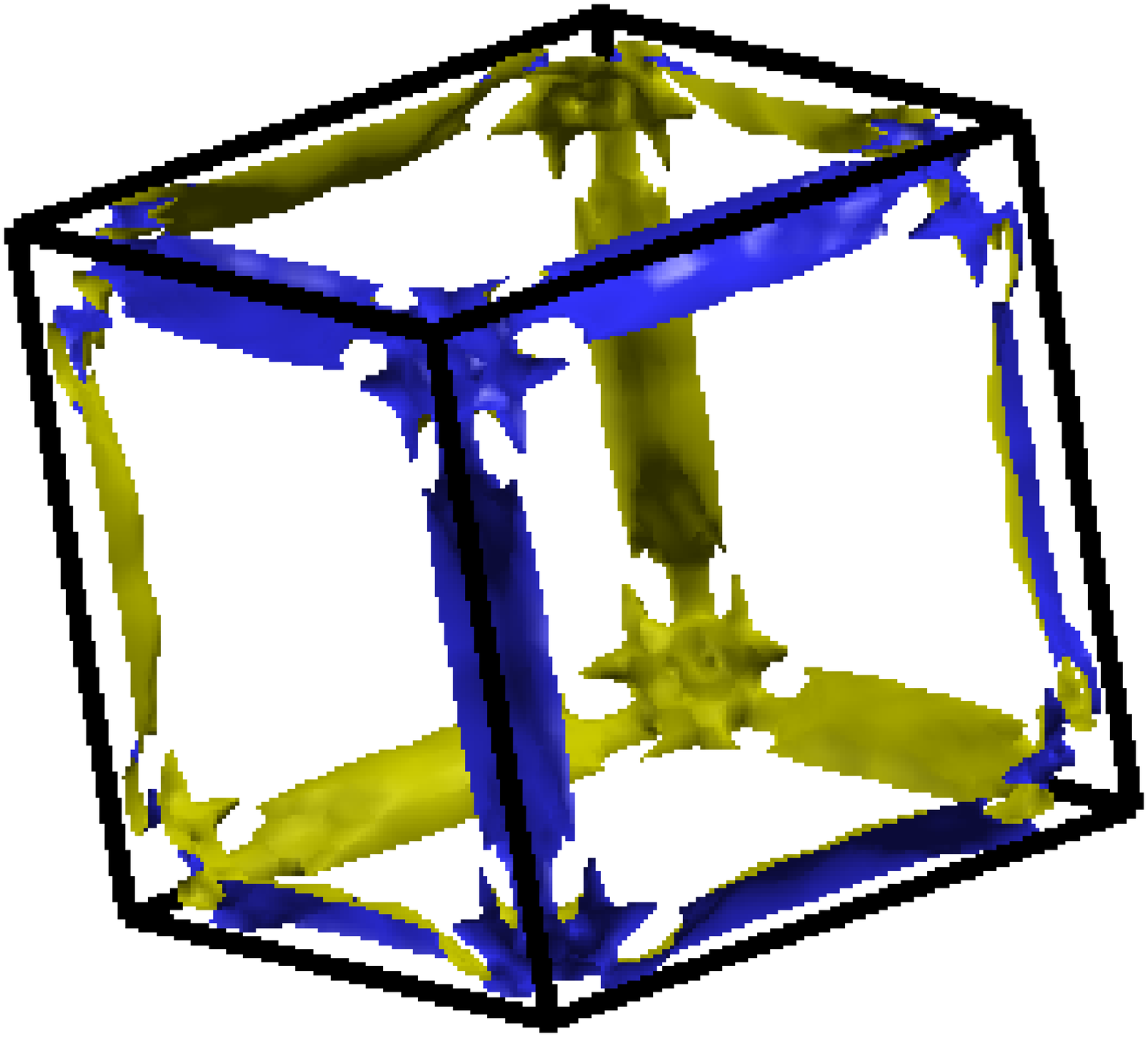}}
\subfigure[]{\includegraphics[width=32mm,height=32mm]{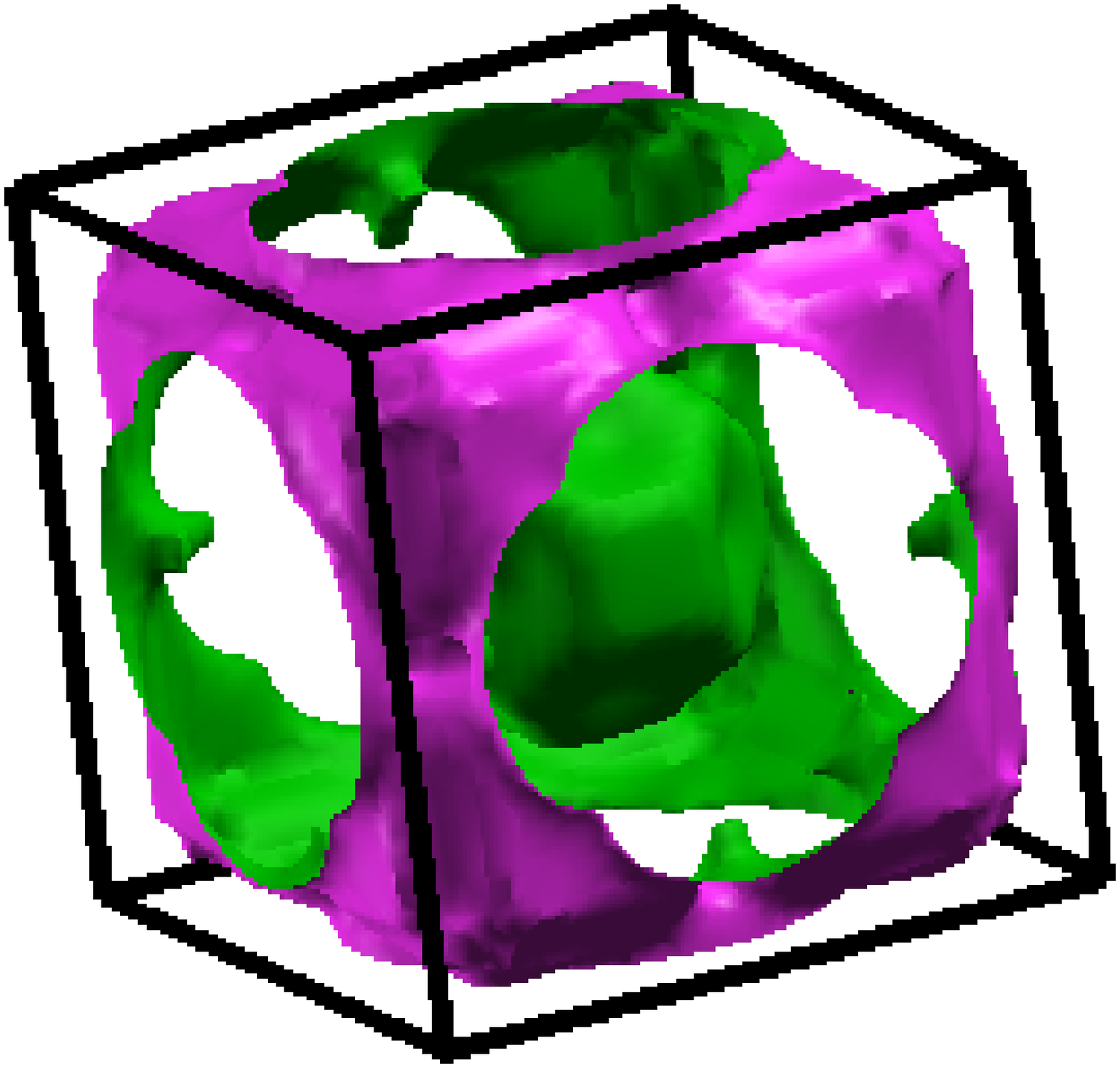}}
\subfigure[]{\includegraphics[width=32mm,height=32mm]{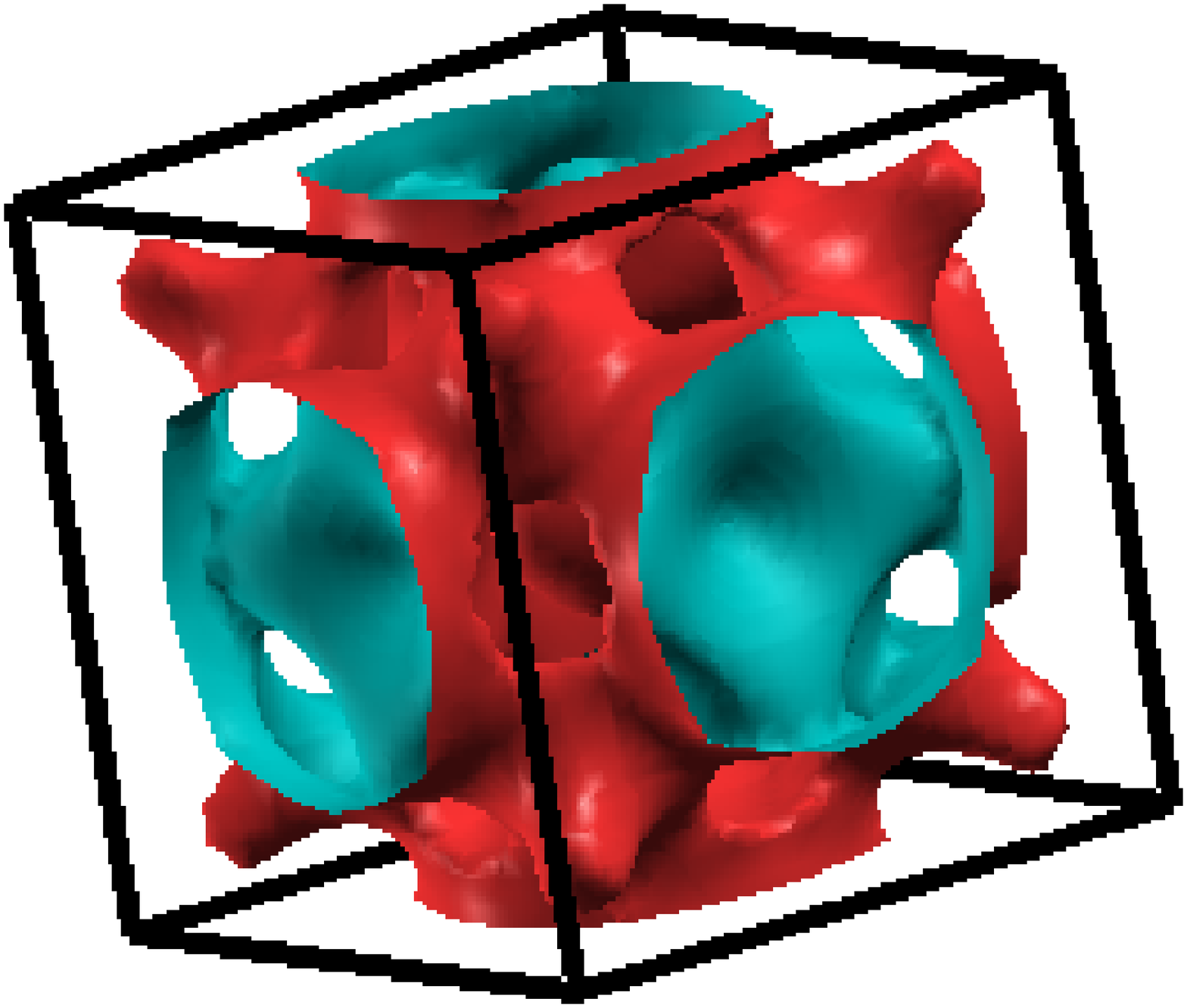}}\\
\subfigure[]{\includegraphics[width=32mm,height=32mm]{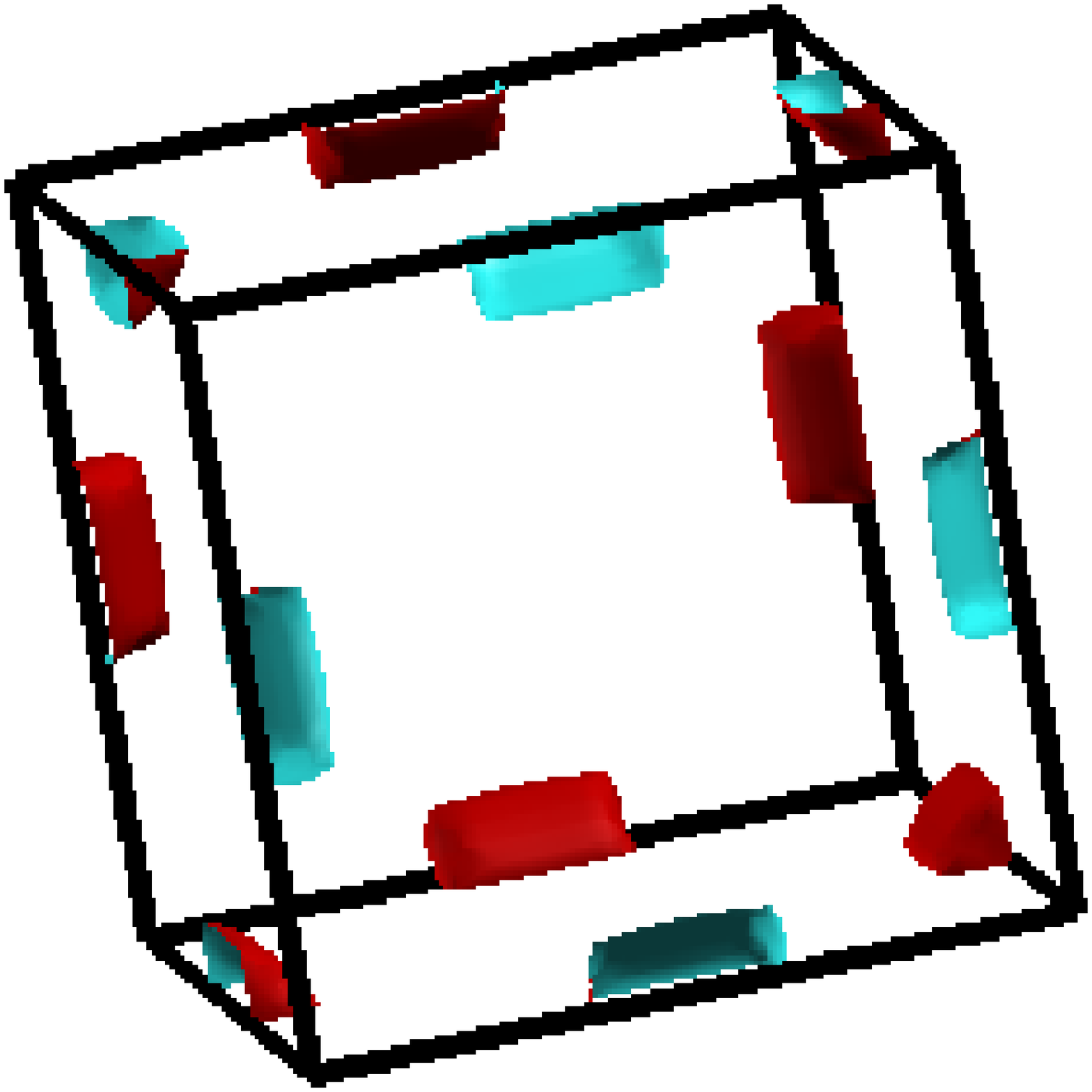}}
\subfigure[]{\includegraphics[width=32mm,height=32mm]{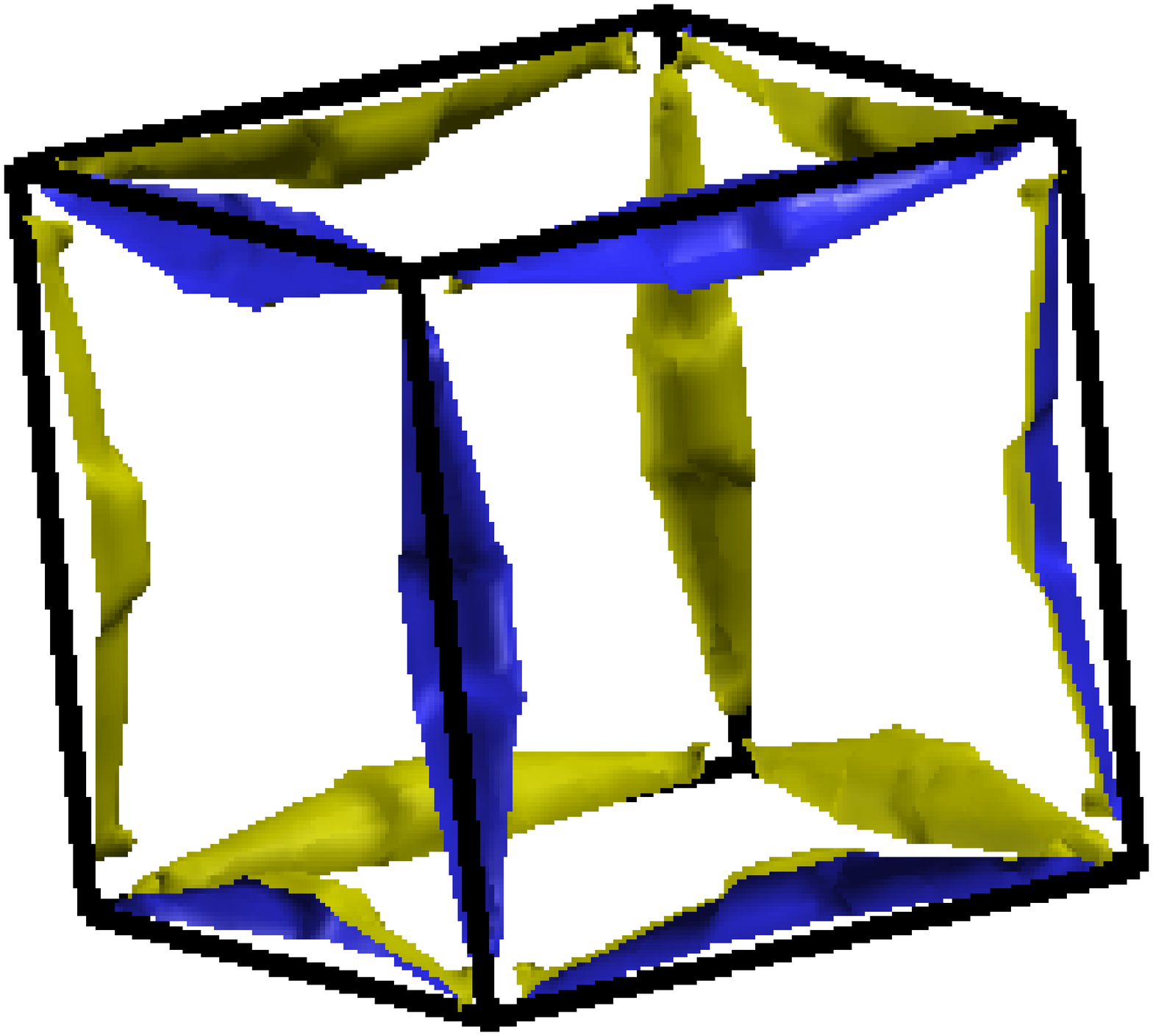}}
\subfigure[]{\includegraphics[width=32mm,height=32mm]{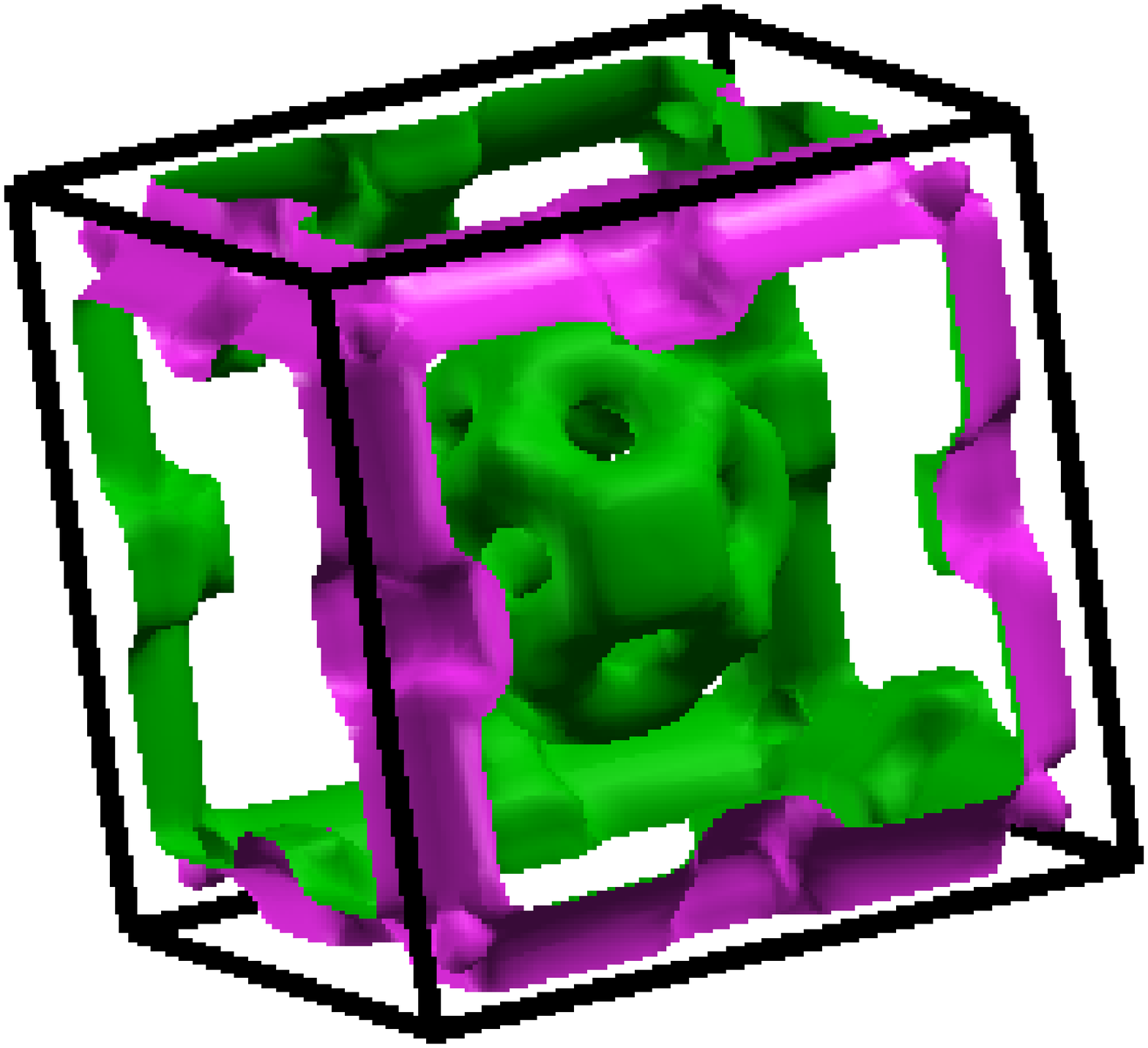}}
\subfigure[]{\includegraphics[width=32mm,height=32mm]{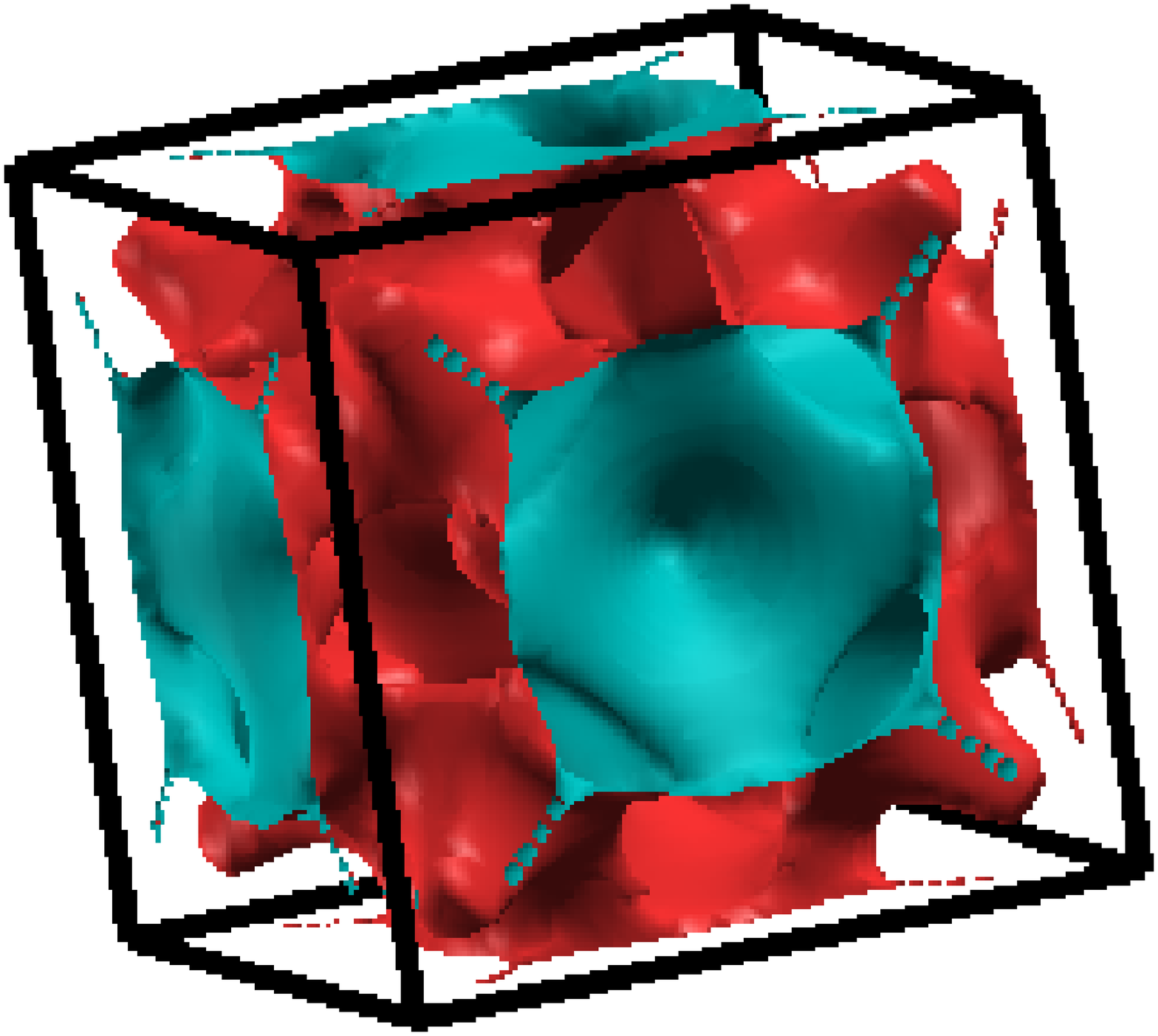}}
\caption{Band structure of Nb$_3$Sn at (a)V/V$_0$ = 1.00, (b)V/V$_0$ = 0.90 (pressure of 25 GPa) and FS for which change in FS is observed at ambient (c), (d), (e), (f) for band no. 51, 52, 53 and 54 and at V/V$_0$ = 0.90 (g), (h), (i)and (j) where change in FS topology is observed.}
\end{center}
\end{figure*}

\clearpage

\begin{figure*}
\begin{center}
\subfigure[]{\includegraphics[width=60mm,height=50mm]{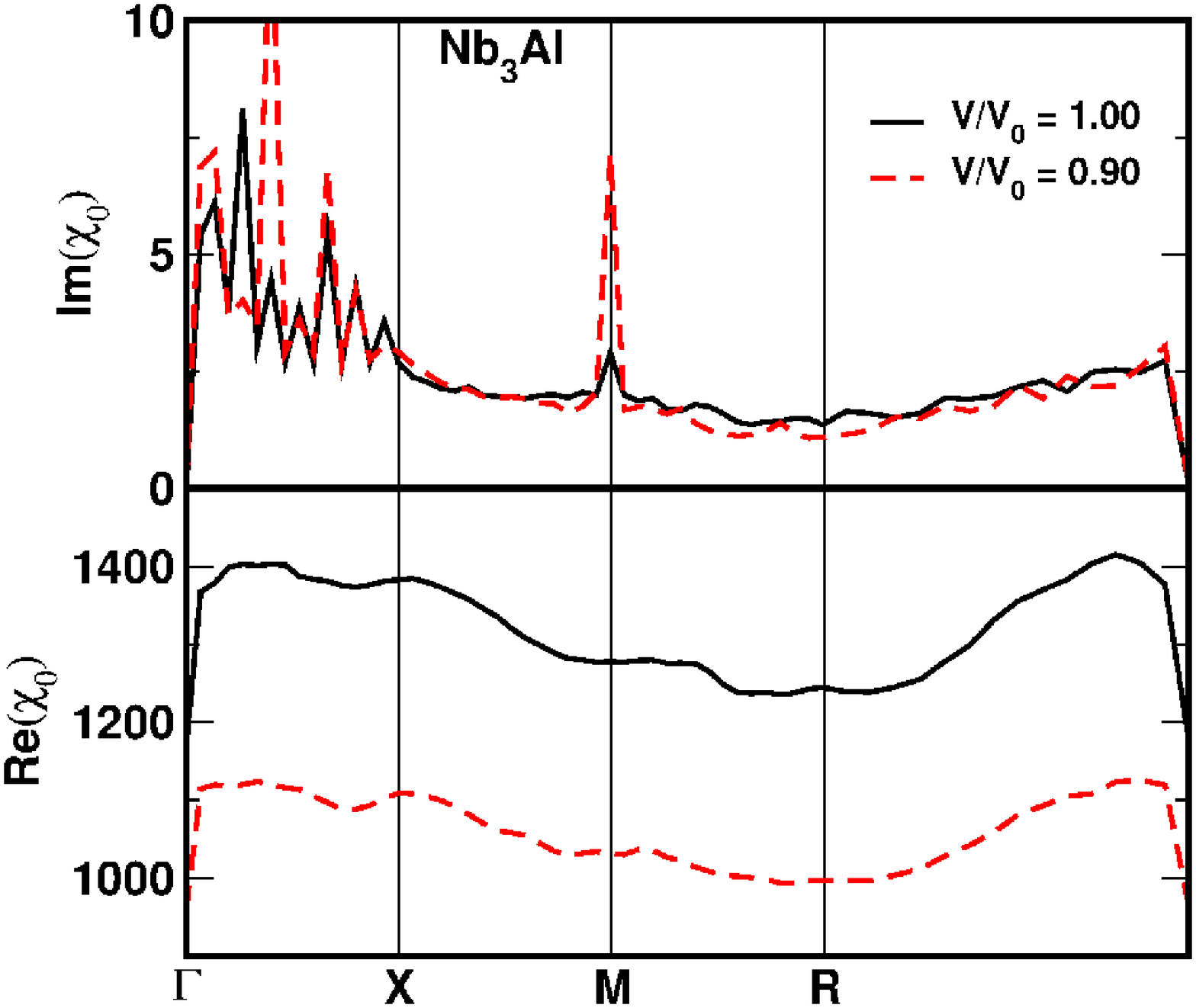}}
\subfigure[]{\includegraphics[width=60mm,height=50mm]{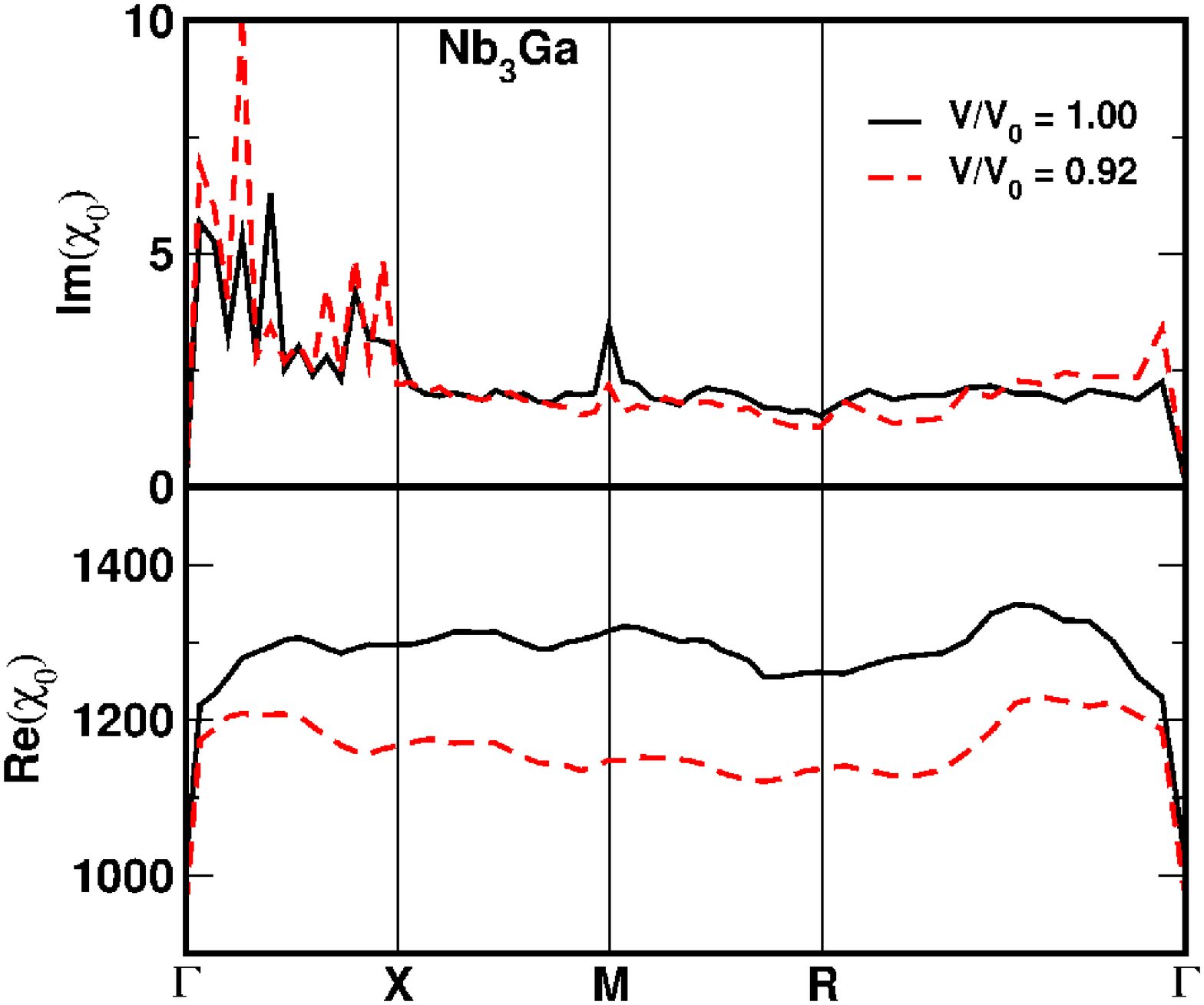}}\\
\subfigure[]{\includegraphics[width=60mm,height=50mm]{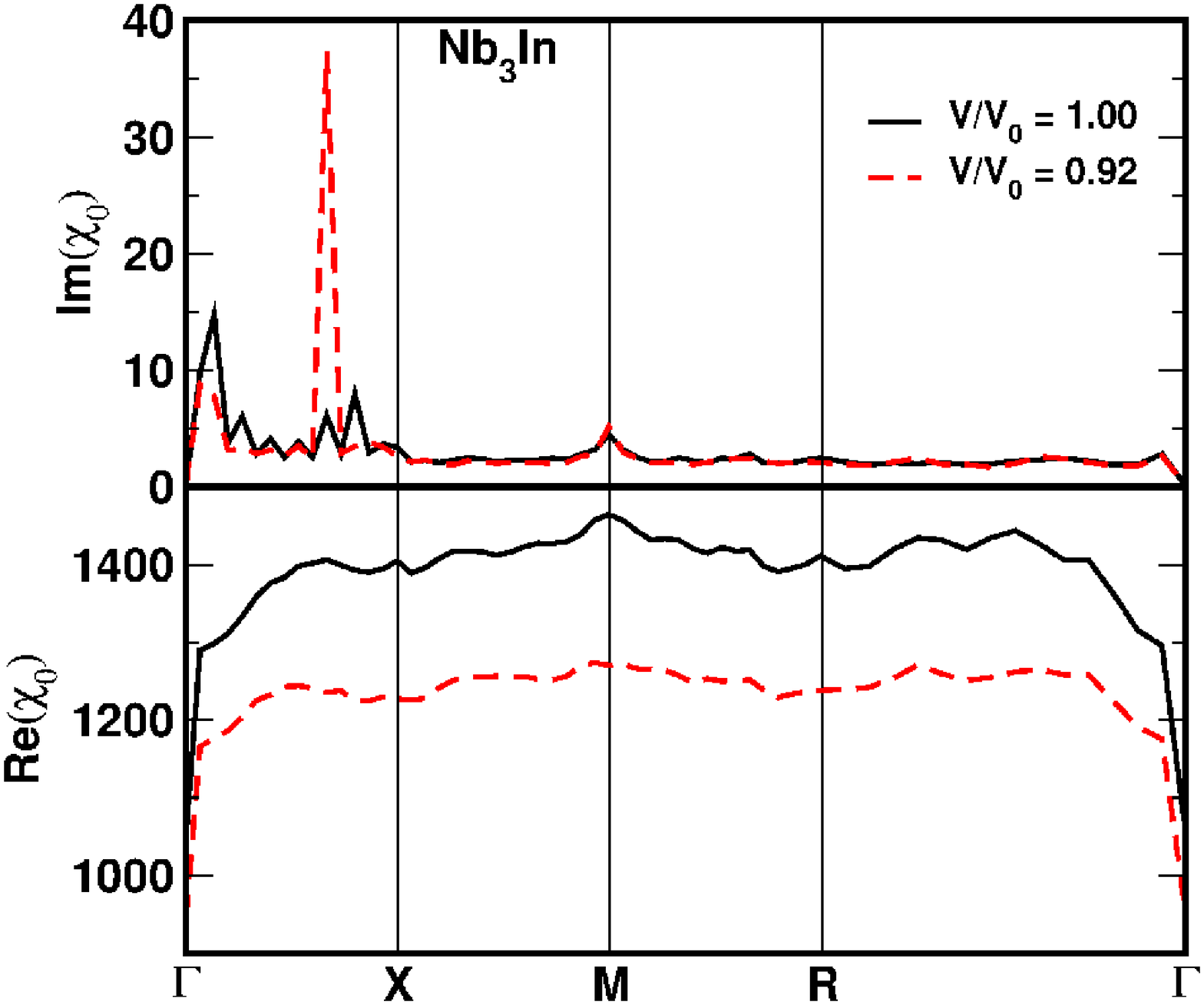}}
\subfigure[]{\includegraphics[width=60mm,height=50mm]{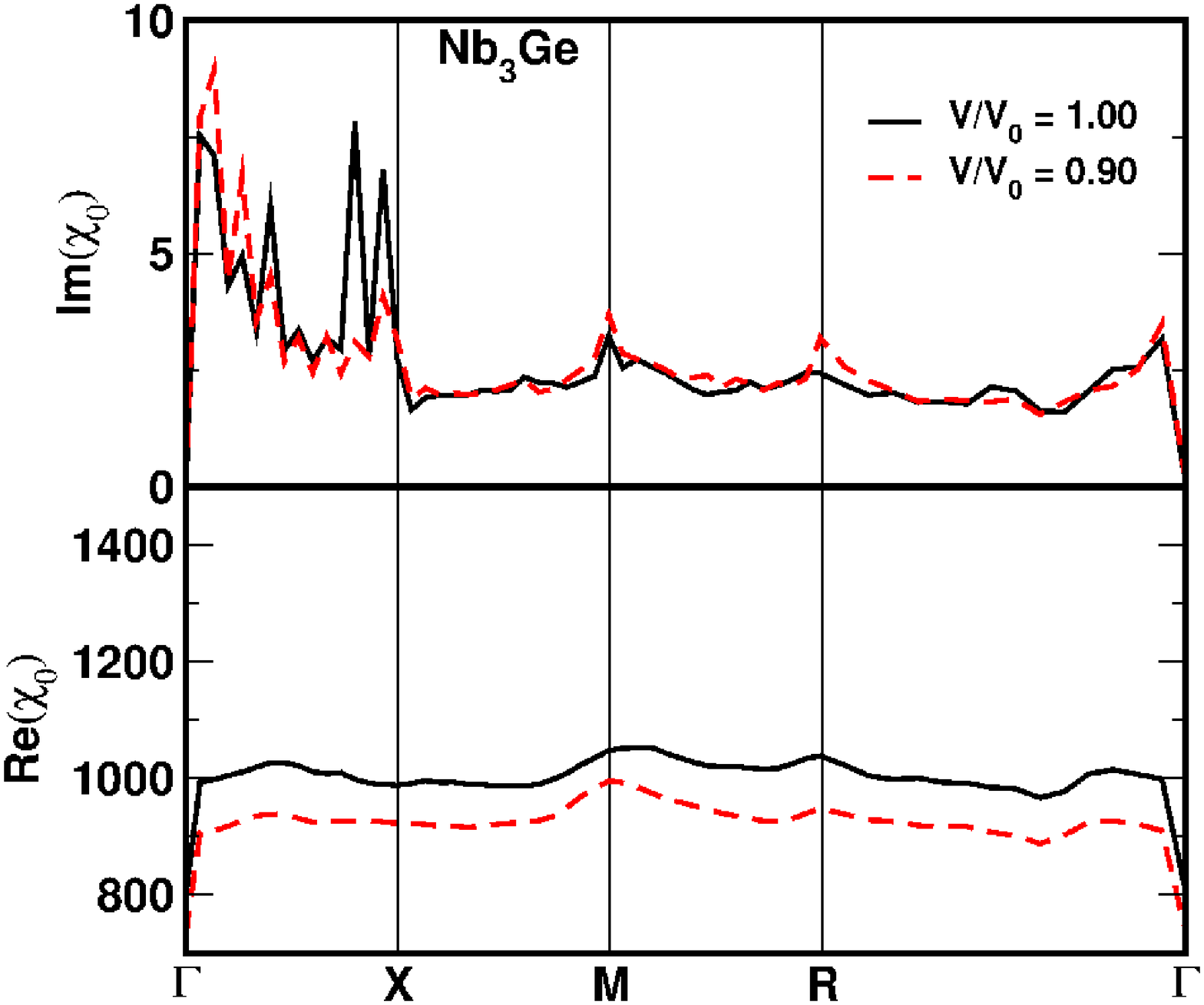}}\\
\subfigure[]{\includegraphics[width=60mm,height=50mm]{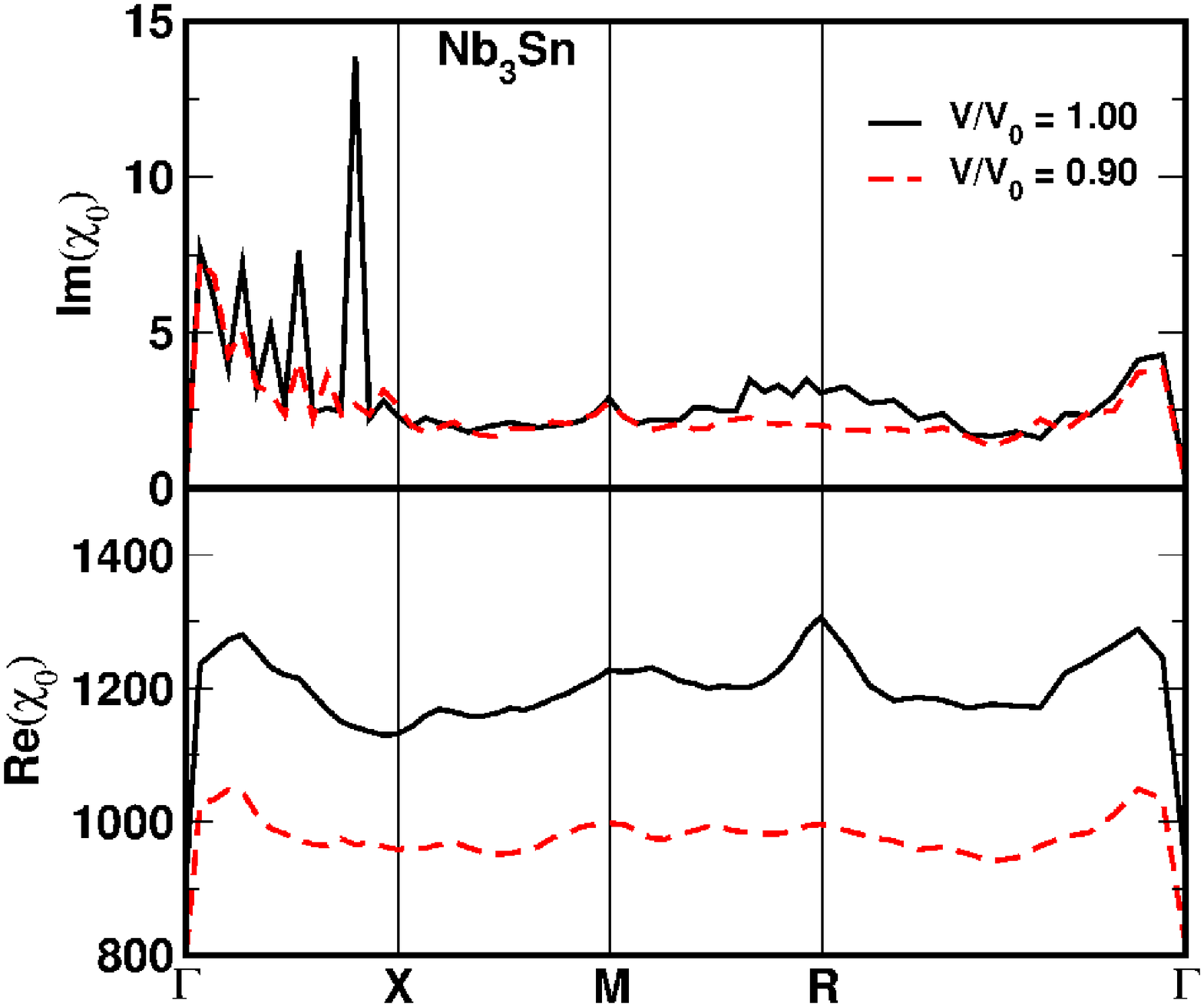}}
\caption{Real and imaginary parts of Lindhard susceptibility plots (a) Nb$_3$Al, (b) Nb$_3$Ga, (c) Nb$_3$Ge, (d) Nb$_3$In and (e) Nb$_3$Sn. }
\end{center}
\end{figure*}

\begin{figure*}
\begin{center}
\includegraphics[width=70mm,height=65mm]{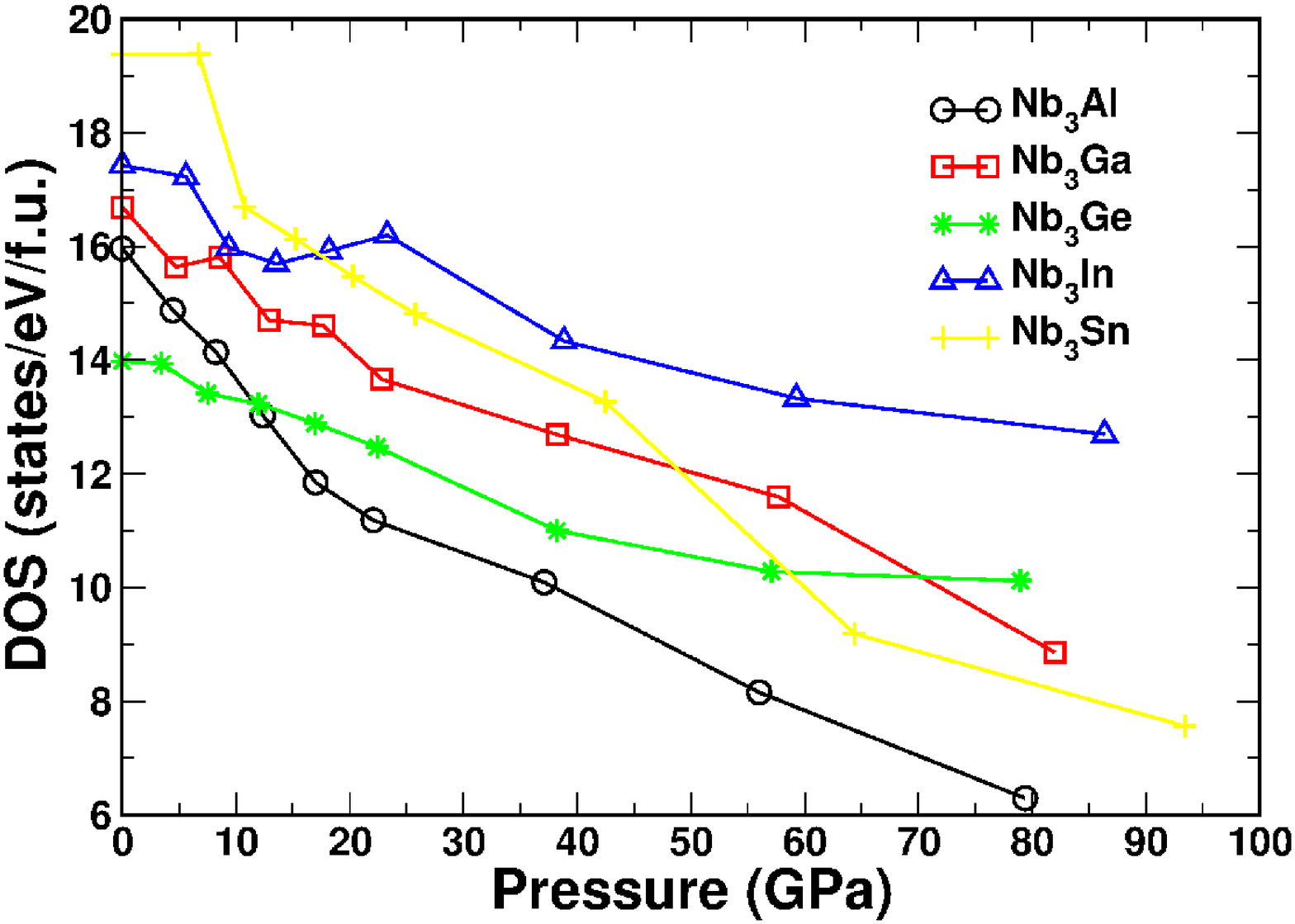}
\caption{Density of states and 
for all the compounds under compression.}
\end{center}
\end{figure*}

\begin{figure*}
\begin{center}
\subfigure[]{\includegraphics[width=60mm,height=60mm]{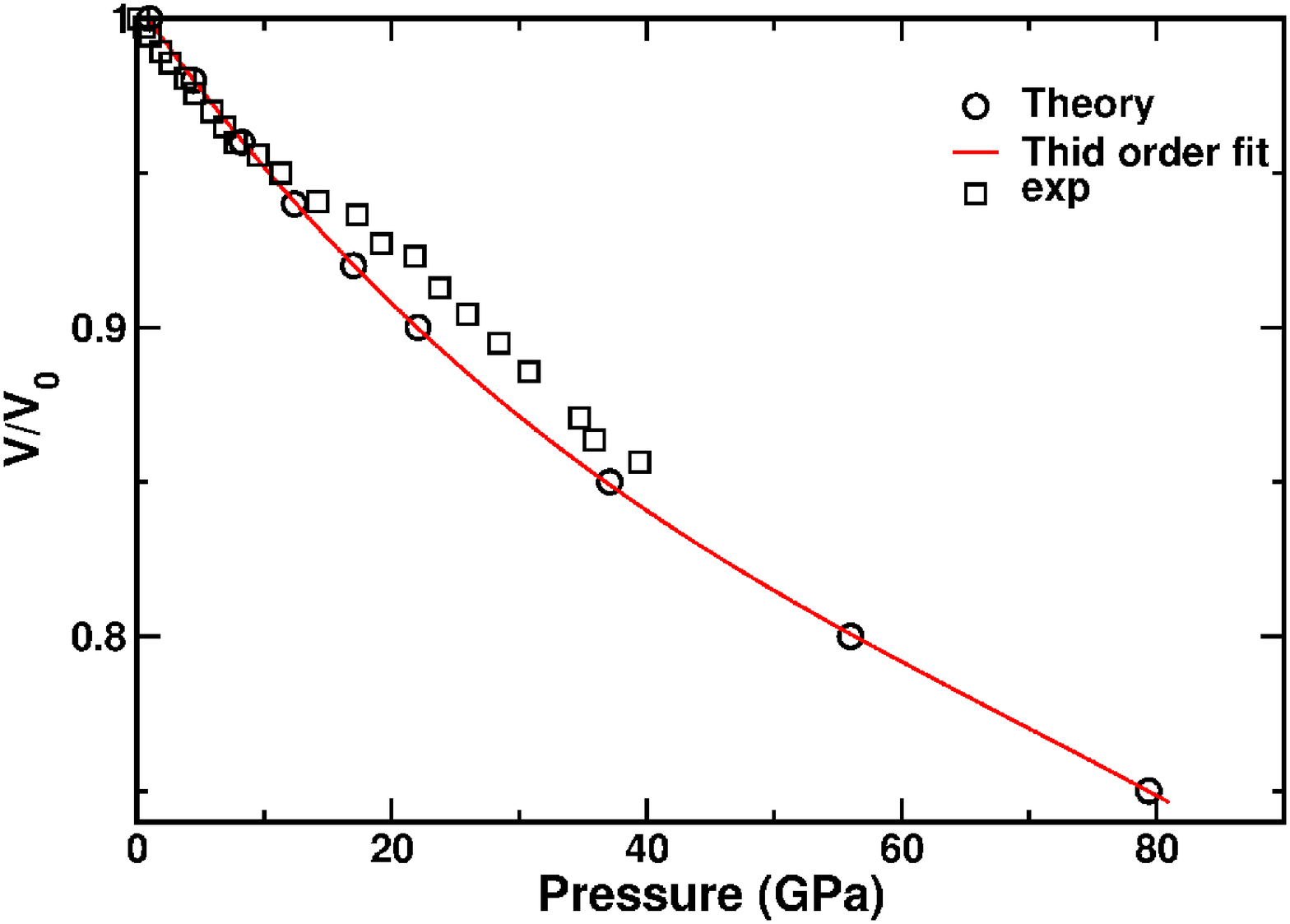}}
\subfigure[]{\includegraphics[width=60mm,height=60mm]{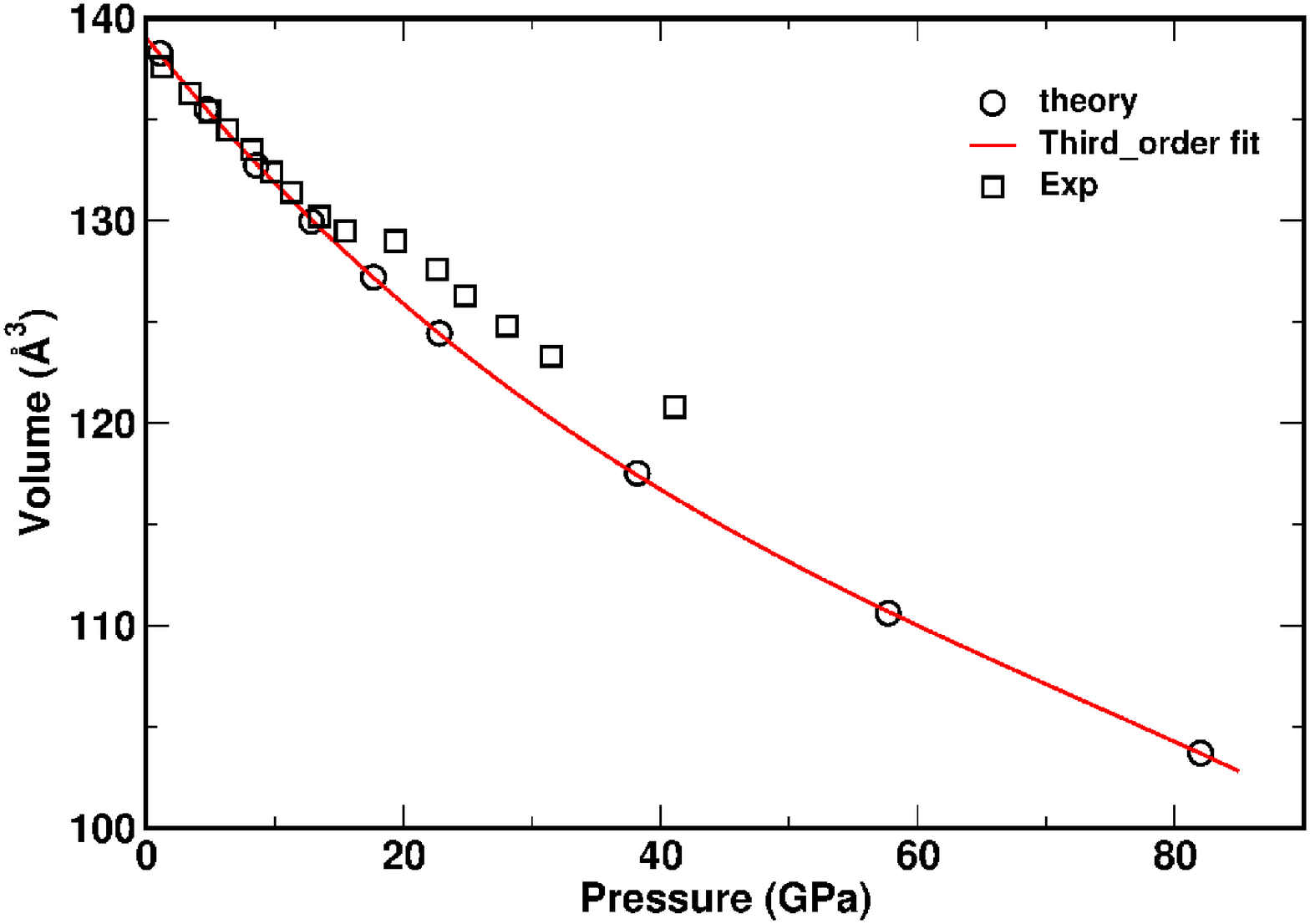}}
\caption{(a)Variation of V/V$_0$ with respect to pressure in Nb$_3$Al and (b) Variation of volume with effect of pressure in Nb$_3$Ga. Here circles are indicated theory values in this work, squares are indicated the experimental values and solid line is a third order fit for the theoretical values.}
\end{center}
\end{figure*}

\begin{figure*}
\begin{center}
\subfigure[]{\includegraphics[width=60mm,height=50mm]{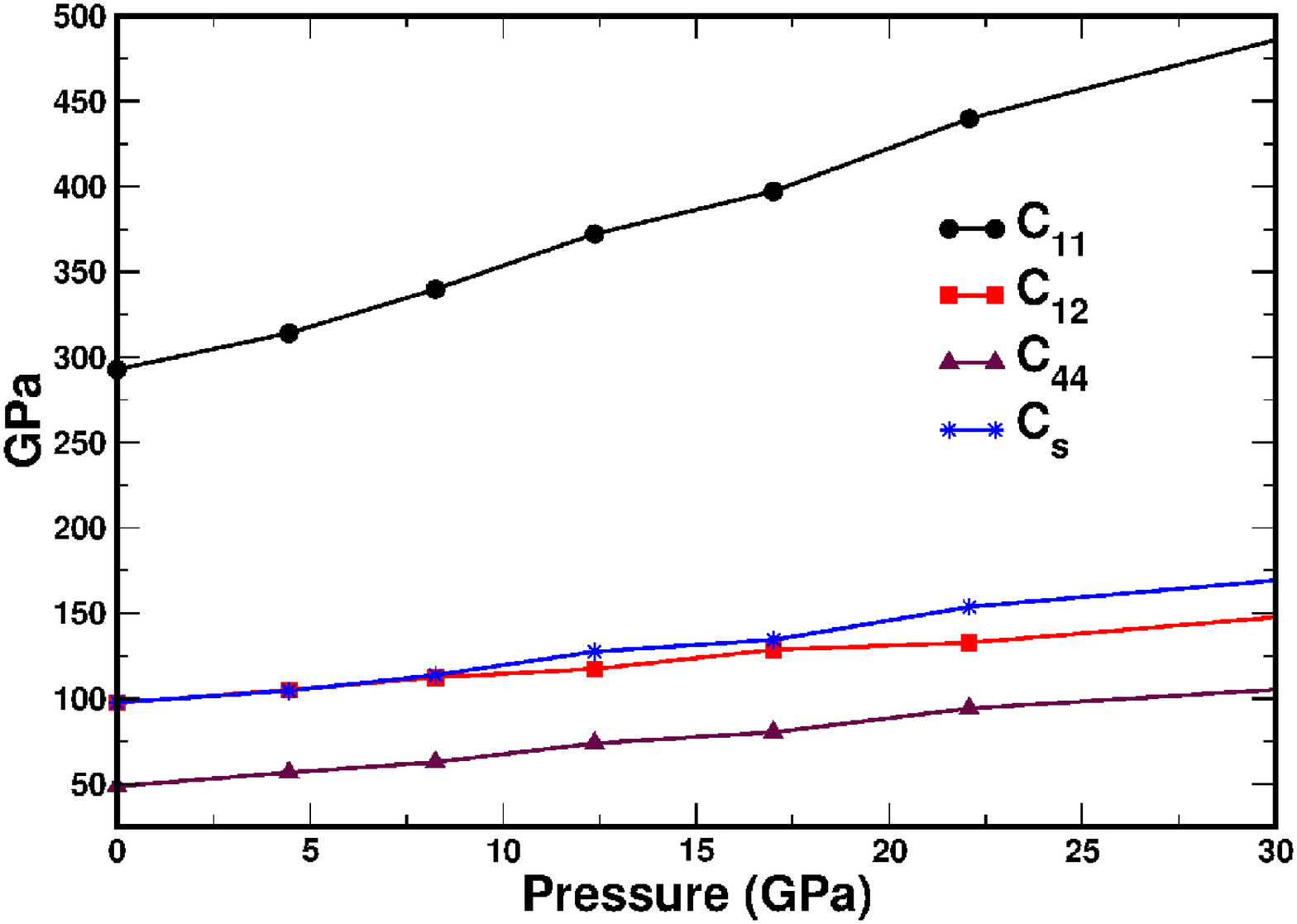}}
\subfigure[]{\includegraphics[width=60mm,height=50mm]{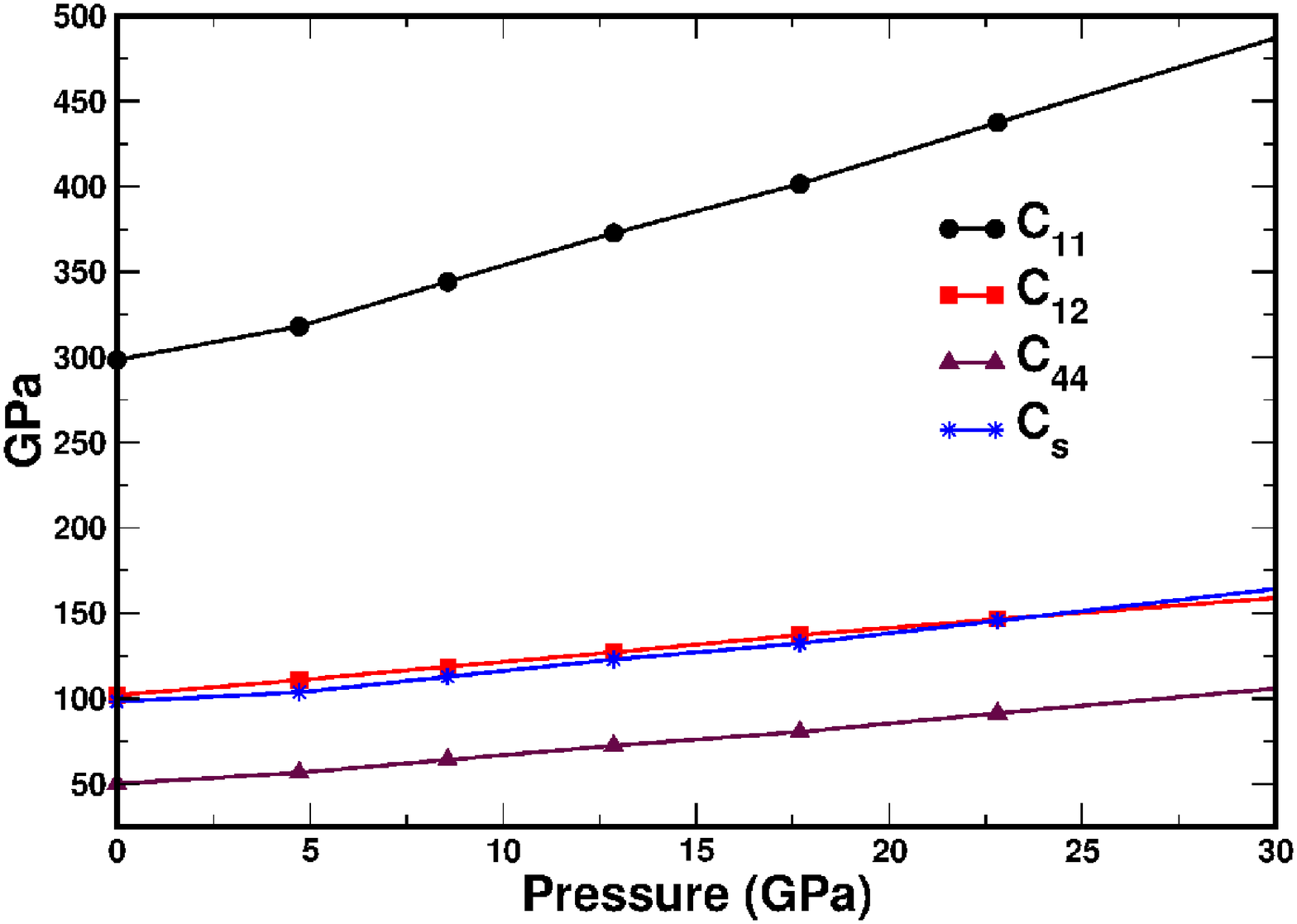}}\\
\subfigure[]{\includegraphics[width=60mm,height=50mm]{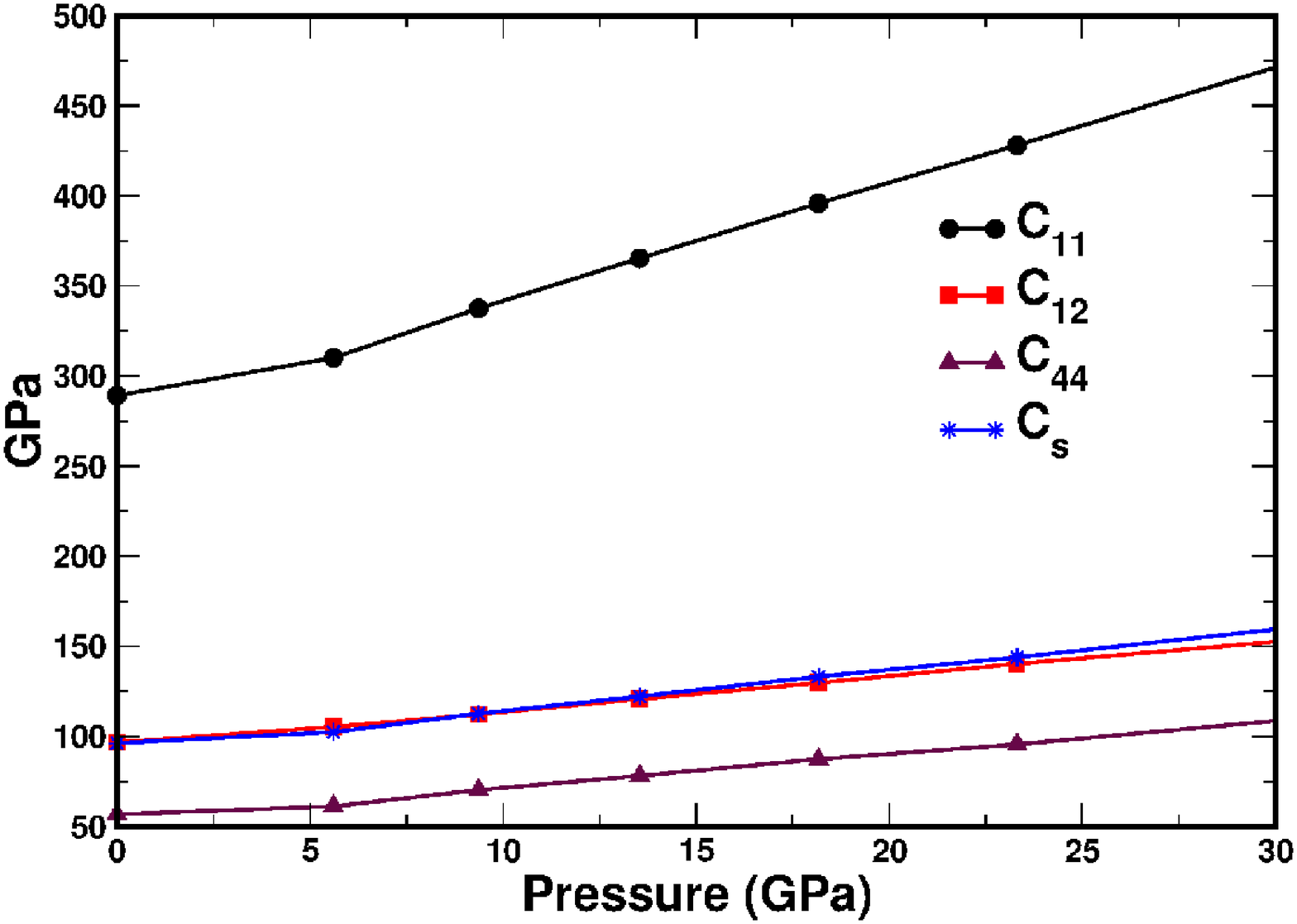}}
\subfigure[]{\includegraphics[width=60mm,height=50mm]{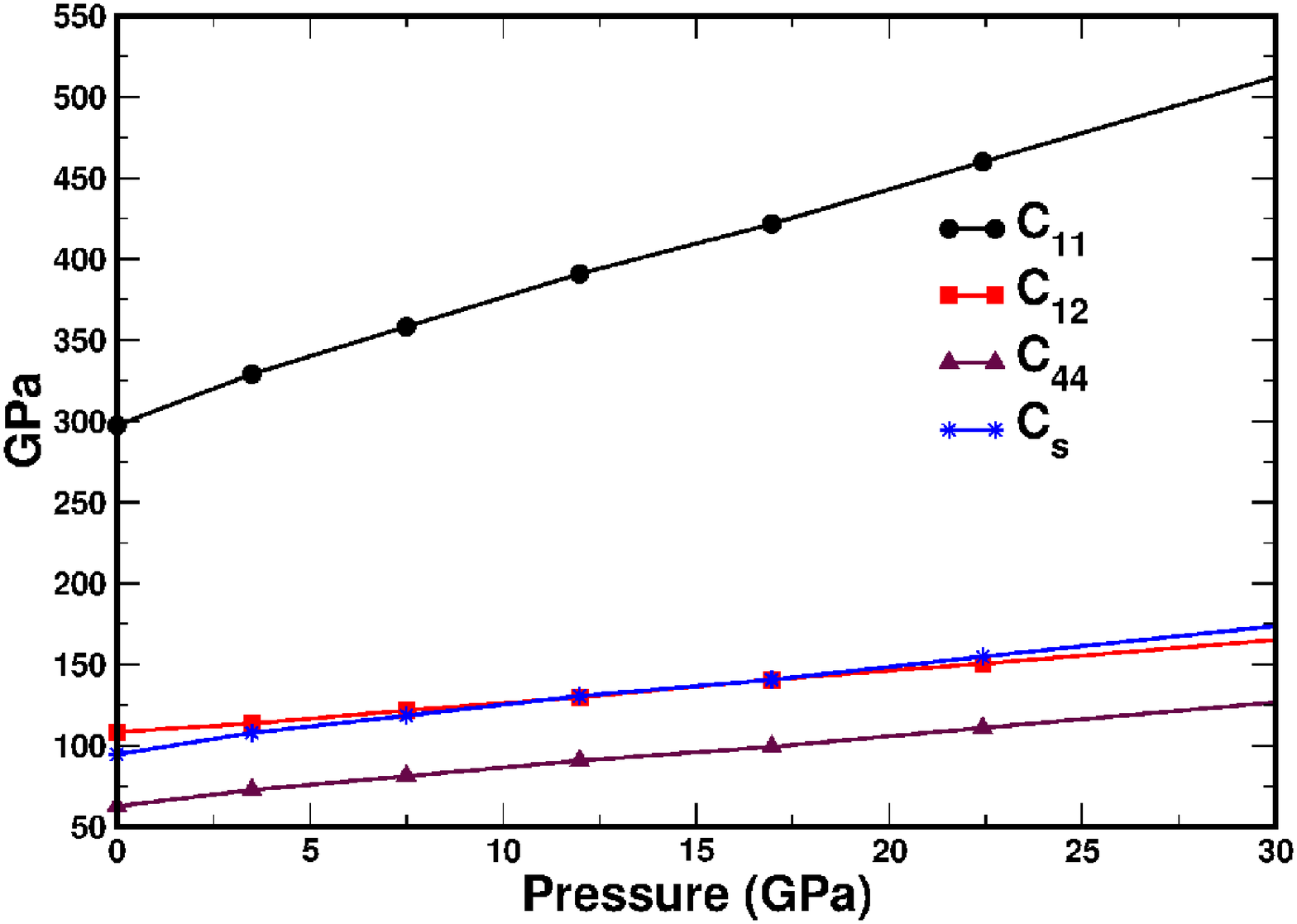}}\\
\subfigure[]{\includegraphics[width=60mm,height=50mm]{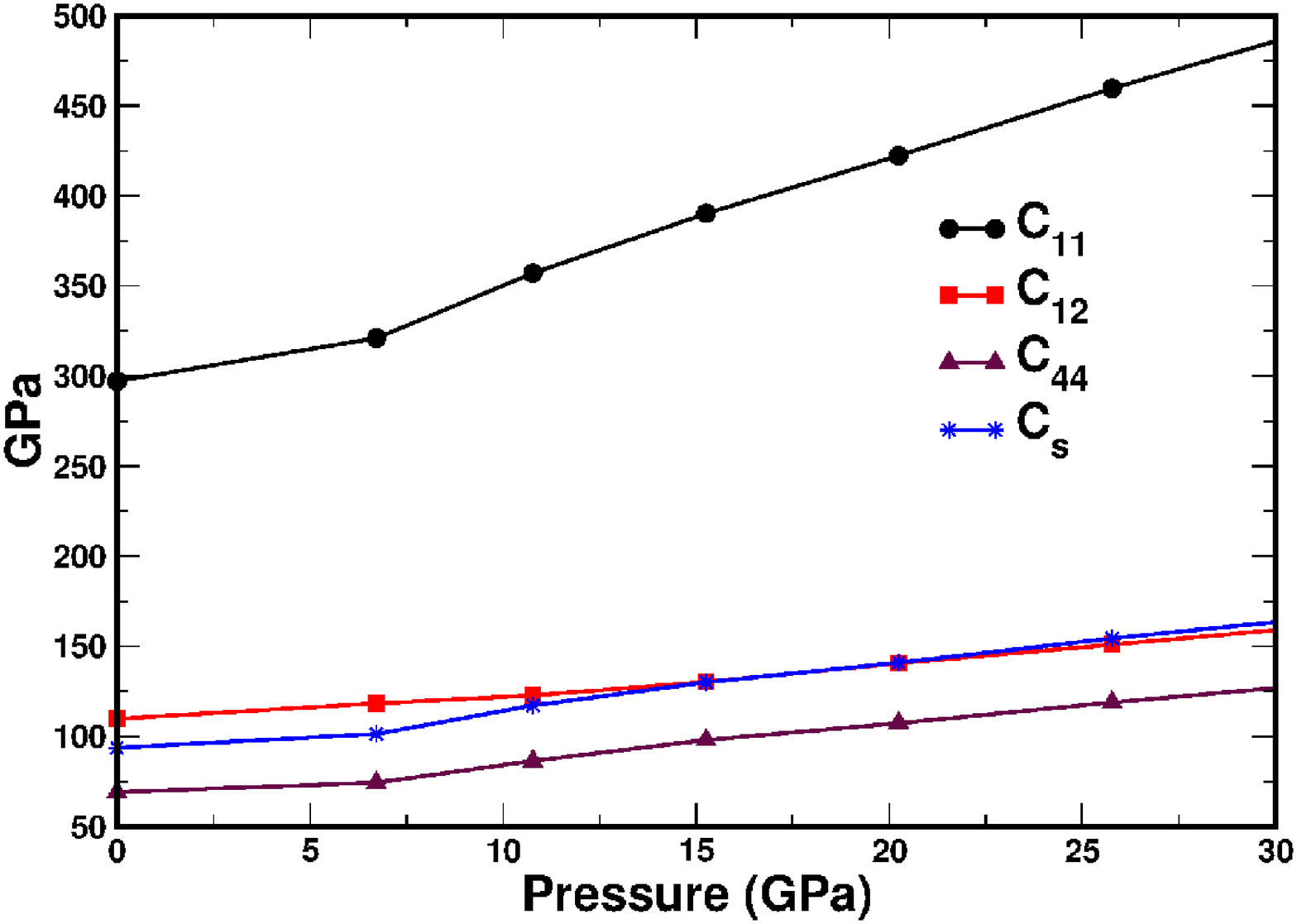}}
\caption{Elastic constants under pressure for (a)Nb$_3$Al, (b)Nb$_3$Ga, (c) Nb$_3$In, (d) Nb$_3$Ge and (e) Nb$_3$Sn.}
\end{center}
\end{figure*}

\end{document}